\documentclass[prd,twocolumn,nofootinbib, superscriptaddress,preprintnumbers]{revtex4-1}
\usepackage{epsfig}
\usepackage{graphicx}
\usepackage{palatino}
\usepackage[english]{babel}
\usepackage{hyphenat}
\usepackage{amsmath}
\DeclareMathOperator\arctanh{arctanh}
\usepackage{amssymb}
\usepackage{mathtools}
\usepackage{mathrsfs}
\usepackage{slashed}
\usepackage{epstopdf}
\usepackage{xcolor}
\usepackage{booktabs}
\definecolor{lcolor}{rgb}{0.,0.0,0.}
\definecolor{citcolor}{rgb}{0,0.,0.5}
\usepackage[breaklinks,colorlinks,urlcolor=blue,citecolor=blue,linkcolor=blue]{hyperref}
\usepackage{multirow}
\usepackage{ltablex}
\newcommand{\beq}{\begin{eqnarray}}
\newcommand{\eeq}{\end{eqnarray}}

\def\dd{{\rm d}}

\newcommand{\bem}{\begin{multline}}
\newcommand{\eem}{\end{multline}}
\newcommand{\beg}{\begin{gather}}
\newcommand{\eeg}{\end{gather}}

\newcommand{\ben}{\begin{eqnarray*}}
\newcommand{\een}{\end{eqnarray*}}

\newcommand{\secn}[1]{Section~1}
\newcommand{\appn}[1]{Appendix~1}

\long\def\comment#1{ }

\def\and{\quad\text{and}\quad}

\newcommand{\rmd}{{\rm d}}

\newcommand{\rme}{{\rm e}}

\begin{document}

\preprint{CERN-TH-2022-164}

\title{Pushing forward jet substructure measurements in heavy-ion collisions}

\author{Daniel Pablos}
 \email{daniel.pablos.alfonso@to.infn.it}
\affiliation{%
 INFN, Sezione di Torino, via Pietro Giuria 1, I-10125 Torino, Italy
}%

\author{Alba Soto-Ontoso}
\email[]{alba.soto.ontoso@cern.ch}
\affiliation{Universit\'e Paris-Saclay, CNRS, CEA, Institut de physique th\'eorique, 91191, Gif-sur-Yvette, France}
\affiliation{Theoretical Physics Department, CERN, 1211 Geneva 23, Switzerland}

\setlength{\belowcaptionskip}{-10pt}

\renewcommand{\arraystretch}{1.4}

\begin{abstract}
Energetic jets that traverse the quark-gluon plasma created in heavy-ion collisions serve as excellent probes to study this new state of deconfined QCD matter.
Presently, however, our ability to achieve a crisp theoretical interpretation of the crescent number of jet observables measured in experiments is hampered by the presence of selection biases.
The aim of this work is to minimise those selection biases associated to the modification of the quark- vs. gluon-initiated jet fraction in order to assess the presence of other medium-induced effects, namely color decoherence, by exploring the rapidity dependence of jet substructure observables. So far, all jet substructure measurements at mid-rapidity have shown that heavy-ion jets are narrower than vacuum jets. We show both analytically and with Monte Carlo simulations that if the narrowing effect persists at forward rapidities, where the quark-initiated jet fraction is greatly increased, this could serve as an unambiguous experimental observation of color decoherence dynamics in heavy-ion collisions. 
\end{abstract}

\maketitle


\section{Introduction}
Ultra-relativistic heavy-ion collisions have succeeded in recreating the extreme temperature and pressure conditions that our Universe experienced during the first microseconds after the Big Bang. Unraveling the microscopic properties of the medium that permeated our Universe during this epoch, namely the quark-gluon plasma (QGP), is one of the long-standing questions of particle physics~\cite{Baym:2016wox,Busza:2018rrf}. 
A widely used approach to this challenge is to study the modification of high-momentum particles, or jets, when traversing the QGP, very much like in the Rutherford experiment. Data recorded during the last two decades at both RHIC~\cite{Adcox:2001jp,Adler:2002xw,STAR:2020xiv} and the LHC~\cite{Adam:2015ewa,CMS:2016uxf,ATLAS:2018gwx,CMS:2021vui} has confirmed that the interaction between jets and the QGP leads to an overall depletion of the jet yield at high-$p_t$. Theoretically, this suppression, commonly known as \textit{jet quenching}~\cite{dEnterria:2009xfs,Majumder:2010qh,Mehtar-Tani:2013pia}, is understood as a result of the wide-angle nature of medium-induced emissions which end up being radiated outside of the jet cone and thus lead to a net energy loss.

Aiming at a more detailed picture of the multi-scale evolution of jets in the presence of a thermal medium, experimental measurements in the last five years have explored jet substructure observables such as the momentum sharing fraction or the opening angle of a pair of \textit{hard} subjets~\cite{CMS:2017qlm,ALICE:2019ykw,ALargeIonColliderExperiment:2021mqf}. We refer the reader to~\cite{Cunqueiro:2021wls} for a comprehensive review of the latest jet measurements. 
These observables can be designed such that the perturbative part of the radiation phase-space dominates and are, consequently, under better theoretical control than global ones. 
Up to now, a varied set of jet substructure measurements has revealed an overall narrowing of the jet core with respect to the vacuum baseline~\cite{ALargeIonColliderExperiment:2021mqf}. However, the current experimental precision is not high enough to discriminate between disparate theoretical models. In this paper, we develop a strategy that shall allow future measurements to identify the actual physical mechanism behind the observed narrowing effect.

The first model, introduced in Ref.~\cite{Spousta:2015fca}, argues that the experimental trend is driven by a larger number of quark-initiated jets, known to be more collimated, after the $p_t$ selection cut in the $Pb+Pb$ sample with respect to $p+p$. An enhanced quark fraction in $Pb+Pb$ could originate from a combination of the color charge dependence of jet quenching and the jet spectrum. That is, since gluon jets radiate more, they will lose more energy and, as a consequence of the steeply falling spectrum, will not pass the jet $p_t$ selection cut. A critical feature of this model is that if one fixes the color charge of the jet initiator, no modifications are expected with respect to vacuum jet evolution (modulo any potential $p_t$ dependence of the observable itself). The natural question is how different the $q/g$ fractions need to be in order to quantitatively describe the data, or equivalently, how much stronger is the quenching that gluon jets experience. The authors of Ref.~\cite{Ringer:2019rfk} achieved a quantitative description of jet substructure observables with a factor of 4 more quark jets in $Pb+Pb$ with respect to $p+p$. Since this number was extracted via a global fit to jet spectrum data~\cite{Qiu:2019sfj}, it is agnostic to the dynamics of energy loss.\footnote{This quark fraction is in tension with the one extracted in Ref.~\cite{Brewer:2020och} that corresponds to $1.5$ increase (see Fig. 1, panels (c) and (f)), although an apples-to-apples comparison is not possible due to the different jet selections used in those studies. Experimentally, no sizeable modification of the quark and gluon fractions has been observed~\cite{CMS:2020plq,Li:2019dre}.} Thus, the physical mechanism that would lead to such a dramatic quenching of gluon jets, far larger than that expected from Casimir scaling~\footnote{Casimir scaling is violated, even in vacuum, beyond leading order as was shown in Ref.~\cite{Apolinario:2020nyw}.}, remains to be settled. Throughout the rest of this paper, we will refer to this hypothesis as `modified $q/g$ fraction model' and we will elaborate more on it in Secs.~\ref{sec:mod-qg} and \ref{sec:toy-template}. 

An alternative explanation to the narrowing effect relies on the existence of a critical resolution angle of the QGP. This angular scale, defined as $\theta_c=2/\sqrt{\hat q L^3}$ with $\hat q$ being the quenching parameter and $L$ the medium length, naturally emerges when considering the soft radiation pattern of an antenna in the multiple soft scattering approximation \cite{Mehtar-Tani:2010ebp,Mehtar-Tani:2011hma,Casalderrey-Solana:2011ule,Mehtar-Tani:2011vlz,Casalderrey-Solana:2012evi,Apolinario:2014csa}, and splits the radiation phase-space into resolved and unresolved emissions.\footnote{A resolution scale also appears when calculating the gluon emission pattern of an antenna using the opacity expansion formalism~\cite{Wiedemann:2000za,Gyulassy:2000fs,Sievert:2018imd}, particularly relevant for thin media~\cite{Mehtar-Tani:2011lic,Casalderrey-Solana:2015bww}.}
In short, if the opening angle of a vacuum-like splitting is larger than $\theta_c$, its two prongs behave as independent emitters of medium-induced gluons. On the contrary, collinear branchings with $\theta<\theta_c$ are not resolved by the medium and thus lose energy coherently as an individual color charge. Therefore, jets with $\theta>\theta_c$ are more quenched, leading to an overall narrowing of the jet sample. Two jet quenching Monte Carlos that incorporate some notion of color coherence, namely \texttt{JetMed}~\cite{Caucal:2018dla,Caucal:2019uvr} and the Hybrid Strong/Weak Coupling Model~\cite{Casalderrey-Solana:2014bpa,Hulcher:2017cpt,Casalderrey-Solana:2019ubu}, are able to quantitatively describe jet substructure data. Naturally, these two models are also sensitive to the different degree of quenching of quark and gluon jets. However, a purely coherent description of energy loss in these models is not sufficient to match the experimental data and thus a resolution criterion in terms of a critical angle/length is required. 

This paper addresses the question on how to experimentally disentangle between the `modified $q/g$ fraction' and `color decoherence' models. Our strategy is to explore the rapidity dependence of jet substructure observables. For simplicity, we focus on the $k_t$-distribution of the hardest splitting in a jet, but our conclusions apply to any jet substructure measurement.\footnote{Since $k_t=z\theta$ and the $z$-distribution is barely modified in heavy-ion collisions, the observed narrowing in terms of $\theta$ is directly translated into a shift towards smaller $k_t$. Preliminary experimental results on $k_t$ can be found in Ref.~\cite{Bossi:2022fpc}.} The idea is based on a simple observation: increasing the jet rapidity enhances the fraction of quark-initiated jets. For a fixed color charge of the jet initiator, the two models under study lead to dramatically different predictions for the $k_t$-distribution. On the one hand, as we have already anticipated, the `modified $q/g$ fraction' heavy-ion result would approach the vacuum one when moving to forward rapidities, since the ensemble is dominated by quark-initiated jets in both collision systems. 
In contrast, if the medium is able to resolve the substructure fluctuations developed during the DGLAP evolution of the jet, the $k_t$-distribution would differ from the $p+p$ baseline for every rapidity bin, as both quark- and gluon-initiated jets feature wide ($\theta>\theta_c$) and narrow ($\theta<\theta_c$) configurations. 

In this work, we study the rapidity dependence of the leading-$k_t$ distribution up to $|y|=4.5$. Experimentally, jet substructure measurements at such forward rapidities are rather challenging and require dedicated instrumentation. For this reason, we deem necessary to address the feasibility of the proposed measurements, both with the current data and in the upcoming high-luminosity phase of the LHC (HL-LHC). The ATLAS collaboration has pioneered the study of the rapidity dependence of jet quenching by measuring both the inclusive jet spectrum~\cite{ATLAS:2018gwx} and the fragmentation function~\cite{ATLAS:2018bvp} at $|y|<2.8$ and $|y|<2.1$, respectively. CMS has explored even more forward kinematics and measured the jet cross-section in the pseudorapidity region $-6.6 <\eta<-5.2$ using CASTOR~\cite{CMS:2018yhi,CMS:2020ldm}, but only in $p+Pb$ collisions. Similarly, the LHCb detector allows for the reconstruction of charged and neutral particles in the very forward rapidity region ($2<|y|<4.5$) but, to this date, heavy-ion measurements were limited to peripheral collisions, where medium effects are expected to be reduced. It is then possible to measure the leading-$k_t$ distribution in both $p+p$ and $Pb+Pb$ collisions with the current technology of both ATLAS and CMS using charged, high-$p_t$ jets up to $|y|=2.5$ with data recorded during Runs 2 and 3. 

As we show below, this rapidity interval could be enough, for jets at sufficiently high $p_t$, to disentangle between the theoretical models studied in this paper. However, more stringent constraints on the two confronting pictures can be obtained by pushing the measurements to even larger rapidities. This can be achieved in the future HL-LHC, where an integrated luminosity of about $\mathcal{L}=$10~$\rm nb^{-1}$ in $Pb+Pb$ collisions at a centre-of-mass of energy of 5.5 TeV per nucleon pair will be delivered~\cite{Citron:2018lsq}. A further increase in the integrated luminosity is expected in case the heavy-ion program is extended up to the end of Run~6, as recently proposed by the ALICE3 experiment~\cite{ALICE:2803563}. In fact, this detector would be ideal for this study since it is designed to cover 8 units in rapidity with very good tracking and $p_t$-resolution of charged particles~\cite{ALICE:2803563}. Upgrades in both ATLAS and CMS will also extend their rapidity coverage~\cite{CMS:2017jpq,ATLAS:2021yvc} and LHCb plans to extend its centrality coverage to semi-central collisions in Run~3 and central collisions in Run~4~\cite{Citron:2018lsq}. To sum up, we consider two experimental settings: 
(i) `current LHC', which includes measurements up to $|y|<2.5$, and (ii) `future LHC', where we study jet rapidities as high as $|y|=4.5$.

\begin{figure*}[t!]
    \centering
    \includegraphics[width=0.49\textwidth]{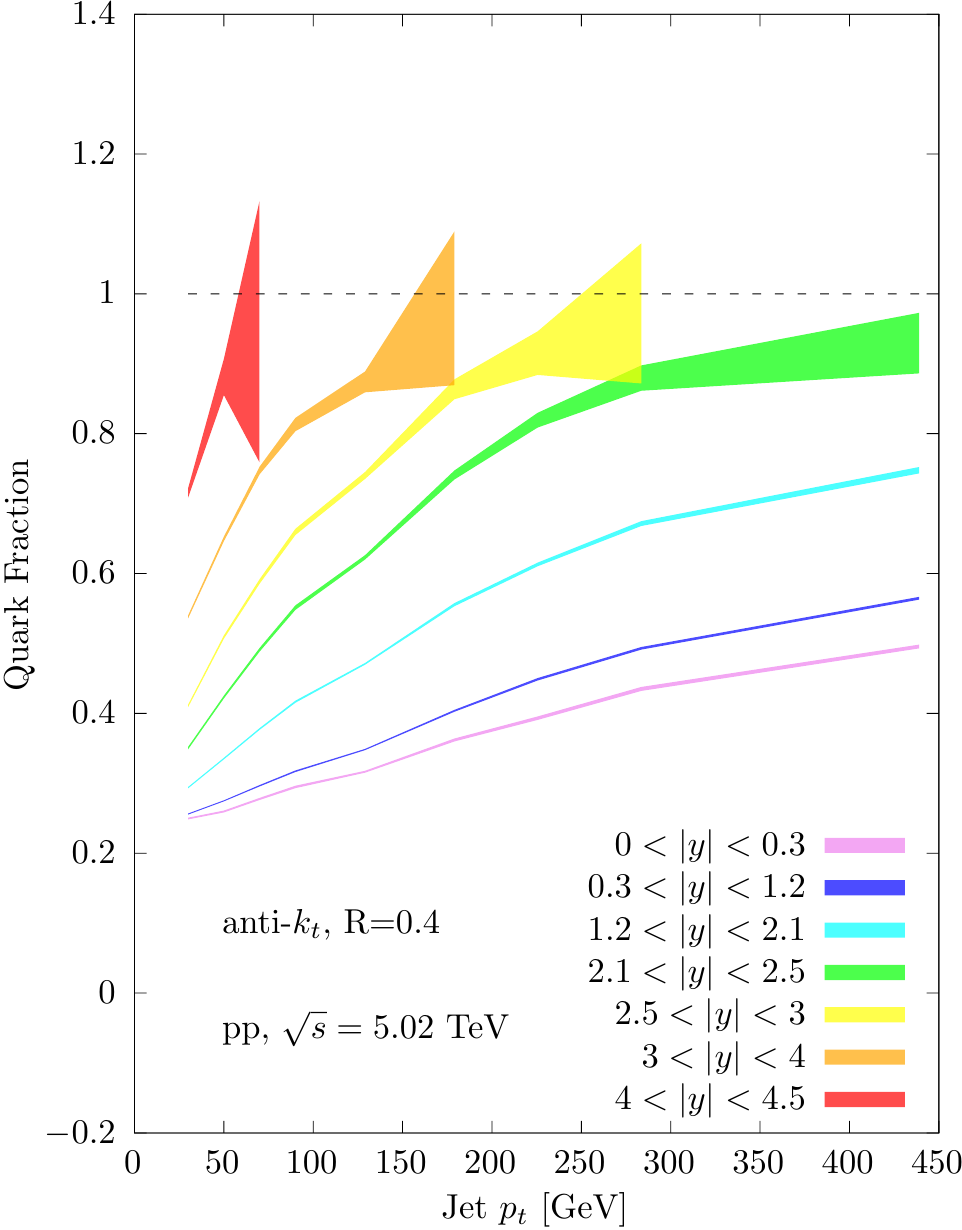}
    \includegraphics[width=0.49\textwidth]{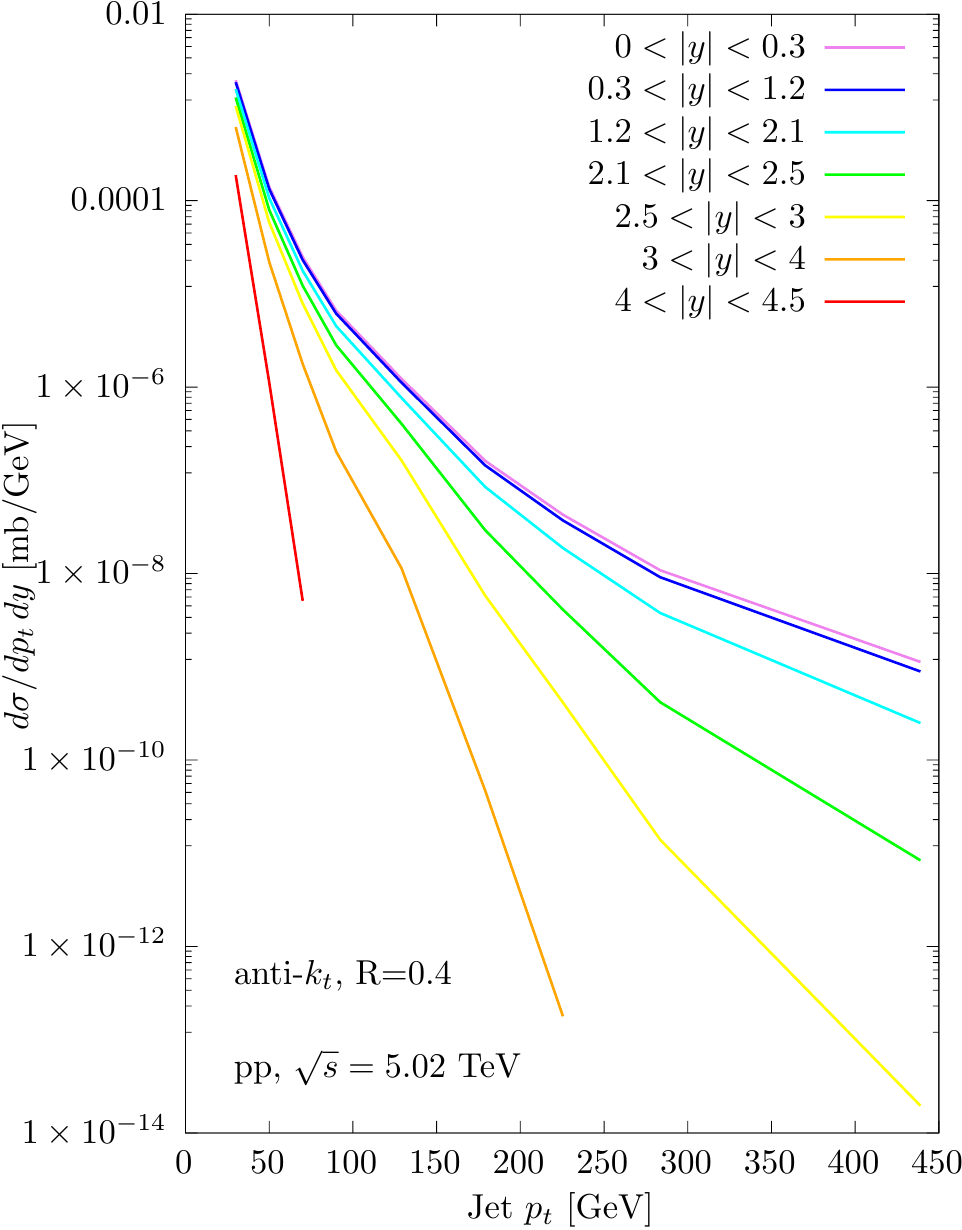}
    \caption{Left: Quark-initiated jet fraction as a function of the jet $p_t$ for different intervals of jet rapidity $y$. Right: Same as left panel but for the inclusive jet spectrum.}
    \label{fig:qfracpp}
\end{figure*}

The rest of the paper is organised as follows. In Section~\ref{sec:vacuum} we consider dijet events in $p+p$ collisions and study the rapidity dependence of (i) quark and gluon fractions, (ii) the jet $p_t$ spectrum, and (iii) the leading-$k_t$ distribution using \texttt{Pythia}~\cite{Bierlich:2022pfr}. We also provide an analytic estimate of the vacuum $k_t$ distribution. Next, in Sec.~\ref{sec:medium} we explore the modification of all these observables when including medium effects, both analytically (Sec.~\ref{sec:analytics}) and with Monte Carlo simulations (Secs.~\ref{sec:hybrid}, \ref{sec:toy-template}). We calculate the medium-modified $k_t$ distribution within two different semi-analytic approaches. One model is based on the modified $q/g$ fraction picture and the results are displayed in Sec.~\ref{sec:mod-qg}. The second model, presented in Sec.~\ref{sec:color-dec}, incorporates the dynamics of color decoherence, on a similar fashion as Ref.~\cite{Caucal:2021cfb}. 
For the numerical results we pursue two paths. In Sec.~\ref{sec:hybrid}, we use the \texttt{Hybrid} Monte Carlo model to study the rapidity dependence of jet suppression and to predict the rapidity dependence of the leading-$k_t$ distribution. Next, in Sec.~\ref{sec:toy-template}, we present an implementation of the modified $q/g$ fraction model in which \texttt{Pythia} is used to obtain the vacuum distributions that are then 
combined within the logic of a fully coherent energy loss model. We conclude with a summary of the main results of this manuscript in Sec.~\ref{sec:conclusions}. 

\section{Vacuum baseline}
\label{sec:vacuum}

\subsection{Engineering the quark-initiated jet fraction}

A key quantity in this paper is the quark-initiated jet fraction, $q$-fraction in short. We begin by analysing its rapidity dependence on a dijet sample in $p+p$ collisions at LHC energies ($\sqrt{s}=5.02$ TeV) using the \texttt{Pythia}~\cite{Bierlich:2022pfr} event generator. This quantity is defined as follows. First, we reconstruct anti-$k_t$~\cite{Cacciari:2008gp} jets of radius $R=0.4$ with \texttt{FastJet}~\cite{Cacciari:2011ma}. The quark- or gluon-initiated tag is assigned according to a coincidence procedure, introduced in Ref.~\cite{Spousta:2015fca}, and valid at leading order (LO): for a given jet, we find the parton produced by the hard-scattering matrix element whose angular distance, $\Delta R\equiv \sqrt{\Delta \phi^2+\Delta \eta^2}$, with respect to the jet axis is minimized. We then assign the quark- or gluon-initiated jet tag depending on the identity of the selected hard parton, considered to be the shower-initiator. 
By selecting different jet momenta and rapidities one effectively changes the value of Bjorken-$x$ explored in the incoming protons, i.e. $x=2 \, p_t \, \textrm{cosh}(y)/\sqrt{s}$, and thus the flavour of the parton that participates in the hard scattering. The quark-initiated jet fraction as a function of both the jet rapidity $y$ and transverse momentum $p_t$ is shown in the left panel of Fig.~\ref{fig:qfracpp}. We observe that the $q$-fraction is enhanced both at forward rapidities and at high jet $p_t$. This is expected since at large values of $x$ the parton distribution function (PDF) of valence quarks dominate. Thus, by selecting jets within a given $p_t$ range and different bins in rapidity, we can engineer the $q$-fraction of a given jet ensemble.

\setlength{\tabcolsep}{7pt}
\begin{table}[t]
    \centering
   \begin{tabular}{c|c|c|c}
    \multicolumn{4}{c}{Number of jets} \\
    \hline
    & \multicolumn{3}{c}{$q$-fraction} \\
    \hline
     Jet $p_t$ & $\lesssim 0.3$ & $\lesssim 0.6$ & $\lesssim 0.9$ \\ \hline
        $20<p_t<80$ GeV & $3.2 \times 10^8$ & $2.7 \times 10^8$ & $1 \times 10^7$ \\
        $100<p_t<150$ GeV & $8 \times 10^5$ & $4 \times 10^5$ &  $5.4 \times 10^4$ \\
        $225<p_t<300$ GeV & $1.6 \times 10^4$ & $3.4 \times 10^4$ & $1.5 \times 10^3$ 
    \end{tabular}
    \caption{Estimated number of jets, $N_{\rm jets}$, to be measured in heavy-ion collisions in the high-luminosity phase of the LHC for a given $q$-fraction range depending on the jet $p_t$. Guiding us with the left panel of Fig.~\ref{fig:qfracpp}, we select the following rapidity windows. For $q$-fraction $\lesssim 0.3$ all jet $p_t$ bins use the window $|y|<0.3$. For $q$-fraction $\lesssim 0.6$, low-$p_t$ uses $2.5<|y|<3$, mid-$p_t$ uses $2.1<|y|<2.5$ and high-$p_t$ uses $1.2<|y|<2.1$. Finally, for $q$-fraction $\lesssim 0.9$, low-$p_t$ uses $4<|y|<4.5$, mid-$p_t$ uses $3<|y|<4$ and high-$p_t$ uses $2.1<|y|<2.5$.}
    \label{tab:njets}
\end{table}

As one moves towards larger rapidities, thereby approaching the kinematic limit determined by the centre-of-mass energy of the hadronic collision, the jet $p_t$ spectrum becomes increasingly steeper, as shown in the right panel of Fig.~\ref{fig:qfracpp}. These two rapidity-dependent properties of jets, namely the evolution of the $q$-fraction and the evolution of the power index of the spectrum, $n$, are the two determining factors to understand jet suppression, or $R_{\rm AA}$, as a function of rapidity, as will be discussed in Section~\ref{sec:hybrid}.

Using the spectra from the right panel of Fig.~\ref{fig:qfracpp} we can estimate the expected number of jets in the HL-LHC corresponding to a range of $q$-fraction values for different windows in jet $p_t$ and $y$.\footnote{The actual centre-of-mass energies will be $\sqrt{s}=5.5$ ATeV, so we expect a slightly larger number of jets than the ones we estimate by using the spectrum at $\sqrt{s}=5.02$ ATeV.} The number of jets is simply given by $N_{\rm jets}(\Delta p_t,\Delta y) \approx \mathcal{L}N_{\rm coll} \int_{\Delta p_t, \Delta y} d\sigma/dp_tdy$, and we set $\mathcal{L}=10 \, \text{nb}^{-1}$ and $N_{\rm coll}\approx 2000$ for central collisions. The results are shown in Table~\ref{tab:njets}. As we will explicitly see when computing medium modifications in Secs.~\ref{sec:hybrid}, \ref{sec:toy-template} these numbers guarantee enough statistics to disentangle the physical mechanisms under consideration in the present work.

\subsection{Analytic results at double-logarithmic accuracy}

 \begin{figure}[t!]
    \centering
    \includegraphics[width=0.5\textwidth]{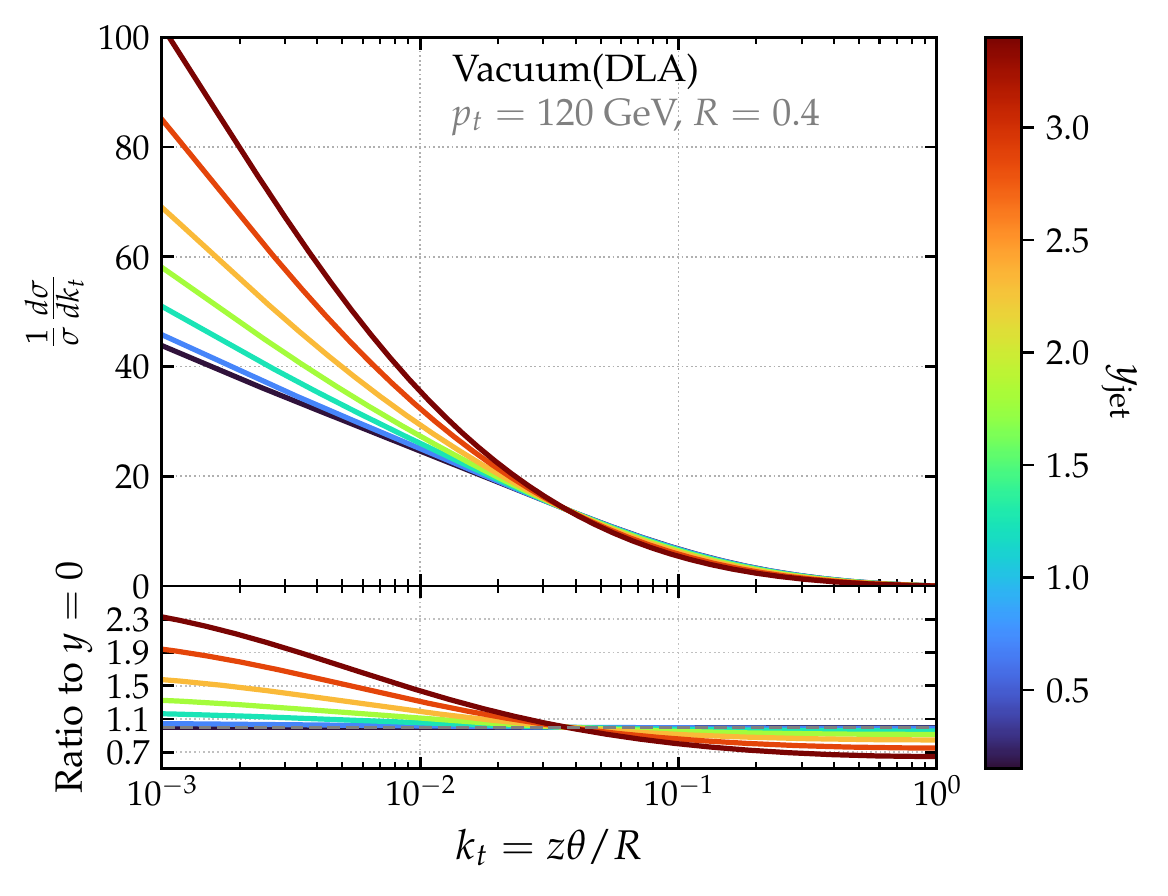}
    \caption{Leading-$k_t$ distribution as a function of rapidity in vacuum at DLA for jets with $p_t=120$ GeV and $R=0.4$. The bottom panel display the ratio to the mid-rapidity result.}
    \label{fig:kt_pp_anal}
\end{figure}

 Next, we focus on the $k_t$-distribution of the \textit{hardest} splitting in the jet clustering sequence, as defined by the Dynamical Grooming procedure~\cite{Mehtar-Tani:2019rrk} with $a=1$.\footnote{Since we are measuring the $k_{t}$ distribution of the splitting with the highest $k_t$, the groomed distribution is equivalent to the plain one.}
 We use the small-angle approximation and define $k_t=z\theta$, where $z$ and $\theta$ are the momentum sharing fraction and opening angle of the splitting, respectively. This observable was calculated at next-to-next-to double logarithmic accuracy and compared to ALICE data in Refs.~\cite{Caucal:2021bae,ALICE:2022hyz}. In the present analytic study, we do not aim at providing precise predictions but rather focus on qualitative features of the distribution. As such, we consider all emissions to be soft and collinear. At this double-logarithmic accuracy (DLA) the self-normalised $k_t$-distribution is given by
 \begin{align}
 \label{eq:kt-vac}
 \nonumber
     \frac{1}{\sigma}\frac{{\rm d}\sigma}{{\rm d} k_t}\Big\vert_{p_t,y} &= 
    \sum_{i\in\lbrace q,g\rbrace} f_i \int_0^1 {\rm d} z \int_0^R{\rm d}\theta P^{\rm vac}(z,\theta)\delta(k_t-z\theta)  \\ \nonumber
    &\times e^{-\int {\rm d}z' \int {\rm d}\theta' P^{\rm vac}(z',\theta')\Theta(z'\theta'-k_t)} \\
    &\stackrel{\text{DLA}}{=} \sum_{i\in\lbrace q,g\rbrace} f_i \frac{2\bar\alpha}{k_t} \ln\frac{R}{k_t} e^{-\bar\alpha\ln^2\frac{R}{k_t}},
 \end{align}
 where we fixed the strong coupling to $\bar\alpha\equiv\alpha_s(p_tR) C_i/\pi$ with $C_i$ the color factor, $f_i$ corresponds to the quark (or gluon) fraction and we have used that the branching kernel reduces to $P^{\rm vac}=\bar\alpha/(z \theta)$ in the soft-and-collinear limit. 
 In Fig.~\ref{fig:kt_pp_anal} we plot Eq.~\eqref{eq:kt-vac} for different rapidities.
 \footnote{Our analytic results, both in vacuum and in the medium, rely in general on the properties of the species- and rapidity-dependent initial jet spectra, such as their power index $n$ and $q$-fraction.
 We have used \texttt{Pythia} results to perform a fit of the jet $p_t$-spectra, for jet-initiator species $k$ within a rapidity window $y$, as $\dd \hat \sigma^{(k,y)}/\dd p_t = \sigma^{(k,y)}_0 \,(p^{(k,y)}_{t,0}/p_t)^{n^{(k,y)}(p_t)}$ and $n^{(k,y)}(p_t) = \sum_{i=0}^2 c^{(k,y)}_i \ln^i (p^{(k,y)}_{t,0}/p_t )$.}
 We observe that the low-$k_t$ regime is enhanced with increasing rapidity due to the more collimated DGLAP evolution of $q$-initiated jets. Note that, at this level of accuracy, the maximum of the distribution is dictated by the strong coupling constant, i.e. $k_{t,{\rm max}} \propto e^{-1/(2\bar\alpha)}$ \cite{Caucal:2021bae}. We cut the plot at $z\theta/R=10^{-3}$ to downplay the region where non-perturbative corrections would dominate.

\subsection{Pythia results}
\begin{figure*}[t!]
    \centering
    \includegraphics[width=0.95\textwidth]{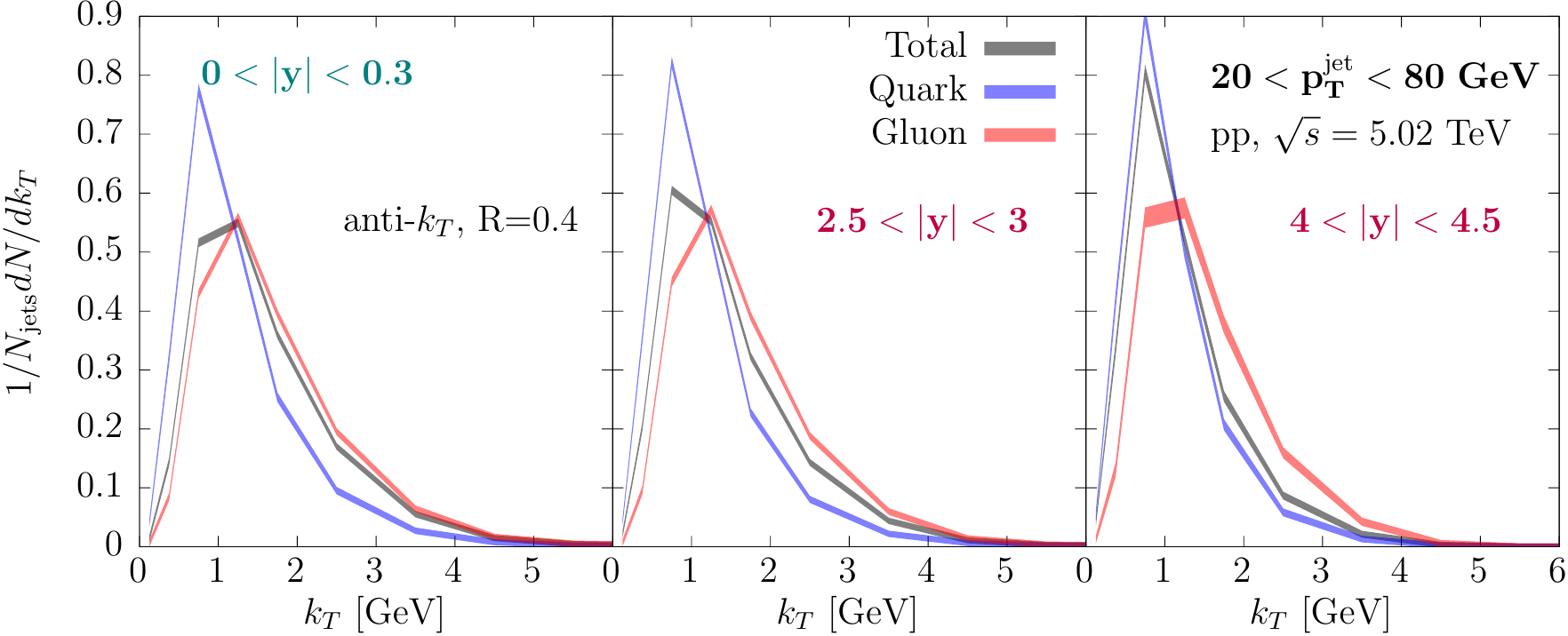}
    \includegraphics[width=0.95\textwidth]{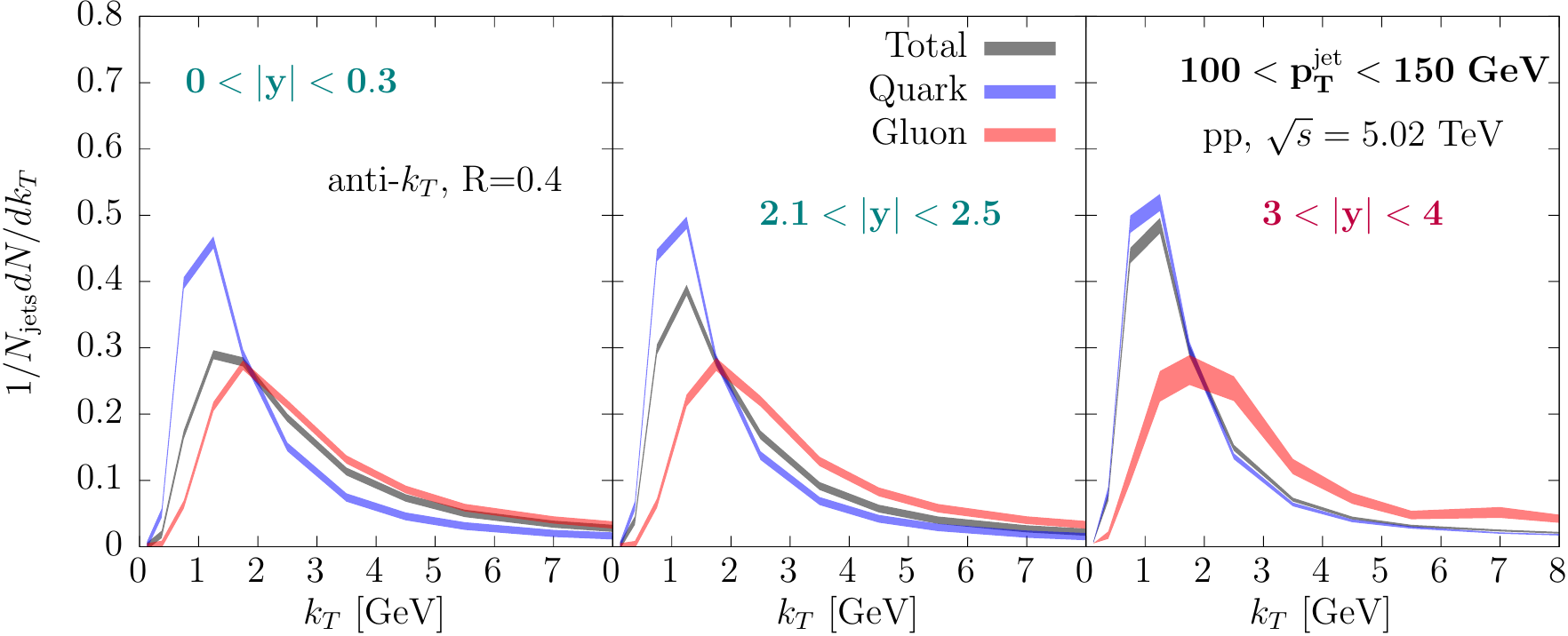}
    \includegraphics[width=0.95\textwidth]{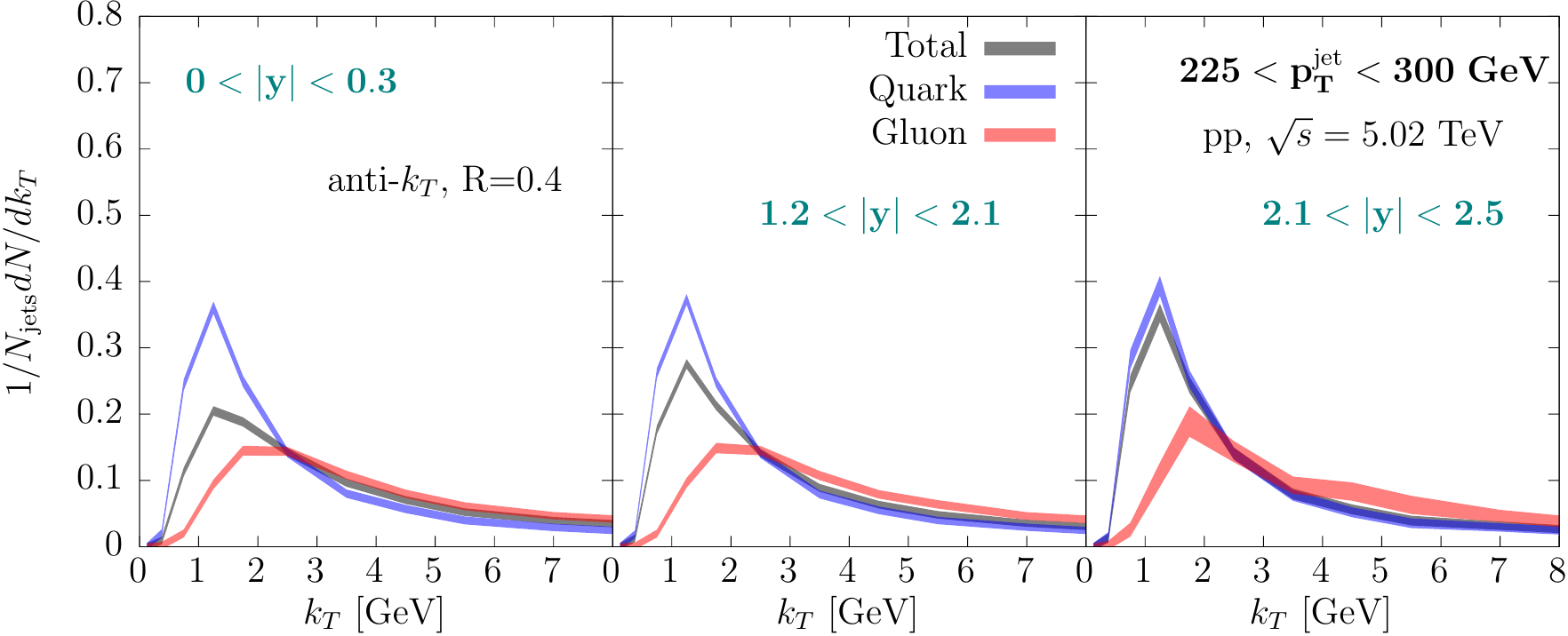}
    \caption{Leading-$k_t$ distribution for quark-initiated and gluon-initiated jets, as well as the total result, where each panel selects jets with different rapidity cuts that tag different $q$-fractions, as specified by Table~\ref{tab:njets}. From top to bottom the jet $p_t$ increases. The $p_t$ and rapidity cuts that are accessible with the current LHC technology are marked in teal, while those of the future LHC runs are highlighted in purple.
    }
    \label{fig:kt_pp}
\end{figure*}

 Since the previous calculation misses several ingredients of a realistic parton shower, we also calculate the leading-$k_t$ distribution using \texttt{Pythia} simulations. The results are shown in Fig.~\ref{fig:kt_pp} for quark- and gluon-initiated jets, as well as the total result. We reconstruct the particles of an anti-$k_t$ jet with $R=0.4$ using the Cambridge/Aachen algorithm~\cite{Dokshitzer:1997in}, and look for the highest $k_t$ throughout its clustering history, starting from the last clustering and following the leading branch. Note that for this particular figure we use a definition of $k_t$ with dimensions of energy, $k_t\equiv z p_t^{\rm parent}\sin\theta$, where $p_t^{\rm parent}$ is the momentum of the parent branch. In consistency with what is shown in Fig.~\ref{fig:qfracpp}, the inclusive results for leading-$k_t$ almost coincide with the gluon results at mid-rapidity, while they are dominated by the quark results when increasing the jet $p_t$ and/or selecting larger rapidities. As expected from the analytics (see Fig.~\ref{fig:kt_pp_anal}), the $k_t$-distribution for quark-initiated jets is shifted towards smaller values compared to that of gluon-initiated jets.

\section{Medium results}
\label{sec:medium}

In the previous section we gained intuition on the shape of the leading-$k_t$ distribution for quark- and gluon-initiated jets, and how we can engineer jet samples with different $q$-fractions by selecting jets that belong to specific $p_t$ ranges and rapidity windows. We now move to the description of the medium effects on this observable. 
Our main goal is to study how the different physical descriptions of the jet-medium interaction (`modified $q/g$ fraction' and `color decoherence') provide distinctive results as we study the medium modifications of jet ensembles with different $q$-fractions.

\subsection{Analytic calculation in the multiple, soft scattering approximation}
\label{sec:analytics}
Let us begin by generalising Eq.~\eqref{eq:kt-vac} to include medium modifications following the approximations introduced in Ref.~\cite{Caucal:2021cfb}. The medium-modified leading-$k_t$ distribution is given by

\begin{align}
    \label{eq:ktg-med}
    \nonumber
    \frac{1}{\sigma}\left.\frac{\dd \sigma}{\dd k_t}\right|_{p_t,y}&=\frac{1}{\mathcal{N}}\sum_{i\in\{q,g\}}f^{\rm n}_i\int_0^1 \dd z \int_0^R \dd \theta P_i^{\rm med}(z,\theta) \\ \nonumber 
    &\times \rme^{-\int \dd z' \int \dd \theta' P^{\rm med}(z',\theta')\Theta(z'\theta'-k_t)} \delta(k_t-z\theta) \\
    &\times \int_0^\infty\dd\varepsilon \mathcal{E}_i(\varepsilon|z,\theta)\rme^{-\frac{n\varepsilon}{p_t}},
\end{align}
where $\mathcal N$ is a normalisation factor, $f^{\rm n}_i$ is the quark/gluon fraction computed using nuclear PDFs (nPDFs) and $n$ is the spectral index of the $p_t$ spectrum.\footnote{Similarly to the vacuum case, we have performed a fit to the jet spectra $\dd \sigma^{(k,y)}/\dd p_t$, with the only difference being the use of the central set of the nPDFs provided by EPPS16~\cite{Eskola:2016oht}.} The two main novelties of Eq.~\eqref{eq:ktg-med} are the in-medium branching kernel, $P^{\rm med}$, and the energy loss probability distribution, $\mathcal{E}$. The former accounts for the fact that the tagged emission can be either vacuum-like or medium-induced, and may include constraints on the radiation phase-space. The latter represents the probability for an $i$-initiated splitting to radiate energy $\varepsilon$ out of the jet cone via a medium-induced cascade. 

In what follows we will specify the exact form of $P^{\rm med}$ and $\mathcal{E}$ in two different models: the modified $q/g$ modification model and another based on color decoherence. We would like to emphasize that we make numerous approximations to simplify the models as much as possible, while maintaining their main physical ingredients. Consequently, the following results serve an illustrative purpose and are not meant to be quantitative predictions. For the color decoherence-based model we follow Ref.~\cite{Caucal:2021cfb} and consider the QGP to be a brick of length $L=4$~fm characterised by the transport coefficient $\hat q=0.3$ GeV$^3$ and a coupling constant for medium-induced emissions of $\alpha^{\rm med}_{s}=0.24$. These values lead to a reasonably good description of the jet yield depletion defined as
\begin{equation}
\label{eq:RAA}
    R_{\rm AA}=\frac{\dd\sigma^{AA}/\dd p_t}{\dd\sigma^{pp}/\dd p_t},
\end{equation}
in heavy-ion collisions, i.e. $R_{\rm{AA}}\sim 0.47$ at $p_t=120$ GeV and $R=0.4$. Keeping the same parameters for the modified $q/g$ fraction model would lead to a different value of $R_{{\rm AA}}$, since the energy loss model is different. Imposing that $R_{{\rm AA}}$ coincides in both models at $p_t=120$ GeV leads to $L=4$ fm, $\hat q =0.7$ GeV$^3$ and $\alpha^{\rm med}_{s}=0.28$.\footnote{Many other combinations of ($L, \hat q, \alpha^{\rm med}_{s}$) would yield the same $R_{{\rm AA}}$.}

We would like to remark that we keep $L$ fixed for all rapidity values. To understand why this is correct, let us consider a particle moving perpendicular to the beam axis that travels a length $x$. If instead the particle moves with some longitudinal momentum, it will traverse, in the centre-of-mass frame, a length $x'=x/\cos \theta$, where $\theta$ is the angle with respect to the direction perpendicular to the beam. Using the definition of rapidity, $y \equiv \arctanh{(p_z/p)}$, with $p_z$ and $p$ being the longitudinal momentum and total momentum of the particle, respectively, one sees that $\cos\theta=1/\cosh y$, so we can express the length travelled by the particle with a finite rapidity in the centre-of-mass frame as $x'=x\cosh y$. Now, the quantity that matters for energy loss is the distanced travelled by a particle in the local fluid rest frame (LFRF). In general, the collision centre-of-mass frame will not coincide with the LFRF, and so a Lorentz transformation needs to be performed to translate $x'$ into its LFRF analogue, that we denote $x'_F$. For the case in which a parton is moving with rapidity $y$ and some transverse velocity through a Bjorken-flow, one gets that $x'_F(y)=x'/\cosh y$ (see, e.g., Ref.~\cite{Casalderrey-Solana:2015vaa}). Then, after taking this factor into account we obtain that $x'_F=x$. That is, regardless of the rapidity of the particle, it will always traverse a length $x$ in the LFRF, as if it was moving transversely to the beam axis. This justifies the use of a rapidity-independent medium length $L$.

\subsubsection{Modified $q/g$ fraction model}
\label{sec:mod-qg}
We consider a description of the in-medium jet evolution that resembles the one presented in Refs.~\cite{Spousta:2015fca,Ringer:2019rfk}. To start with, the branching kernel is taken to be the vacuum one, i.e. the leading-$k_t$ condition is always met by a vacuum emission and therefore we set $P^{\rm med}\to \bar\alpha/(z\theta)$ in Eq.~\eqref{eq:ktg-med}. Regarding the energy loss distribution, we assume that the intrajet activity is irrelevant and thus every jet loses energy as if it was a single color charge. That is, we follow the quenching weights paradigm~\cite{Baier:2001yt} and calculate the probability for a single parton of flavor $i$ to lose energy $\varepsilon$ as
\begin{align}
\label{eq:qw-single-parton}
   \mathcal Q_i(p_t,R)&\equiv \int_0^\infty\dd\varepsilon \mathcal{E}_i(\varepsilon|z,\theta)\rme^{-\frac{n\varepsilon}{p_t}} \\ \nonumber  
     &= \exp\left[\int_R^\infty\rmd\theta\int_0^1\rmd z P_i^{\rm mie}(z,\theta)\left(\rme^{-nz}-1\right)\right],
\end{align}
where we adopt the multiple, soft scattering approximation to describe the spectrum of medium-induced emissions, i.e.
\begin{align}
\label{eq:mie-kernel}
P^{\rm mie}(z,\theta) & = \bar \alpha_{s,\rm med} \sqrt{\frac{2\omega_c}{z^3 p_t}}\Theta(\omega_c-z p_t)  \\ \nonumber
&\times 2\theta\frac{z^2p^2_t}{Q^2_s}\Gamma\left(0,\frac{z^2 p^2_t\theta^2}{Q^2_s}\right) \, ,
\end{align}
with $\omega_c=\hat q L^2/2$ being the maximum frequency that a medium-induced emission can acquire and $Q_s=\sqrt{\hat q L}$ its typical transverse momentum. The first line in Eq.~\eqref{eq:mie-kernel} corresponds to the energy spectrum while the second line describes transverse momentum broadening. Note that Eq.~\eqref{eq:mie-kernel} is an approximation of the fully differential medium-induced spectrum that is only valid in the $k_t\ll Q_s$ and $\omega \ll \omega_c$ limit, i.e. in the soft-and-collinear limit.  

 \begin{figure}[t!]
    \centering
    \includegraphics[width=0.5\textwidth]{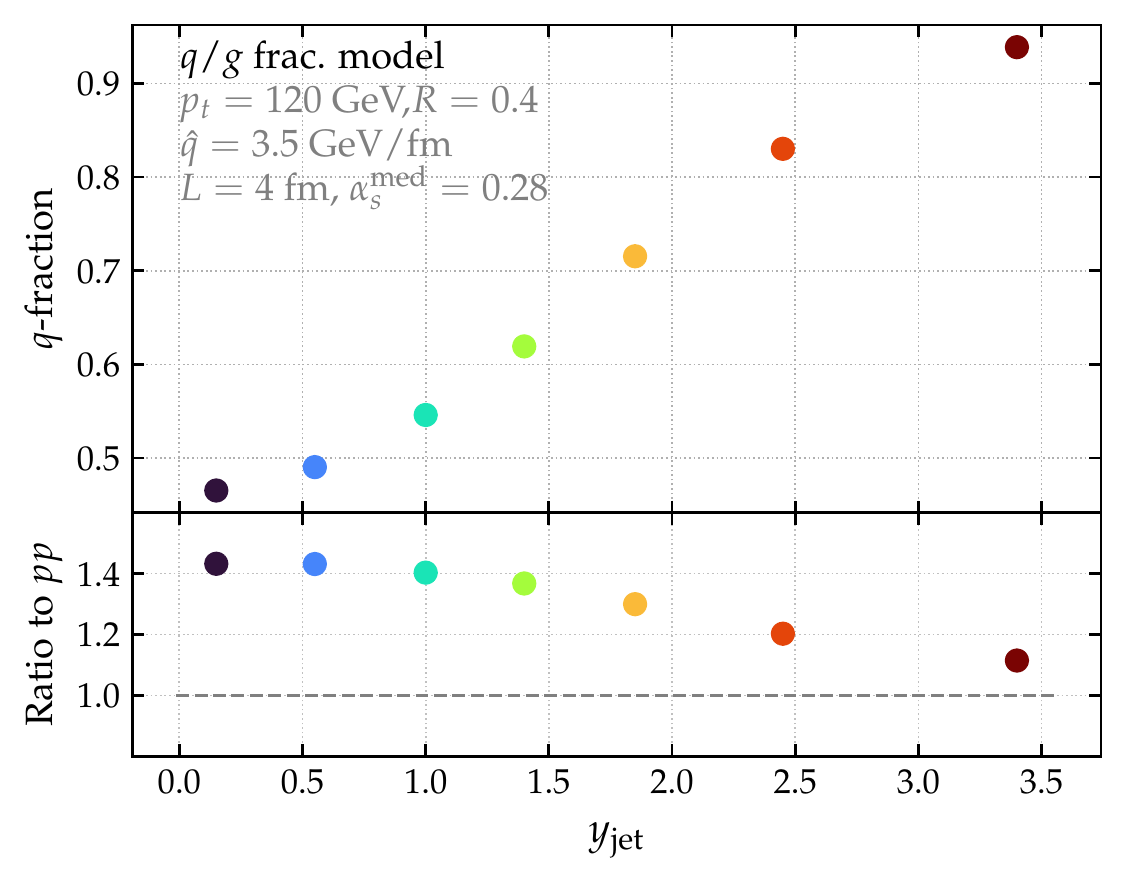}
    \caption{Quark fraction as a function of rapidity for jets with $p_t=120$ GeV calculated in the modified q/g fraction model. The bottom panel display the ratio to the $p+p$ result.}
    \label{fig:qgfrac-anal}
\end{figure}

Having specified the ingredients of the model, an important remark is in order. The shape of the energy loss distribution impacts the $q/g$ fraction of the jet sample after quenching.\footnote{The $q$-fraction can be obtained by fixing the flavour in Eq.~\eqref{eq:ktg-med}, integrating over $k_{t}$ and dividing by the total jet cross section $\mathcal N$.} Our specific choice, given by Eq.~\eqref{eq:qw-single-parton}, is comparable in spirit to that of Ref.~\cite{Spousta:2015fca}, but it is expected to differ from Ref.~\cite{Ringer:2019rfk}. The $q$-fractions obtained with our toy model are displayed in Fig.~\ref{fig:qgfrac-anal}. At mid-rapidity, our $q$-fraction, driven by a Casimir scaling of energy loss, is similar to that of Ref.~\cite{Brewer:2020och} but substantially lower than that of Ref.~\cite{Ringer:2019rfk}. This implies that our results for the $k_t$ distribution correspond to a conservative version of the modified $q/g$ fraction model. We will explore $q$-fraction values similar to those of Ref.~\cite{Ringer:2019rfk} using Monte Carlo simulations in Sec.~\ref{sec:toy-template}. The evolution with rapidity of the ratio of the $q$-fraction between $Pb+Pb$ and $p+p$ is displayed in the lower panel of Fig.~\ref{fig:qgfrac-anal}. This marked evolution, getting close to no modification at all at forward rapidities, compactly elucidates the motivations of the proposed rapidity scan.

\begin{figure}[t!]
    \centering
    \includegraphics[width=0.5\textwidth]{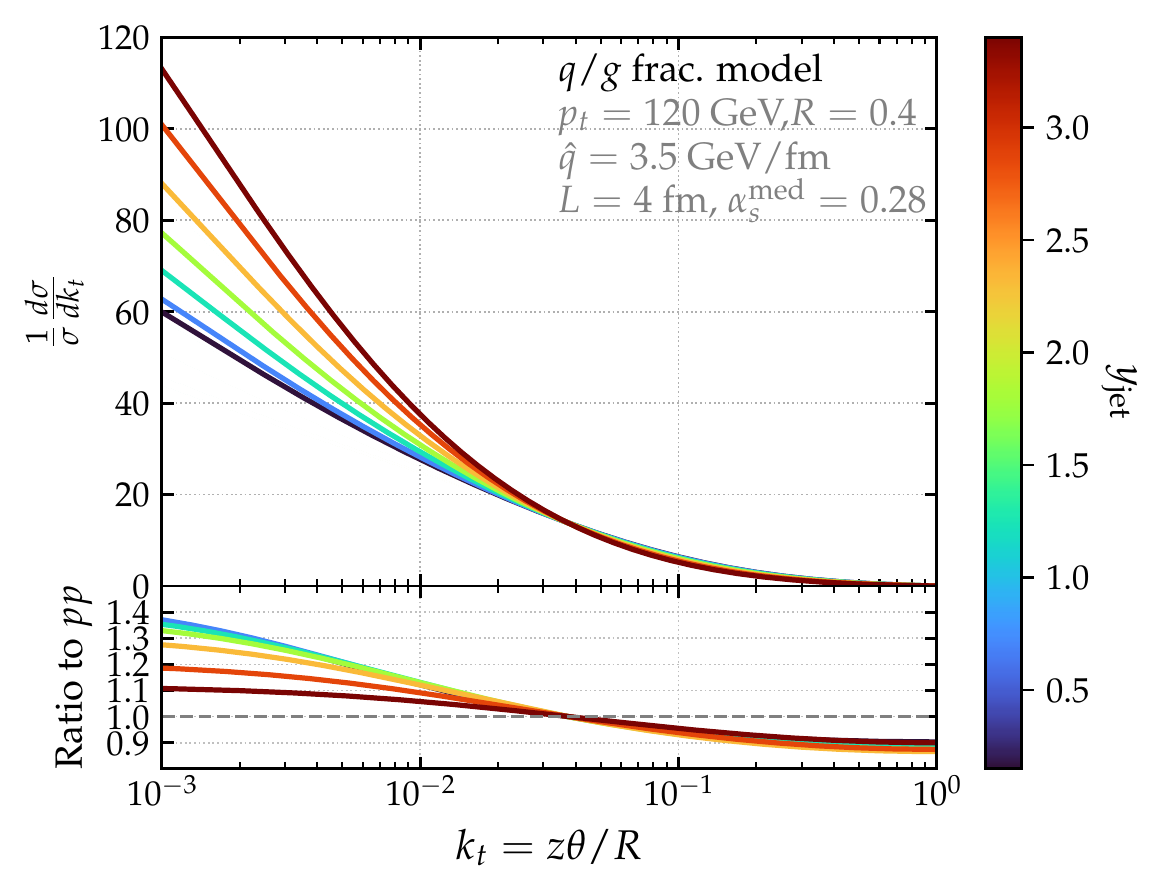}
    \caption{Leading-$k_t$ distribution in the medium using the modified $q/g$ model for jets with $p_t = 120$ GeV. The bottom panel displays the ratio to the $p+p$ result.}
    \label{fig:kt_pbpb-qg}
\end{figure}

In Fig.~\ref{fig:kt_pbpb-qg} we show results for the coherent modification of the leading-$k_t$ distribution as a function of the jet rapidity and take the ratio with respect to the vacuum baseline. Attending to the message provided by Fig.~\ref{fig:qgfrac-anal}, the interpretation of these results is quite transparent: the visible narrowing of the $k_t$ distribution at mid-rapidities is greatly reduced when the $q$-fraction is so high that the depletion of gluon jets in the measured $Pb+Pb$ ensemble becomes irrelevant. Naturally, this model predicts a ratio of the $k_t$-distribution with respect to $p+p$ close to one at large rapidities. Note that, if following Ref.~\cite{Ringer:2019rfk}, one started with a larger value of the $q$-fraction in $Pb+Pb$ at mid-rapidity, the approach of the $k_t$-distribution towards the vacuum result with increasing rapidities would be more abrupt.

\subsubsection{Color decoherence model}
\label{sec:color-dec}
 We now turn to a theoretical description of the leading-$k_t$ distribution based on Ref.~\cite{Caucal:2021cfb}. Here we give a brief summary of the main ingredients of the model, while a more detailed description can be found in Ref.~\cite{Caucal:2021cfb}. This calculation, grounded in perturbative QCD, accounts for the factorization in time between vacuum-like and medium-induced processes in the double-logarithmic approximation. In this case, the medium branching kernel reads
 \begin{equation}
     \label{eq:pmed-caucal}
     P^{\rm med}(z,\theta) = P^{\rm vac}(z,\theta)\Theta_{\notin \rm veto}(z,\theta) + P^{\rm mie}(z,\theta),
 \end{equation}
where $\Theta_{\notin \rm veto}$ constraints the phase-space for vacuum like emissions to be~\cite{Caucal:2018dla}
\begin{equation}
\label{eq:veto}
\Theta_{\notin \rm veto}(z,\theta)= 1 - \Theta(\theta-\theta_c)\Theta(k^2_t-k^2_{t,\rm med}) \Theta(L-t_f),
\end{equation}
with the formation time of an emission given by $t_f = 2/(k_t\theta)$ and $k^2_{t,\rm med} = \hat q t_f$ is the minimum transverse momentum acquired by the emission via multiple soft collisions during its formation. In a nutshell, Eq.~\eqref{eq:veto} imposes that vacuum emissions inside the medium are only allowed for sufficiently high values of $k_t$ or, equivalently, short formation times such that they are not affected by medium dynamics. This separation has been rigorously proven at DLA in Ref.~\cite{Caucal:2018dla}.  

Note that compared to the model presented in Sec.~\ref{sec:mod-qg}, Eq.~\eqref{eq:pmed-caucal} includes two new ingredients: (i) the possibility that a medium-induced emission is the one with the highest-$k_t$, and (ii) a restriction in the phase-space for vacuum emissions. Regarding the first point, since we describe the interactions between the hard propagating parton and the medium in the multiple, soft scattering approximation, we use Eq.~\eqref{eq:mie-kernel} to describe the medium-induced branching probability. This is an important aspect of the calculation since the leading-$k_t$ distribution has been proposed as a potential observable to search for QCD Molière scattering in the medium~\cite{DEramo:2018eoy,Ehlers:2020piz,Hulcher:2022kmn}. Importantly, our analytic estimates provide the multiple, soft scattering baseline for future studies which will account for rare, hard scatterings. Note that the transverse diffusion term in Eq.~\eqref{eq:mie-kernel} can lead to a broadening of the $k_t$ distribution.

The calculation of the energy loss probability distribution in this model reduces the whole jet to just the tagged splitting. Then, depending on whether the angle of the splitting is smaller or larger than the QGP resolution angle $\theta_c$ the jet will lose less or more energy. That is, we replace the last line of Eq.~\eqref{eq:ktg-med} with
\begin{align}
    \label{eq:eloss-caucal}
    \int_0^\infty\dd\varepsilon \mathcal{E}_i(\varepsilon|z,\theta)\rme^{-\frac{n\varepsilon}{p_t}} &= (1-\Theta_{\rm res})\mathcal Q_i(p_t,R) \\ \nonumber
    & +\Theta_{\rm res}\mathcal Q_g(p_t,R)\mathcal Q_i(p_t,R),
\end{align}
where the resolution criterion reads
\begin{equation}
    \Theta_{\rm res}(z,\theta)=\Theta(\theta-\theta_c)\Theta(k_t-k_{t,\rm med}).
\end{equation}
The physical meaning of Eq.~\eqref{eq:eloss-caucal} is rather simple. The first term accounts for the case in which the splitting is unresolved and thus the jet loses energy as a global color charge, just like in the modified $q/g$ fraction model. In turn, the second term in Eq.~\eqref{eq:eloss-caucal} describes the energy loss of a splitting resolved by the medium as the product of the quenching weight of both prongs. We would like to emphasize that this model is a rather crude simplification of the dynamics of color coherence. We plan to explore the use of a resummed quenching weight~\cite{Mehtar-Tani:2017web} for jet substructure calculations together with a determination of the phase-space for vacuum-like emissions beyond the double-logarithmic approximation in a separate publication~\cite{dyg-ioe:paper}.  

 \begin{figure}[t!]
    \centering
    \includegraphics[width=0.5\textwidth]{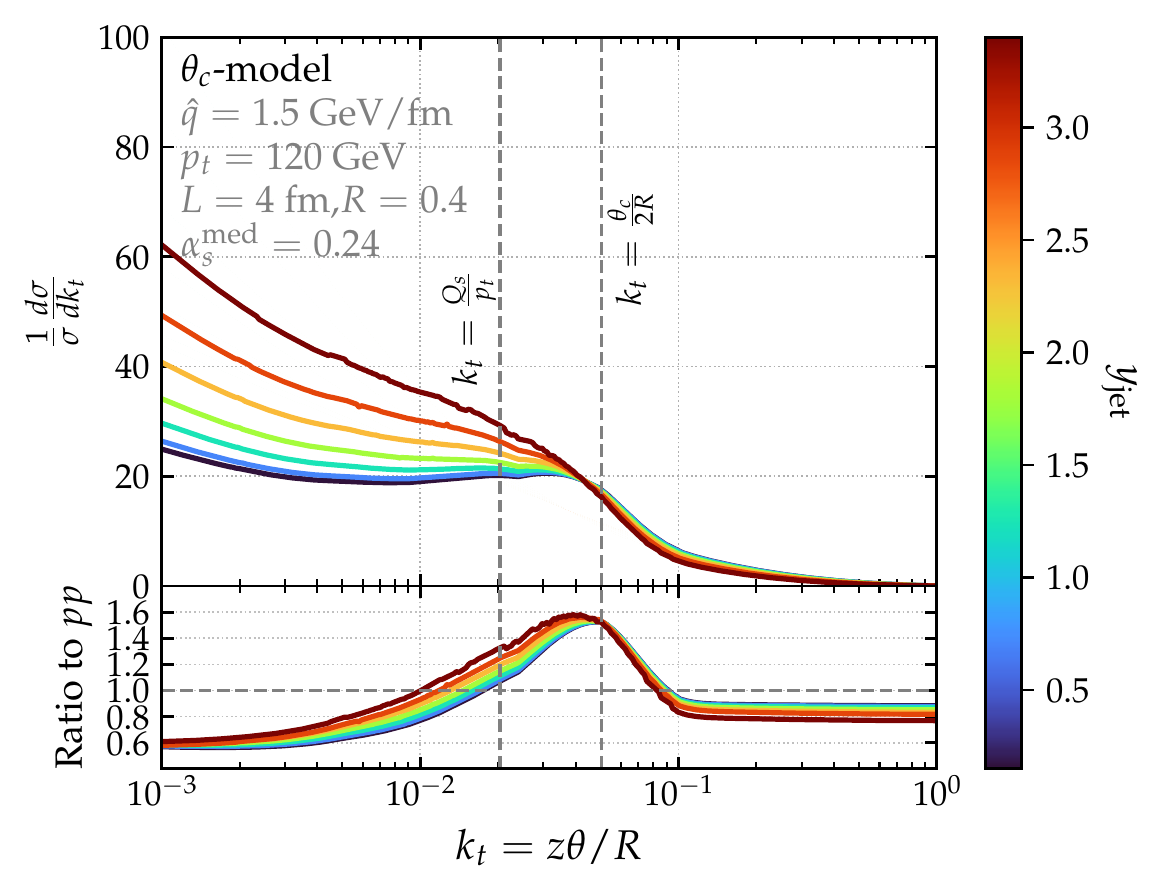}
    \caption{Leading $k_t$-distribution in the medium using the color decoherence model for jets with $p_t = 120$ GeV. The bottom panel displays the ratio to the $p+p$ result. We have highlighted two particularly relevant scales that are further discussed in the main text.}
    \label{fig:kt_pbpb-med}
\end{figure}

The leading-$k_t$ distribution for this model is displayed in Fig.~\ref{fig:kt_pbpb-med}. We observe a drastically different result compared to Fig.~\ref{fig:kt_pbpb-qg}: medium modifications on the $k_t$-distribution are enhanced at forward rapidities. In fact, we observe, for all rapidities, an enhancement of $k_t$ values around the scale $\theta_c/(2R)$. That is, when the energy is shared democratically between the two prongs and the opening angle is identical to the critical one. The only scale in this model is $\theta_c$ and it is thus natural that the $k_t$-distribution exhibits a great sensitivity to it, as it was also the case in the study performed in Ref.~\cite{Caucal:2021cfb} for the opening angle of the splitting. We also observe that this enhancement at $k_t=\theta_c/(2R)$ is more pronounced at forward rapidities. In this plot, we have highlighted as well the typical scale for medium induced-emissions, i.e. $k_t=Q_s/p_t$.

Let us now focus on the rapidity dependence of this model. 
At asymptotically large rapidities, quarks dominate the sample and the quenching weight corresponding to the emitter in Eq.~\eqref{eq:eloss-caucal} always reduces to that of a quark. All those splittings with $\theta>\theta_c$ are resolved and, therefore, suppressed, leading to an enhancement of the relative yield of jets featuring small-$k_t$ (equivalently small-angle) splittings.
This filtering mechanism becomes less effective at mid-rapidity since there one has an admixture of quarks and gluons. The fact that the sample is not pure is important since energy loss depends both on the color factor and the opening angle of the splitting. Due to their vacuum-like evolution, gluon-initiated jets are on average broader than quark-initiated jets, 
so it is more likely that they can be resolved by the medium. However, gluon-initiated jets with unresolved, small-$k_t$ splittings can be more quenched than quark-initiated ones, since $\mathcal{Q}_g \propto \mathcal{Q}_q^{C_A/C_F}$. 
The competition between these two effects, i.e. the color charge and the opening angle of the splitting, yields an overall milder narrowing at mid-rapidity than in the forward regime. 

To sum up this analytic section, we have identified an experimental measurement that could pin down the origin of the narrowing effect observed for mid-rapidity jets. These analytic estimates have allowed us to highlight the core ideas underlying the potential of jet substructure measurements at forward rapidities to disentangle between different jet quenching dynamics. In what follows, we turn to a more quantitative approach and present predictions both with the \texttt{Hybrid} Monte Carlo model and with a modified $q/g$ fraction model that uses \texttt{Pythia} as its vacuum baseline.  
 
\subsection{The Hybrid Strong/Weak Coupling Model}
\label{sec:hybrid}
The hybrid strong/weak coupling model~\cite{Casalderrey-Solana:2014bpa,Casalderrey-Solana:2015vaa} combines a perturbative high-$Q^2$ evolution of the parton shower together with a non-perturbative description of the dynamics between the jet partons and the strongly coupled QGP. The amount of hydrodynamized energy per unit length has been computed in a strongly coupled $\mathcal{N}=4$ SYM plasma at large-$N_c$ and infinite coupling~\cite{Chesler:2014jva,Chesler:2015nqz}:
\begin{equation}
\label{eq:elossrate}
   \left. \frac{\rmd E}{\rmd x}\right|_{\rm strongly~coupled}= - \frac{4}{\pi} E_{\rm in} \frac{x^2}{x_{\rm stop}^2} \frac{1}{\sqrt{x_{\rm stop}^2-x^2}} \quad ,
\end{equation}
where $x_{\rm stop}\equiv  E_{\rm in}^{1/3}/(2{T}^{4/3}\kappa_{\rm sc})$ is the distance an energetic parton with initial energy $E_{\rm in}$ will travel within the strongly-coupled medium before completely hydrodynamizing. $\kappa_{\rm sc}$ is an $\mathcal{O}(1)$ parameter fitted to hadron and jet heavy-ion data measured at the LHC~\cite{Casalderrey-Solana:2018wrw}. The energy and momentum lost by the colored charges, cf.~Eq.~(\ref{eq:elossrate}), hydrodynamizes and 
excites a wake in the flowing plasma, described hydrodynamically, 
which later decays into soft hadrons at the freeze-out hypersurface. We estimate the distributions of those hadrons by applying the Cooper-Frye prescription to the jet-induced perturbations, where we assume that the background fluid is characterized by Bjorken-flow and that the perturbations stay close in rapidity around the jet~\cite{Casalderrey-Solana:2016jvj}. Improvements to this approximated description of medium response are in development~\cite{Casalderrey-Solana:2020rsj}.

All results in this work will include the wake, and comparisons against the results without the wake will not be presented. Even though the wake has an important effect for jets with larger cones~\cite{Tachibana:2017syd,Pablos:2019ngg,Pablos:2020wnp,Mehtar-Tani:2021fud}, its impact on moderate $R\sim 0.4$ cones is milder, and does not play an important role for the kind of jet ensembles here considered. The effect of the soft hadrons from the wake on jet substructure, for inclusive jet ensembles at high enough jet $p_t$, is also subleading, due to a selection bias effect~\cite{Casalderrey-Solana:2019ubu,Casalderrey-Solana:2020jbx} in which the narrower, less quenched jets, which naturally have associated a smaller wake, represent a large fraction of the measured jet population after imposing $p_t$-cuts~\cite{Rajagopal:2016uip}.\footnote{The effect of the wake on jet substructure observables can be enhanced by selecting very quenched jets using boson-jet samples~\cite{Brewer:2021hmh} or machine learning techniques~\cite{Du:2020pmp}.} Note that alternative descriptions of medium response in which the recoils from the elastic scattering processes stay relatively close to the jet axis can present more sizeable effects on, e.g., groomed observables of inclusive jet samples~\cite{Milhano:2017nzm}.
\begin{figure*}[t!]
    \centering
    \includegraphics[width=1\textwidth]{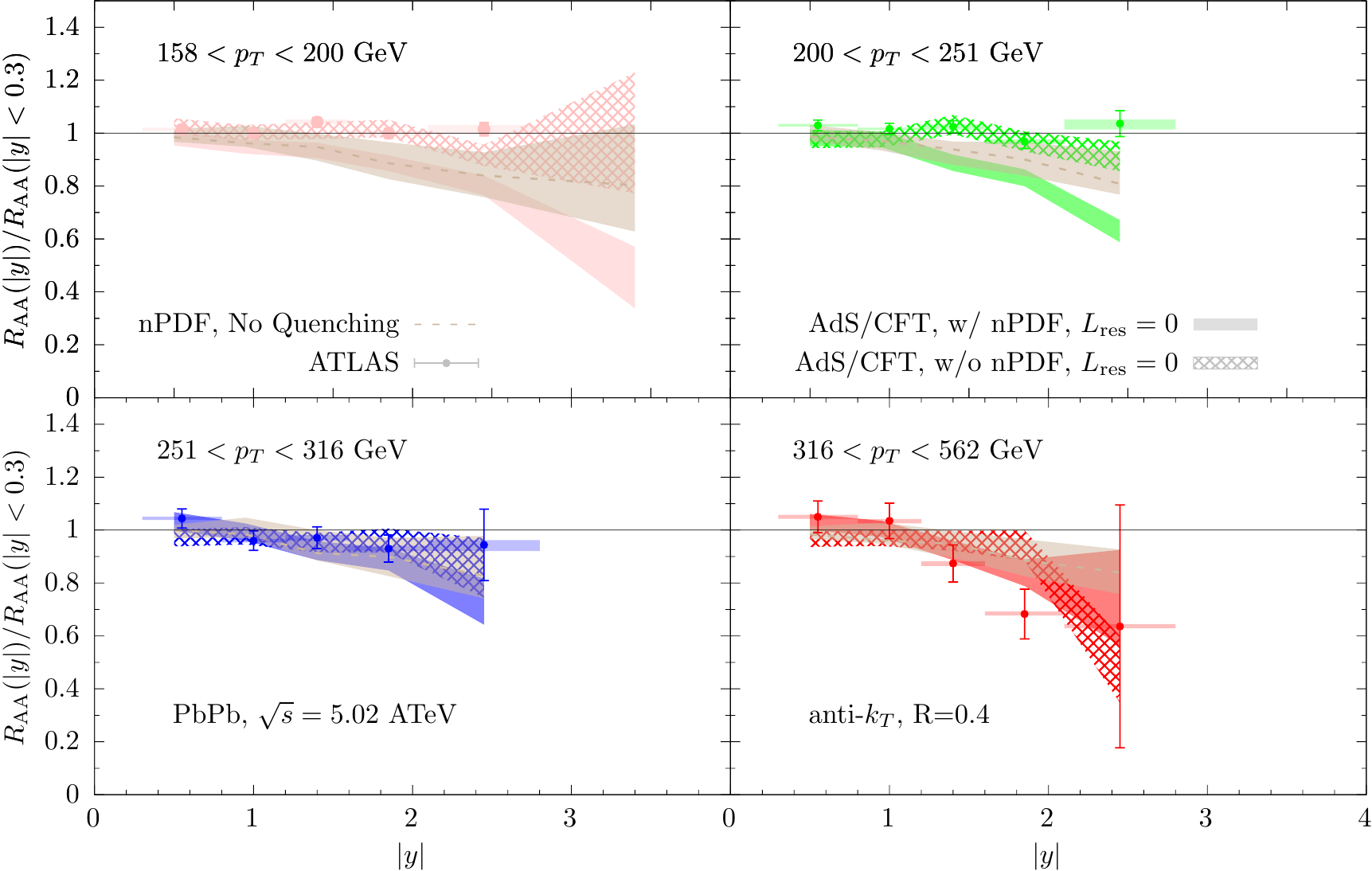}
    \caption{$R_{\rm AA}$ ratio between results at a given rapidity $|y|$ and those at mid-rapidity ($|y|<0.3$). Jets are reconstructed with anti-$k_t$ and $R=0.4$. Results are shown for different jet $p_t$ bins.}
    \label{fig:raa_double_ratio_eta}
\end{figure*}

Finite resolution effects are incorporated in the hybrid model~\cite{Hulcher:2017cpt} in analogy to the notion of screening: the resolution length $L_{\rm res}$ determines the minimal distance between two color charges such that they engage with the QGP independently. We will explore two limiting regimes: (i) $L_{\rm res}=0$, when partons are resolved the instant after they are formed, and (ii) $L_{\rm res}=\infty$, that corresponds to the fully unresolved scenario in which the QGP is sensitive to the global color charge of the jet only. Compared to our analytic results from the previous section, $L_{\rm res}=\infty$ resembles the coherent energy loss implemented for the modified $q/g$ fraction model, while a finite value of $L_{\rm res}$ (not explored here) would play the role of $\theta_c$ in the color decoherence model. 

In view of the goals of the present study, the distinctive color-charge dependence of energy loss in this model deserves some comment. Within an holographic energy loss scenario, the different quenching between quarks and gluons scales like $\mathcal{C}\equiv(C_A/C_F)^{1/3} \simeq 2^{1/3}$ at large-$N_c$~\cite{Gubser:2008as}, in contrast to the linear $C_A/C_F$ scaling expected from perturbative arguments. Since the energy loss rate Eq.~\eqref{eq:elossrate} was derived for the dual of a parton in the fundamental representation, a quark, this means that $\kappa_{\rm sc}^G=\kappa_{\rm sc}\, \mathcal{C}$. Therefore, the pure $q/g$ fraction effect is by construction milder than in pQCD-inspired energy loss scenarios. However, when the traversed distance is much smaller than the stopping distance, i.e. $\epsilon \equiv x/x_{\rm stop}\ll 1$, the energy loss rate Eq.~\eqref{eq:elossrate} can be expanded in powers of $\epsilon$ and, to leading order, the dependence on the Casimirs becomes linear, as in pQCD. Deviations from the full expression start to become significant ($\mathcal{O}(10\%)$) at around $\epsilon \approx 0.7$, which for a $T=0.25$ GeV and a traversed length of $x=5$ fm means that for partons with $E_{\rm in}> 100$ GeV, the color charge dependence is well described by a linear Casimir scaling.

Before moving on to analyze the modification of the leading-$k_t$ distribution with the \texttt{Hybrid} model, and specially its rapidity dependence, it is important to understand whether the rapidity dependence of jet suppression itself is reasonably described. 

\subsubsection{Rapidity dependence of jet suppression}
The energy loss rate in Eq.~\eqref{eq:elossrate} was actually derived for a parton moving through a fluid in the local fluid rest frame. Following the discussion by the end of the introduction to Sec.~\ref{sec:analytics}, given that the QGP is well approximated as a Bjorken-flow, energy loss of a given single parton using Eq.~\eqref{eq:elossrate} is in practice very mildly dependent on rapidity. Therefore, any rapidity dependence of jet suppression found in observables has to have its origin elsewhere.

In Fig.~\ref{fig:raa_double_ratio_eta} we show results for jet suppression, for different cuts in rapidity using $L_{\rm res}=0$ (using $L_{\rm res}=\infty$ yields a very similar picture). We present the results as ratios to the $|y|<0.3$ result for different jet $p_t$ bins and compare against ATLAS data~\cite{ATLAS:2018gwx}. In addition, we calculate this observable with and without nuclear PDFs using the EPPS16~\cite{Eskola:2016oht} set. We discuss the results without nPDF first. As the quark fraction increases with rapidity (cf. left panel of Fig.~\ref{fig:qfracpp}), so does $R_{\rm AA}$, meaning less suppression, as quark jets tend to be less quenched. On the other hand, by increasing rapidity, the initial jet spectrum becomes steeper, specially at higher jet $p_t$ (cf. right panel of Fig.~\ref{fig:qfracpp}), which translates into a reduction of $R_{\rm AA}$. These competing effects yield an evolution of $R_{\rm AA}$ that is almost independent of rapidity within $|y|\leq2$ (as was also observed in Refs.~\cite{Spousta:2015fca,He:2018xjv}). For very large rapidities, where one is approaching the kinematic limit at increasingly lower jet $p_t$, the steepness of the spectrum is the dominant effect, notably reducing $R_{\rm AA}$. Overall, there is good agreement between the Hybrid model without nPDF and the experimental data.

In turn, the results of the Hybrid Model with the inclusion of nPDF (using only the central set for computing time reasons) are quite different. The nPDF-induced modification of the initial jet spectrum at moderate rapidities yields a reduction of $R_{\rm AA}$ with respect to mid-rapidity that is notable for a wide range in jet $p_t$. To highlight the sizeable effect of nPDF on this observable, we show in a dotted bisque-colored line the results that use nPDF, but without quenching. 
For this `no quenching' baseline we have also included the uncertainties associated to the 40 error sets, as prescribed in Ref.~\cite{Eskola:2016oht}.
We can appreciate how the deviation from unity of the full results with nPDF are actually dominated by the nPDF effect itself, yielding these inconsistent with data in the first two $p_t$ bins displayed in the top panels of Fig.~\ref{fig:raa_double_ratio_eta}. While the results in the second to highest jet $p_t$ bin, in the bottom left panel, are the least sensitive to the effects of the nPDF, the highest jet $p_t$ bin, in the bottom right panel, indicates that data seems to slightly prefer those that include nPDF. For this jet $p_t$ bin, both nPDF and quenching effects are needed to achieve the best description of experimental data. The results of Fig.~\ref{fig:raa_double_ratio_eta} provide strong motivation to revisit the impact of nPDF in other jet quenching models. Along with the previous observations on the sizeable effects of nPDF in jet $R_{\rm AA}$ at high-$p_t$~\cite{Pablos:2019ngg,Caucal:2020uic}, which questions the direct interpretation of $R_{\rm AA}\simeq 1$ as a signal of the recovery of lost energy for large-$R$ jets~\cite{Pablos:2019ngg}, these results contribute to build a case for a thorough consideration of the role of nPDF in jet quenching physics, specially in view of the future precision studies programmed at RHIC and the LHC.

\subsubsection{Predictions for a rapidity scan of the leading-$k_t$ distribution}

\begin{figure}[t!]
    \centering
    \includegraphics[width=0.48\textwidth]{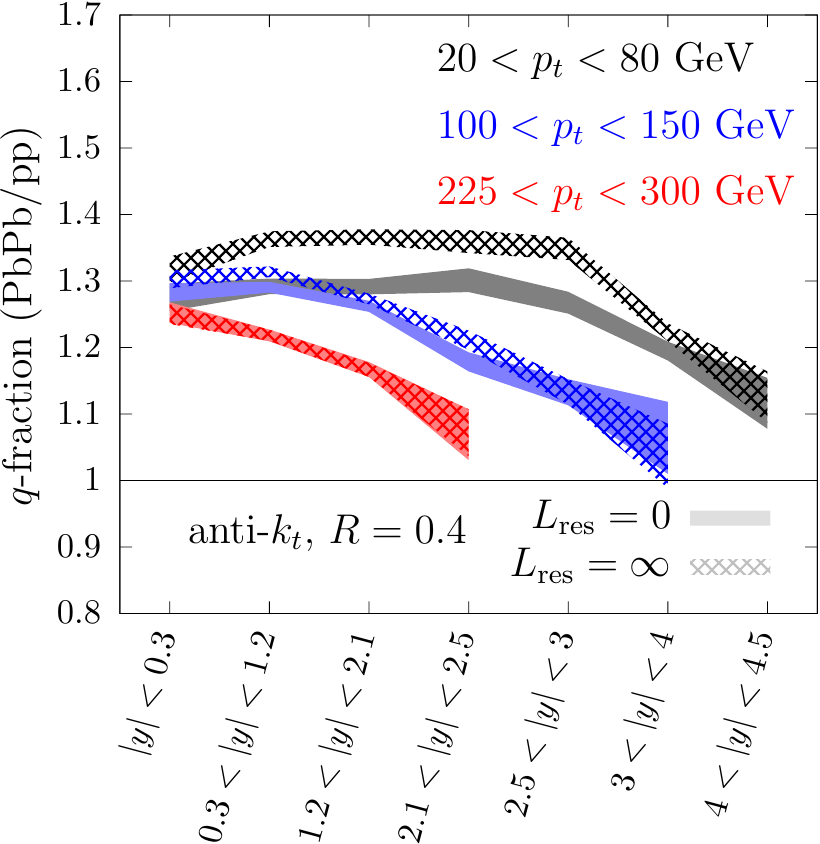}
    \caption{Ratio of the $q$-fraction between $Pb+Pb$ and $p+p$ for different jet $p_t$ intervals and rapidity windows, comparing $L_{\rm res}=0$ ($L_{\rm res}=\infty$) in solid (grid pattern) bands.}
    \label{fig:hybrid_qfracmed}
\end{figure}

In the previous section, we have demonstrated that the rapidity dependence of jet $R_{\rm AA}$ can be reasonably described by the Hybrid Model. We are now in a position to perform predictions on the rapidity dependence of jet substructure observables, such as the leading-$k_t$ distribution. Before presenting our results for this observable, it is relevant to discuss the modification of the $q$-fraction due to medium effects in the Hybrid Model. We show in Fig.~\ref{fig:hybrid_qfracmed} the ratio between the $q$-fraction in $Pb+Pb$ and in $p+p$ both for the fully resolved case, $L_{\rm res}=0$, in solid bands, and for the fully unresolved case, $L_{\rm res}=\infty$, in grid pattern bands. In general, this ratio tends to decrease with increasing rapidities, as we saw in Fig.~\ref{fig:qgfrac-anal}, because the value of the $p+p$ baseline in the denominator steadily increases.\footnote{Discrepancies between the results in Figs.~\ref{fig:qgfrac-anal} and \ref{fig:hybrid_qfracmed} are expected since the energy loss model differs.} There is an exception to this trend, observed in the different behaviour for the results of the lowest jet $p_t$ bin, in black. For jets with $20<p_t<80$ GeV, the ratio of the $q$-fraction can be somewhat larger for rapidity values around $|y|\gtrsim 2.5$ than the one at mid-rapidity. The reason for this is the strong jet $p_t$ dependence in the value of the $q$-fraction for jets with $p_t\lesssim 100$ GeV at forward rapidities, as seen in the left panel of Fig.~\ref{fig:qfracpp}. Shifting the jet $p_t$ has a notable effect on the value of the $q$-fraction at $|y|\approx 2.5$, while the impact is much smaller at mid-rapidity. Therefore, due to medium-induced energy loss, jet $p_t$ bin migration leads to a sizeable increase in the number of quark-initiated jets for those kinematic ranges in which the $q$-fraction features such a strong dependence with $p_t$. 

In Fig.~\ref{fig:hybrid_qfracmed} we also observe that the scenario with $L_{\rm res}=\infty$ produces larger differences in the $q$-fraction between $Pb+Pb$ and $p+p$. The differences with respect to the case with $L_{\rm res}=0$ are approximately independent of the rapidity window, and decrease in magnitude as jet $p_t$ increases. 
We can understand this based on simple arguments. The ratio of the $q$-fraction can be expressed in terms of jet suppression as $f_q^{\rm ratio}\equiv R_{\rm AA}^q/R_{\rm AA}^{\rm total}$, this is, the suppression of quark-initiated jets divided by the total jet suppression.
One can approximate $R_{\rm AA}\approx 1-\varepsilon n/p_t$, see Eq.~\eqref{eq:qw-single-parton}, where $\varepsilon$ is the average energy lost per jet, $n$ is the power index of the spectrum and $p_t$ is the jet transverse momentum. Then, 
\begin{align}
&R_{\rm AA}^q\approx 1-\varepsilon_q \frac{n}{p_t}, \\ 
&R_{\rm AA}^{\rm total}\approx 1-[\varepsilon_q f_q+\varepsilon_g(1-f_q)]\frac{n}{p_t}, 
\end{align}
where $f_q$ represents the vacuum $q$-fraction, and where we have assumed that quark and gluon jets have similar values of $n$. To leading power in $\varepsilon_i n/p_t$, we write 
\begin{equation}
\label{eq:fqratio}
f_q^{\rm ratio}\approx 1+(1-f_q)(\varepsilon_g-\varepsilon_q)\frac{n}{p_t} + \mathcal{O}((\varepsilon n/p_t)^2).
\end{equation}

In order to express the energy loss for a jet with a given flavour, we make use of the resummed quenching weights formalism in the small energy loss approximation~\cite{Mehtar-Tani:2017web}. In this limit, we can simply write
\begin{align}
\label{eq:eloss-parton}
&\varepsilon_q\sim C_F \hat\varepsilon[1+C_A \mathcal{A}(p_t,R)], \\
&\varepsilon_g\sim C_A \hat\varepsilon[1+C_A \mathcal{A}(p_t,R)]\, ,
\end{align}
where $\mathcal{A}(p_t,R)$ refers to the phase-space of extra energy loss sources (assumed to be all vacuum-like gluons) resolved by the medium, and $\hat\varepsilon$ is now the average energy lost \emph{per parton} (stripped of the color charge dependence). 
In the $L_{\rm res}=\infty$ case, the medium cannot resolve the phase-space by definition and so $\mathcal{A}_{\infty} \to 0$. Instead, when $L_{\rm res}=0$, the medium resolves all emissions that are formed within the medium. At fixed coupling, $\mathcal{A}_{0}$ is thus given by
\begin{align}
    \mathcal{A}_0 &= \frac{\alpha_s}{\pi}\int \frac{\dd z}{z}\int\frac{\dd \theta}{\theta} \Theta(t_f<L)\Theta(\theta<R) \\
    &= \frac{\alpha_s}{4\pi}\ln^2\left(\frac{p_tR^2L}{2}\right), \nonumber
\end{align}
where we have used $t_f = 2/(z\theta^2 p_t)$. Finally, we can obtain the difference in the $q$-fraction ratio between these two scenarios as
\begin{align}
\label{eq:fqratio-final}
    f^{\rm ratio}_{q,\infty}-f^{\rm ratio}_{q,0}\approx & (1-f_q)\frac{n}{p_t} (C_A-C_F) \\
    \times & \, \hat\varepsilon_{\infty}[ 1-\frac{\hat\varepsilon_0}{\hat\varepsilon_{\infty}}(1+C_A \mathcal A_0)]\, . \nonumber
\end{align}

Similarly to the fitting procedure adopted within the Hybrid Model, we impose a comparable $R_{\rm AA}\approx 0.5$ for both scenarios for jets with $p_t\approx 150$ GeV and $R=0.4$. Using $L=4$ fm, $f_q=0.5$, $\alpha_s=0.2$ and $n=5$ we get that $\hat\varepsilon_{\infty}\approx 7$ GeV and $\hat\varepsilon_{0}\approx 2.8$ GeV. By numerically evaluating Eq.~\eqref{eq:fqratio-final}, we obtain $\mathcal{O}(10\%)$ in the difference of $f_q^{\rm ratio}$ at $p_t\approx 80$ GeV, decreasing down to less than 2\% at high jet $p_t$, around $p_t\approx 250$ GeV. These results are in qualitative agreement with what is observed in Fig.~\ref{fig:hybrid_qfracmed}.

\begin{figure*}[t!]
    \centering
    \includegraphics[width=0.95\textwidth]{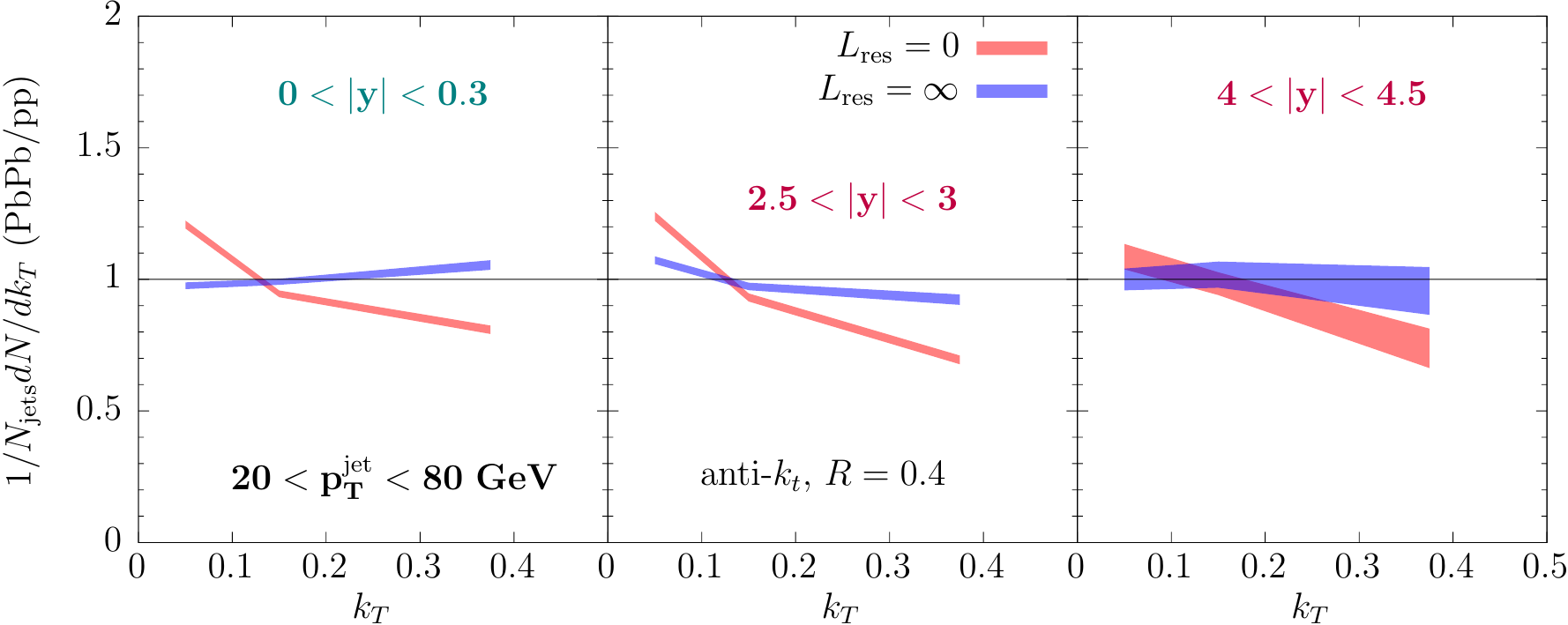}
    \includegraphics[width=0.95\textwidth]{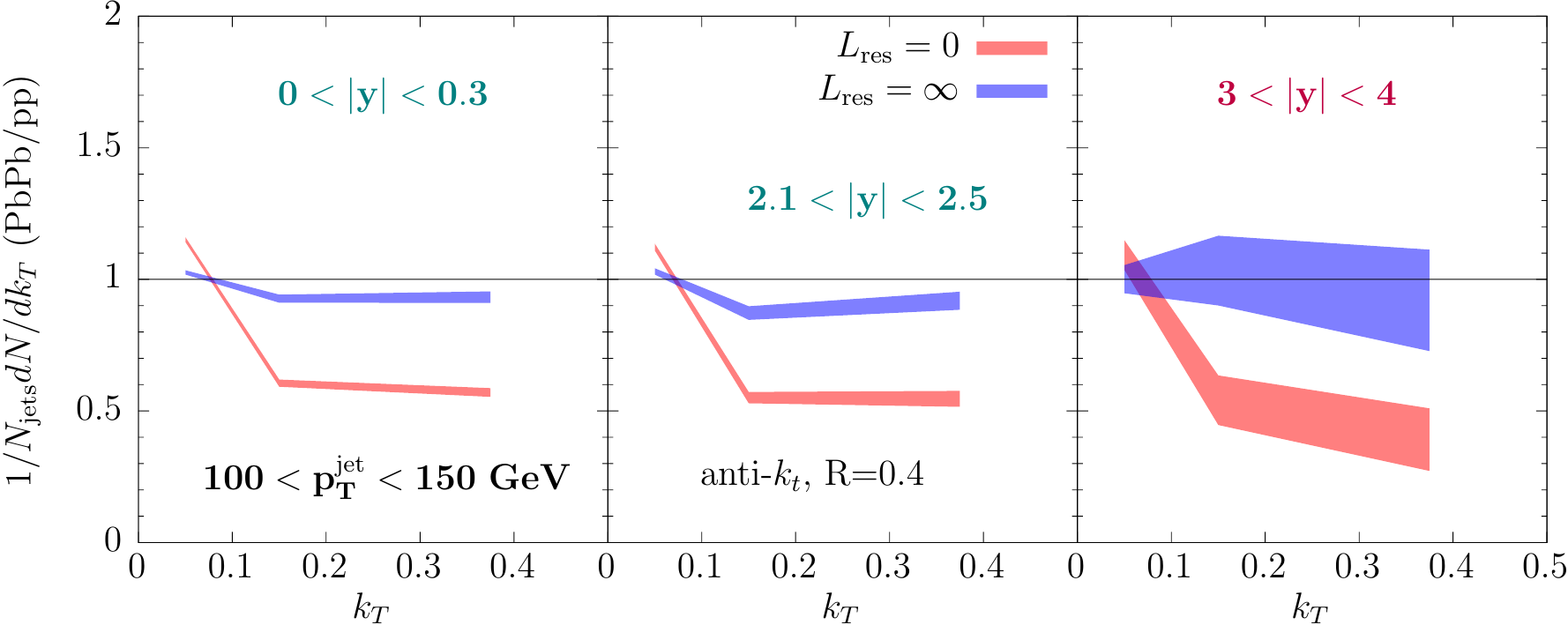}
    \includegraphics[width=0.95\textwidth]{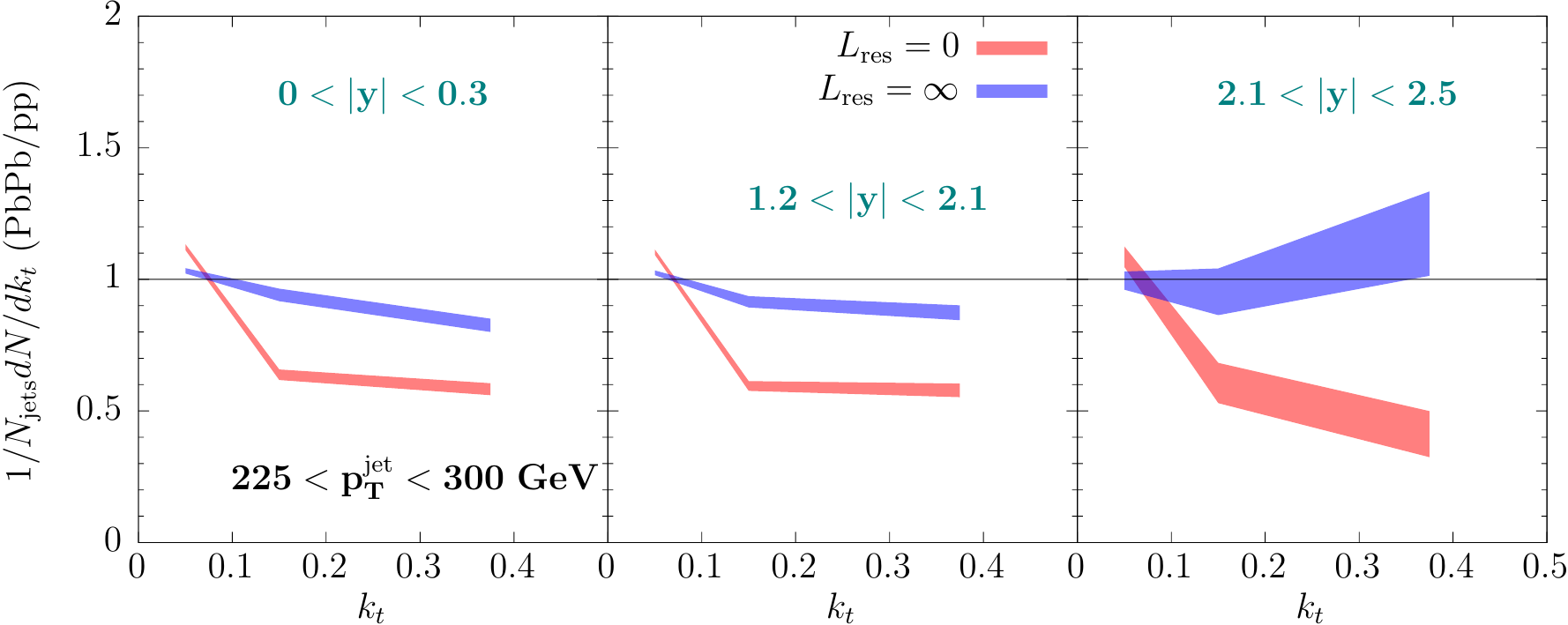}
    \caption{Leading-$k_t$ distribution calculated with the Hybrid model for different rapidity and transverse momentum intervals. Note that we use a dimensionless expression for $k_t$, defined as $k_t \equiv z (p_t^{\rm parent}/p_t^{\rm jet})\sin \theta/R$.}
    \label{fig:kt_PbPboverpp}
\end{figure*}

We discuss next the results of the medium-modified $k_t$ distribution. In Fig.~\ref{fig:kt_PbPboverpp}, we show the ratio between $Pb+Pb$ and $p+p$ for the leading-$k_t$ distribution, using the same cuts as in Fig.~\ref{fig:kt_pp}. For each of the individual plots, the main message is clearly the presence of a sizeable narrowing of the leading-$k_t$ distribution when the jet substructure can be resolved by the QGP, represented in this model by $L_{\rm res}=0$.  Note that the observed narrowing is notably stronger than the one obtained analytically in Fig.~\ref{fig:kt_pbpb-med}. This is most likely due to the fact that, as expressed in Eq.~\eqref{eq:eloss-caucal}, quenching of a resolved jet simply consists in the energy loss from the two tagged prongs; there is no additional quenching for each of the possible subsequent emissions off these partons, as we do in the present Monte Carlo analysis. If one were to include the resummation of the quenching weights~\cite{Mehtar-Tani:2017web,Mehtar-Tani:2021fud,Takacs:2021bpv} in the analytic model, the quenching of the wider structures would be stronger, and the narrowing more pronounced. Another important remark is that the narrowing in the $L_{\rm res}=0$ case persists for all rapidities and that the differences between $L_{\rm res}=\infty$ and $L_{\rm res}=0$ are sufficiently large at the kinematic regions that already are experimentally accessible with current detector capabilities at the LHC.

Focusing now on the $L_{\rm res}=\infty$ curves in Fig.~\ref{fig:kt_PbPboverpp}, we can observe that the ratio becomes increasingly flat as we increase the vacuum $q$-fraction by moving towards larger rapidities. This is fully consistent with the picture obtained in Fig.~\ref{fig:kt_pbpb-qg}. In addition, the evolution of this flattening becomes milder with decreasing jet $p_t$ since the $q$-fraction ratio does not evolve as strongly with rapidity, as we saw in Fig.~\ref{fig:hybrid_qfracmed}. In the lowest jet $p_t$ bin there are a few interesting features that we would like to comment on. In this case, since the ratio of the $q$-fraction can actually increase (cf. Fig.~\ref{fig:hybrid_qfracmed}) we find that the leading-$k_t$ distribution is actually narrower for the $2.5<|y|<3$ bin than for the $|y|<0.3$ bin. In fact, the mid-rapidity result is even slightly wider than in vacuum. We have checked that this last point is an effect of the wake, as the soft particles at large angles produced via jet-energy thermalisation can contribute to generate, or enhance, structures sitting at the jet boundary.\footnote{This effect is irrelevant at higher jet $p_t$ since they are not affected by the relatively small energy injection from the thermal soft particles.}

\subsection{Modified $q/g$ fraction model with Pythia templates}
\label{sec:toy-template}
\begin{figure*}[t!]
    \centering
    \includegraphics[width=0.95\textwidth]{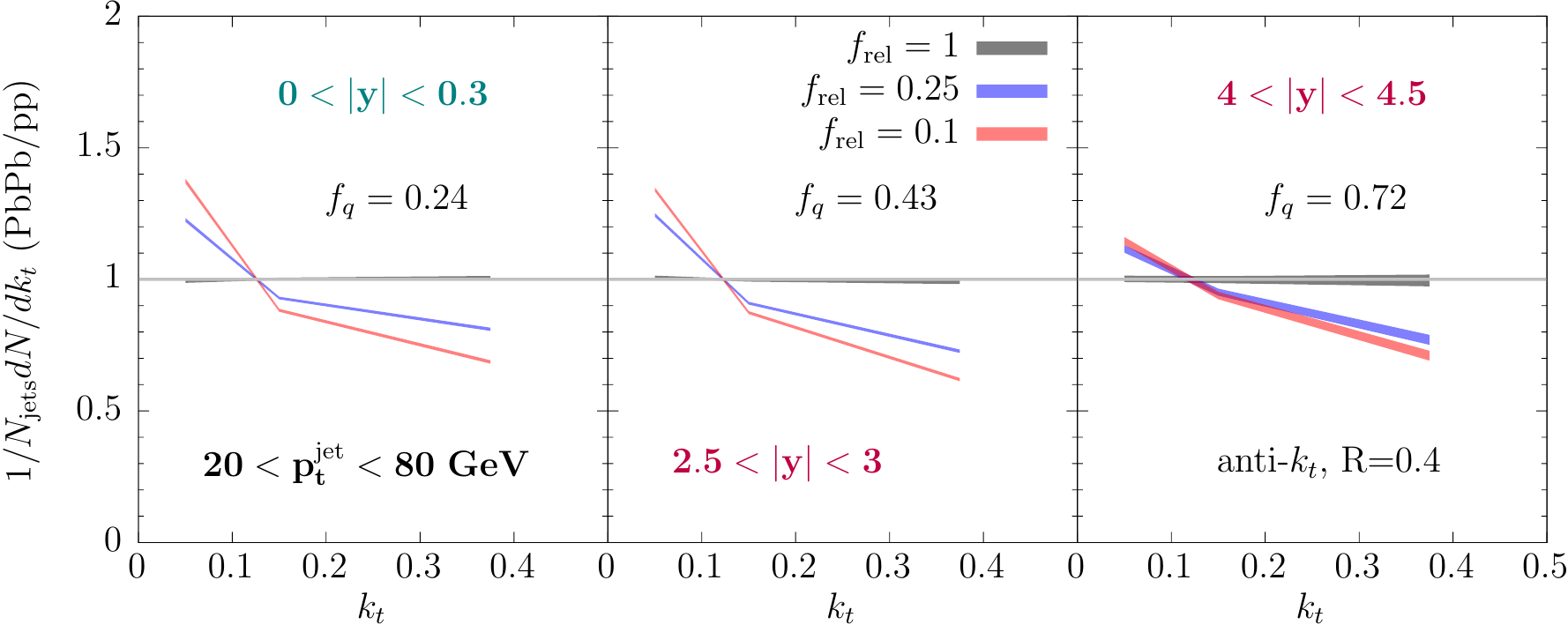}
    \includegraphics[width=0.95\textwidth]{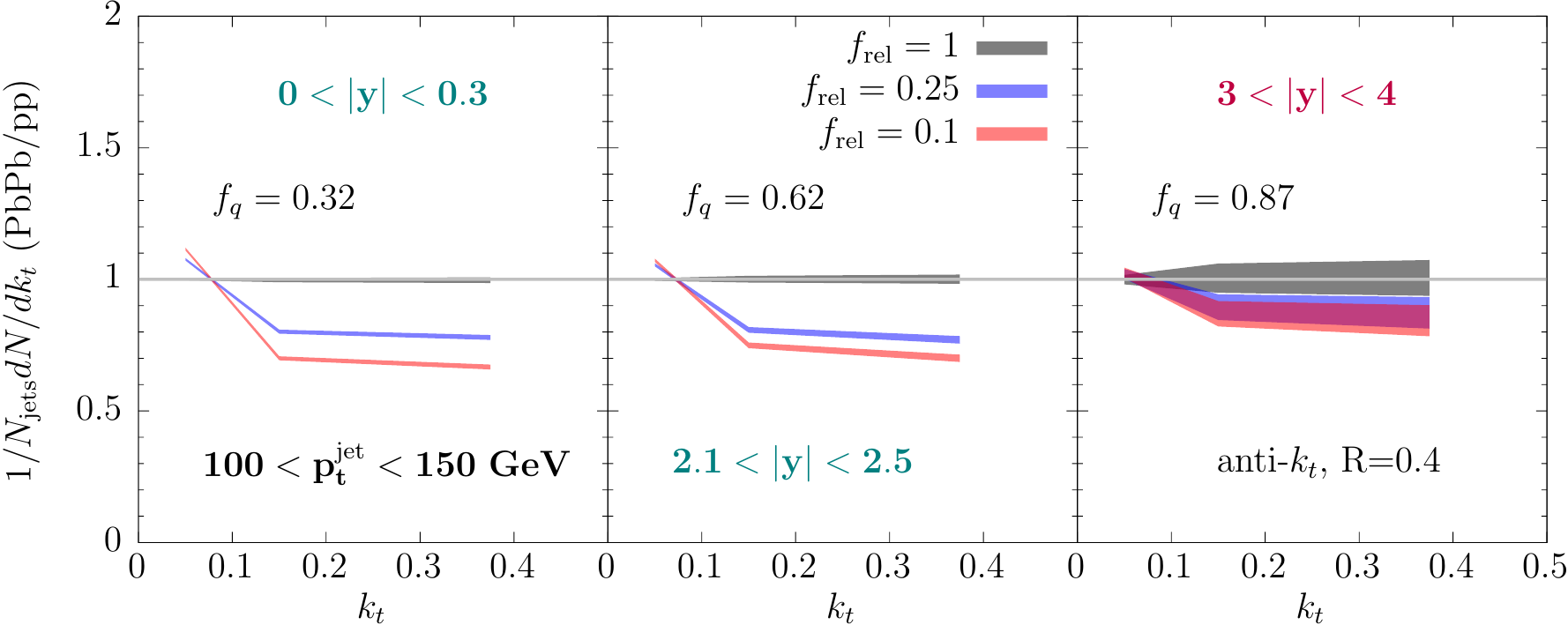}
    \includegraphics[width=0.95\textwidth]{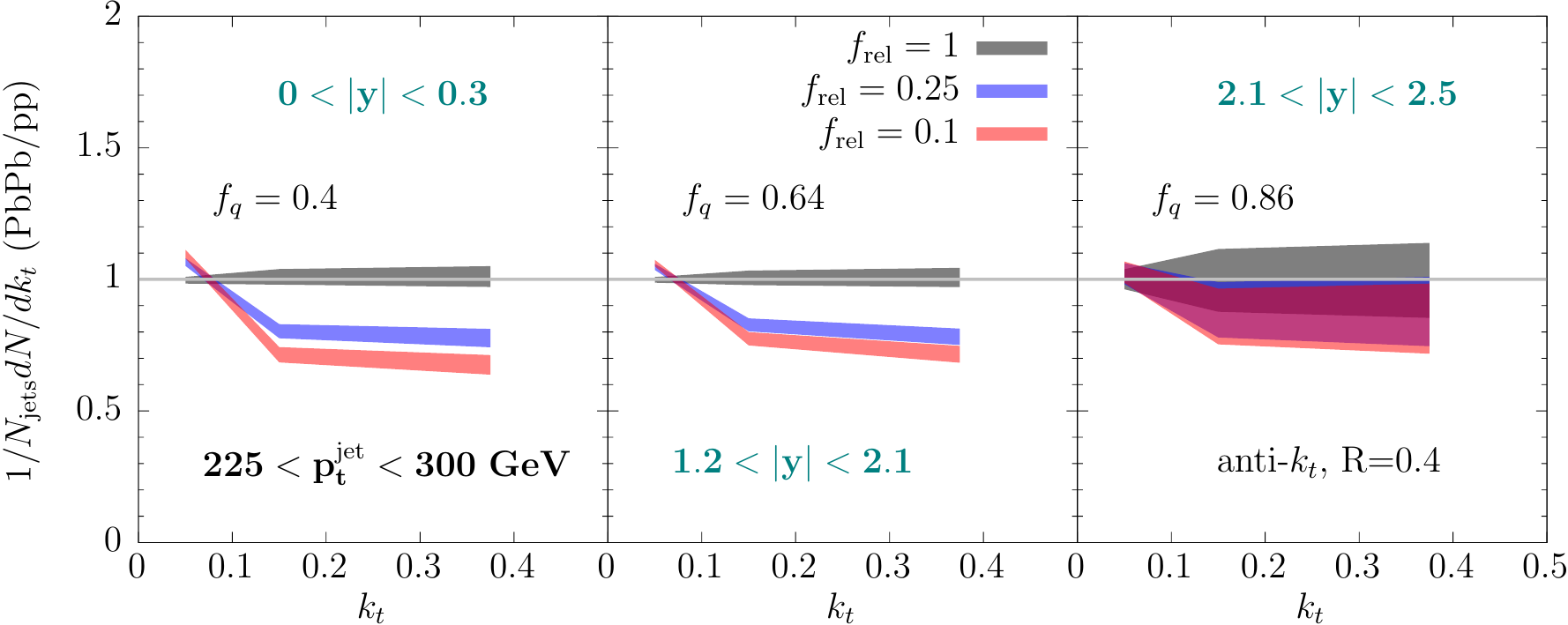}
    \caption{Leading-$k_t$ distribution calculated with the toy model for modified $q/g$ fraction for different rapidity and transverse momentum intervals. Same definition of $k_t$ as in Fig.~\ref{fig:kt_PbPboverpp} is used. }
    \label{fig:kt_toyPbPboverpp}
\end{figure*}
The goal of this section is to extend the analytic results presented in Sec.~\ref{sec:mod-qg} by using as a vacuum baseline templates from \texttt{Pythia} with the estimated statistics of the future HL-LHC run. We do so as to realistically assess the feasibility of future experiments to pin down the differences between the results from a modified $q/g$ fraction hypothesis and those obtained when the medium is capable of resolving the fluctuating substructure of each individual jet, i.e. $L_{\rm res}=0$ results of the Hybrid model presented in the previous section. 

Our starting point consists in expressing $R_{\rm AA}$ in terms of the suppression experienced by quark- and gluon-initiated jets. The definition of jet suppression, Eq.~\eqref{eq:RAA}, can be recast into
\begin{equation}
\label{eq:raafrac}
    R_{\rm AA}=f_q \mathcal Q_q +(1-f_q)\mathcal Q_g \, ,
\end{equation}
where $f_q$ is the quark-initiated jet fraction in vacuum, and $\mathcal Q_q$ and $\mathcal Q_g$ are the (bare) quenching weights of quark-initiated and gluon-initiated jets, respectively. These bare quenching weights can be computed as in Eq.~\eqref{eq:qw-single-parton} within the BDMPS-Z approach. We can further use the fact that $\mathcal Q_i$ scales with the Casimir of the color charge exponentially~\cite{Baier:2001yt}, so that $\mathcal Q_g=\mathcal Q_q^{C_A/C_F}$. For the purposes of the present section, it is convenient to express Eq.~\eqref{eq:raafrac} in terms of the relative suppression of gluon versus quark jets, $f_{\rm rel}$, as
\begin{align}
\label{eq:raafracrel}
    R_{\rm AA}&=\mathcal Q_q[f_q +(1-f_q)\mathcal Q_q^{C_A/C_F-1}] \nonumber \\
    &= \mathcal Q_q[f_q+(1-f_q)f_{\rm rel}] \, .
\end{align}
Let us discuss a few meaningful values of $f_{\rm rel}$. To start with, choosing $f_{\rm rel}=\mathcal Q_q^{C_A/C_F-1}$ corresponds to the relative suppression motivated by the BDMPS-Z physics encapsulated in Eq.~\eqref{eq:pmed-caucal}, which for a reasonable value of $\mathcal Q_q=0.6$ leads to $f_{\rm rel}\approx 0.5$, i.e. a factor of two increase in the relative number of quark jets over gluon jets, due to medium effects, when compared to that same number in vacuum. As mentioned in the introduction, the value $f_{\rm rel}=0.25$ coincides with the one extracted via global fits to the jet fragmentation function~\cite{Qiu:2019sfj}, and used in further quenching studies of the modification of the groomed radius~\cite{Ringer:2019rfk,ALargeIonColliderExperiment:2021mqf}. Finally, we also explore $f_{\rm rel}=0.1$ to test the maximal effects from the modified $q/g$ fraction hypothesis. 

The medium-modified leading-$k_t$ distribution, in close analogy to the formulas presented in Sec.~\ref{sec:mod-qg}, then reads
\begin{equation}
\label{eq:pythia-qgmodel}
    \frac{1}{\sigma}\frac{\dd\sigma}{\dd k_t}\Big\vert_{AA}  = \mathcal{N}^{-1} \left[f_q \frac{\dd\sigma_q}{\dd k_t}\Big\vert_{pp} + f_{\rm rel}(1-f_q)\frac{\dd\sigma_g}{\dd k_t}\Big\vert_{pp} \right],
\end{equation}
where $\mathcal{N}=f_q+f_{\rm rel}(1-f_q)$. Note that the individual distributions are assumed to be unmodified with respect to vacuum. That is, they correspond to templates obtained using \texttt{Pythia} and shown in Fig.~\ref{fig:kt_pp}. An important caveat of Eq.~\eqref{eq:pythia-qgmodel} is that we keep $f_{\rm rel}$ fixed for all values of jet $p_t$ and rapidity. As a consequence, this approach ignores the strong $p_t$-dependence of the $q$-fraction for lower jet $p_t$ at higher rapidities, as highlighted when discussing Fig.~\ref{fig:hybrid_qfracmed}. This $p_t$-dependence becomes important when spectra shifts, due to energy loss, modify the average jet $p_t$ of a given jet ensemble. Keeping $f_{\rm rel}$ fixed also neglects any intrinsic $p_t$-dependence of the substructure observable in consideration. Accounting for all these effects would require $f_{\rm rel}\to f_{\rm rel}(p_t,y)$.

In Fig.~\ref{fig:kt_toyPbPboverpp} we show results for the modification of the leading-$k_t$ distribution using the toy model encapsulated in Eq.~\eqref{eq:pythia-qgmodel}. They are presented as ratios with respect to the vacuum result, using the same kinematic cuts as in Figs.~\ref{fig:kt_pp} and \ref{fig:kt_PbPboverpp}. The presence of the grey band responds to a sanity check that Eq.~\eqref{eq:pythia-qgmodel} reproduces the total vacuum result, $f_{\rm rel}=1$, given the correct $f_q$. The values of $f_q$ are estimated from Fig.~\ref{fig:qfracpp}, and then more finely tuned to get an exact ratio of 1. The statistics used for the $Pb+Pb$ samples approximately correspond to those estimated in Table~\eqref{tab:njets}. In turn, the $p+p$ reference has a factor $\sim 10$ higher statistics.

The first observation concerning Fig.~\ref{fig:kt_toyPbPboverpp} is that this toy model is able to generate a visible narrowing of the leading-$k_t$ distribution at mid-rapidity for all $p_t$ intervals, as it was the case in the comparison to the groomed jet radius data in Ref.~\cite{ALargeIonColliderExperiment:2021mqf}. In agreement with the analytic results of Sec.~\ref{sec:mod-qg}, the ratio becomes close to one as one evolves towards more forward rapidities. Note that for the low-$p_t$ bin the narrowing persists even at the most forward rapidity bin since the value of $f_q$ is smaller than for the other $p_t$ bins. The key outcome of this exercise is that we can now quantify for which values of jet $p_t$ and rapidity can one discriminate between our two predictions: the one based on the modified $q/g$ fraction, and that of the Hybrid model with $L_{\rm res}=0$. By comparing Figs.~\ref{fig:kt_PbPboverpp} and \ref{fig:kt_toyPbPboverpp} we identify two jet selection cuts that achieve such goal: (i) $2.1<|y|<2.5$ and $225<p_t<300$ GeV, and (ii) $3<|y|<4$ and $100<p_t<150$ GeV. The former is potentially accessible with the current technology of ATLAS and CMS experiments. The latter requires detector upgrades and higher statistics, so it can be targeted by all four LHC experiments beyond Run 3. This is the main result of this paper. 

\section{Conclusions}
\label{sec:conclusions}
Jet substructure measurements at LHC energies have revealed that the hard core of the jet is narrower in heavy-ion collisions than in $p+p$. A lack of consensus exists in the theoretical interpretation of this narrowing effect.
On the one hand, fully coherent energy loss models propose
that a relative increase in the number of quark-initiated jets, known to possess a narrower fragmentation, with respect to vacuum suffices to explain current measurements. This increase in the number of quark-initiated jets would be based on the fact that they lose less energy than gluon-initiated jets due to their different color charge, and are more likely to pass the jet $p_t$ cut required to enter the inclusive jet ensemble. It is noteworthy that in order to quantitatively describe the data, such a model requires a quark fraction 4 times larger than in $p+p$, a number far beyond any estimate based on the Casimir scaling of energy loss. The second category of models argue that the jet sample is mainly dominated by unresolved splittings. That is, only sufficiently wide splittings $\theta>\theta_c$, either quark- or gluon- initiated, will be resolved by the medium, lose more energy and make the energy of the jet they belong to fall below the $p_t$-selection cut. 

This paper proposes a rapidity scan of jet substructure measurements to pin down the origin of the narrowing effect. We show that a gradual increase of the jet rapidity enhances the fraction of quark-initiated jets in $p+p$. Then, in a fully coherent picture of energy loss the narrowing effect would vanish at asymptotically forward rapidites, where the flavour of the jet-initiator is fixed. In turn, the resolution scale of the medium $\theta_c$ is independent of the jet rapidity and, thus, if the narrowing effect is driven by this phenomenon it should persist for all rapidity values. To make these statements more quantitative we have studied the relative transverse momentum distribution of the hardest splitting in a jet via analytic toy models and Monte Carlo simulations, the latter using the expected statistics of the high-luminosity run of the LHC. We demonstrate that the the current rapidity reach of LHC experiments at high-$p_t$, i.e. $225<p_t<300$ GeV, would already have strong discriminating power between these two theoretical interpretations of the narrowing effect. Even stronger constraints can be achieved with detector upgrades capable of measuring jet substructure at $|y|>3$ for moderate jet $p_t$, i.e. $100<p_t<150$ GeV. This kinematic regime is well within the expected acceptance of all LHC experiments in Run 4 and beyond.  

Complementary constraints on the interaction of quark-initiated and gluon-initiated jets with the QGP can be obtained by the comparison of the suppression patterns of inclusive jets and (i) heavy-flavour tagged jets~\cite{CMS:2013qak,ALICE:2022phr,Li:2017wwc}, which are expected to be initiated by either a charm or a bottom quark, or (ii) $Z/\gamma$+jet events~\cite{Gallicchio:2011xc,Wang:2013cia,ATLAS:2019dsv,CMS:2018mqn,Kang:2017xnc,Yang:2021qtl,Brewer:2021hmh,Takacs:2021bpv}, where the quark fraction is also enhanced due to the Born level matrix element. In our opinion, measurements at forward rapidities are cleaner than these two alternative options since the latter are subject to multiple sources of contamination. For example, a heavy-flavour tagged jet can originate from a $g\to q\bar q$ splitting, and so a  non-vanishing fraction of $g$-initiated jets makes its way into the sample~\cite{Attems:2022otp}. Equivalently, sophisticated isolation cuts are required in $\gamma$+jet events to guarantee that the reconstructed photon was generated in the hard scattering process and not in the aftermath of the collision via electroweak radiation -- and even after applying these cuts a sizeable number of these secondary photons still persist in the final sample. Finally, $Z$-tagged events suffer from low statistics~\cite{CMS:2021otx}. It would be interesting to perform a more systematic comparison of the pros and cons of these alternative methods with respect to the measurements at forward rapidities proposed in this work. We leave this task for future studies.

\acknowledgements

We are grateful to Benjamin Audurier, Cristian Baldenegro, Gian Michele Innocenti, Dennis Perepelitsa, Martin Rybar and Ricardo V\'azquez for fruitful exchanges on the experimental feasibility of rapidity-dependent jet substructure measurements. We also thank Carlota Andrés for guidance on the use of the nPDF error sets.
The authors would like to thank the organizers of the XXIXth International Conference on Ultra-relativistic Nucleus-Nucleus Collisions, Quark Matter 2022, which took place in Krakow, Poland, where this collaboration started. 
DP has received funding from the European Union's Horizon 2020 research and innovation program under the Marie Skłodowska-Curie grant agreement No. 754496. A.S.O.’s work was supported by the European Research Council (ERC) under the European Union’s Horizon 2020 research and innovation programme (grant agreement No. 788223, PanScales). She would also like to thank the Munich Institute for Astro-, Particle and BioPhysics (MIAPbP), funded by the Deutsche Forschungsgemeinschaft (DFG, German Research Foundation) under Germany´s Excellence Strategy – EXC-2094 – 390783311, for hospitality while this publication was finalised.

\bibliography{biblio}

\begin{thebibliography}{92}%
\makeatletter
\providecommand \@ifxundefined [1]{%
 \@ifx{#1\undefined}
}%
\providecommand \@ifnum [1]{%
 \ifnum #1\expandafter \@firstoftwo
 \else \expandafter \@secondoftwo
 \fi
}%
\providecommand \@ifx [1]{%
 \ifx #1\expandafter \@firstoftwo
 \else \expandafter \@secondoftwo
 \fi
}%
\providecommand \natexlab [1]{#1}%
\providecommand \enquote  [1]{``#1''}%
\providecommand \bibnamefont  [1]{#1}%
\providecommand \bibfnamefont [1]{#1}%
\providecommand \citenamefont [1]{#1}%
\providecommand \href@noop [0]{\@secondoftwo}%
\providecommand \href [0]{\begingroup \@sanitize@url \@href}%
\providecommand \@href[1]{\@@startlink{#1}\@@href}%
\providecommand \@@href[1]{\endgroup#1\@@endlink}%
\providecommand \@sanitize@url [0]{\catcode `\\12\catcode `\$12\catcode
  `\&12\catcode `\#12\catcode `\^12\catcode `\_12\catcode `\%12\relax}%
\providecommand \@@startlink[1]{}%
\providecommand \@@endlink[0]{}%
\providecommand \url  [0]{\begingroup\@sanitize@url \@url }%
\providecommand \@url [1]{\endgroup\@href {#1}{\urlprefix }}%
\providecommand \urlprefix  [0]{URL }%
\providecommand \Eprint [0]{\href }%
\providecommand \doibase [0]{http://dx.doi.org/}%
\providecommand \selectlanguage [0]{\@gobble}%
\providecommand \bibinfo  [0]{\@secondoftwo}%
\providecommand \bibfield  [0]{\@secondoftwo}%
\providecommand \translation [1]{[#1]}%
\providecommand \BibitemOpen [0]{}%
\providecommand \bibitemStop [0]{}%
\providecommand \bibitemNoStop [0]{.\EOS\space}%
\providecommand \EOS [0]{\spacefactor3000\relax}%
\providecommand \BibitemShut  [1]{\csname bibitem#1\endcsname}%
\let\auto@bib@innerbib\@empty
\bibitem [{\citenamefont {Baym}(2016)}]{Baym:2016wox}%
  \BibitemOpen
  \bibfield  {author} {\bibinfo {author} {\bibfnamefont {G.}~\bibnamefont
  {Baym}},\ }\href {\doibase 10.1016/j.nuclphysa.2016.03.007} {\bibfield
  {journal} {\bibinfo  {journal} {Nucl. Phys. A}\ }\textbf {\bibinfo {volume}
  {956}},\ \bibinfo {pages} {1} (\bibinfo {year} {2016})},\ \Eprint
  {http://arxiv.org/abs/1701.03972} {arXiv:1701.03972 [nucl-ex]} \BibitemShut
  {NoStop}%
\bibitem [{\citenamefont {Busza}\ \emph {et~al.}(2018)\citenamefont {Busza},
  \citenamefont {Rajagopal},\ and\ \citenamefont {van~der
  Schee}}]{Busza:2018rrf}%
  \BibitemOpen
  \bibfield  {author} {\bibinfo {author} {\bibfnamefont {W.}~\bibnamefont
  {Busza}}, \bibinfo {author} {\bibfnamefont {K.}~\bibnamefont {Rajagopal}}, \
  and\ \bibinfo {author} {\bibfnamefont {W.}~\bibnamefont {van~der Schee}},\
  }\href {\doibase 10.1146/annurev-nucl-101917-020852} {\bibfield  {journal}
  {\bibinfo  {journal} {Ann. Rev. Nucl. Part. Sci.}\ }\textbf {\bibinfo
  {volume} {68}},\ \bibinfo {pages} {339} (\bibinfo {year} {2018})},\ \Eprint
  {http://arxiv.org/abs/1802.04801} {arXiv:1802.04801 [hep-ph]} \BibitemShut
  {NoStop}%
\bibitem [{\citenamefont {Adcox}\ \emph {et~al.}(2002)\citenamefont {Adcox}
  \emph {et~al.}}]{Adcox:2001jp}%
  \BibitemOpen
  \bibfield  {author} {\bibinfo {author} {\bibfnamefont {K.}~\bibnamefont
  {Adcox}} \emph {et~al.} (\bibinfo {collaboration} {PHENIX}),\ }\href
  {\doibase 10.1103/PhysRevLett.88.022301} {\bibfield  {journal} {\bibinfo
  {journal} {Phys. Rev. Lett.}\ }\textbf {\bibinfo {volume} {88}},\ \bibinfo
  {pages} {022301} (\bibinfo {year} {2002})},\ \Eprint
  {http://arxiv.org/abs/nucl-ex/0109003} {arXiv:nucl-ex/0109003} \BibitemShut
  {NoStop}%
\bibitem [{\citenamefont {Adler}\ \emph {et~al.}(2002)\citenamefont {Adler}
  \emph {et~al.}}]{Adler:2002xw}%
  \BibitemOpen
  \bibfield  {author} {\bibinfo {author} {\bibfnamefont {C.}~\bibnamefont
  {Adler}} \emph {et~al.} (\bibinfo {collaboration} {STAR}),\ }\href {\doibase
  10.1103/PhysRevLett.89.202301} {\bibfield  {journal} {\bibinfo  {journal}
  {Phys. Rev. Lett.}\ }\textbf {\bibinfo {volume} {89}},\ \bibinfo {pages}
  {202301} (\bibinfo {year} {2002})},\ \Eprint
  {http://arxiv.org/abs/nucl-ex/0206011} {arXiv:nucl-ex/0206011} \BibitemShut
  {NoStop}%
\bibitem [{\citenamefont {Adam}\ \emph {et~al.}(2020)\citenamefont {Adam} \emph
  {et~al.}}]{STAR:2020xiv}%
  \BibitemOpen
  \bibfield  {author} {\bibinfo {author} {\bibfnamefont {J.}~\bibnamefont
  {Adam}} \emph {et~al.} (\bibinfo {collaboration} {STAR}),\ }\href {\doibase
  10.1103/PhysRevC.102.054913} {\bibfield  {journal} {\bibinfo  {journal}
  {Phys. Rev. C}\ }\textbf {\bibinfo {volume} {102}},\ \bibinfo {pages}
  {054913} (\bibinfo {year} {2020})},\ \Eprint
  {http://arxiv.org/abs/2006.00582} {arXiv:2006.00582 [nucl-ex]} \BibitemShut
  {NoStop}%
\bibitem [{\citenamefont {Adam}\ \emph {et~al.}(2015)\citenamefont {Adam} \emph
  {et~al.}}]{Adam:2015ewa}%
  \BibitemOpen
  \bibfield  {author} {\bibinfo {author} {\bibfnamefont {J.}~\bibnamefont
  {Adam}} \emph {et~al.} (\bibinfo {collaboration} {ALICE}),\ }\href {\doibase
  10.1016/j.physletb.2015.04.039} {\bibfield  {journal} {\bibinfo  {journal}
  {Phys. Lett. B}\ }\textbf {\bibinfo {volume} {746}},\ \bibinfo {pages} {1}
  (\bibinfo {year} {2015})},\ \Eprint {http://arxiv.org/abs/1502.01689}
  {arXiv:1502.01689 [nucl-ex]} \BibitemShut {NoStop}%
\bibitem [{\citenamefont {Khachatryan}\ \emph {et~al.}(2017)\citenamefont
  {Khachatryan} \emph {et~al.}}]{CMS:2016uxf}%
  \BibitemOpen
  \bibfield  {author} {\bibinfo {author} {\bibfnamefont {V.}~\bibnamefont
  {Khachatryan}} \emph {et~al.} (\bibinfo {collaboration} {CMS}),\ }\href
  {\doibase 10.1103/PhysRevC.96.015202} {\bibfield  {journal} {\bibinfo
  {journal} {Phys. Rev. C}\ }\textbf {\bibinfo {volume} {96}},\ \bibinfo
  {pages} {015202} (\bibinfo {year} {2017})},\ \Eprint
  {http://arxiv.org/abs/1609.05383} {arXiv:1609.05383 [nucl-ex]} \BibitemShut
  {NoStop}%
\bibitem [{\citenamefont {Aaboud}\ \emph
  {et~al.}(2019{\natexlab{a}})\citenamefont {Aaboud} \emph
  {et~al.}}]{ATLAS:2018gwx}%
  \BibitemOpen
  \bibfield  {author} {\bibinfo {author} {\bibfnamefont {M.}~\bibnamefont
  {Aaboud}} \emph {et~al.} (\bibinfo {collaboration} {ATLAS}),\ }\href
  {\doibase 10.1016/j.physletb.2018.10.076} {\bibfield  {journal} {\bibinfo
  {journal} {Phys. Lett. B}\ }\textbf {\bibinfo {volume} {790}},\ \bibinfo
  {pages} {108} (\bibinfo {year} {2019}{\natexlab{a}})},\ \Eprint
  {http://arxiv.org/abs/1805.05635} {arXiv:1805.05635 [nucl-ex]} \BibitemShut
  {NoStop}%
\bibitem [{\citenamefont {Sirunyan}\ \emph {et~al.}(2021)\citenamefont
  {Sirunyan} \emph {et~al.}}]{CMS:2021vui}%
  \BibitemOpen
  \bibfield  {author} {\bibinfo {author} {\bibfnamefont {A.~M.}\ \bibnamefont
  {Sirunyan}} \emph {et~al.} (\bibinfo {collaboration} {CMS}),\ }\href
  {\doibase 10.1007/JHEP05(2021)284} {\bibfield  {journal} {\bibinfo  {journal}
  {JHEP}\ }\textbf {\bibinfo {volume} {05}},\ \bibinfo {pages} {284} (\bibinfo
  {year} {2021})},\ \Eprint {http://arxiv.org/abs/2102.13080} {arXiv:2102.13080
  [hep-ex]} \BibitemShut {NoStop}%
\bibitem [{\citenamefont {d'Enterria}(2010)}]{dEnterria:2009xfs}%
  \BibitemOpen
  \bibfield  {author} {\bibinfo {author} {\bibfnamefont {D.}~\bibnamefont
  {d'Enterria}},\ }\href {\doibase 10.1007/978-3-642-01539-7_16} {\bibfield
  {journal} {\bibinfo  {journal} {Landolt-Bornstein}\ }\textbf {\bibinfo
  {volume} {23}},\ \bibinfo {pages} {471} (\bibinfo {year} {2010})},\ \Eprint
  {http://arxiv.org/abs/0902.2011} {arXiv:0902.2011 [nucl-ex]} \BibitemShut
  {NoStop}%
\bibitem [{\citenamefont {Majumder}\ and\ \citenamefont
  {Van~Leeuwen}(2011)}]{Majumder:2010qh}%
  \BibitemOpen
  \bibfield  {author} {\bibinfo {author} {\bibfnamefont {A.}~\bibnamefont
  {Majumder}}\ and\ \bibinfo {author} {\bibfnamefont {M.}~\bibnamefont
  {Van~Leeuwen}},\ }\href {\doibase 10.1016/j.ppnp.2010.09.001} {\bibfield
  {journal} {\bibinfo  {journal} {Prog. Part. Nucl. Phys.}\ }\textbf {\bibinfo
  {volume} {66}},\ \bibinfo {pages} {41} (\bibinfo {year} {2011})},\ \Eprint
  {http://arxiv.org/abs/1002.2206} {arXiv:1002.2206 [hep-ph]} \BibitemShut
  {NoStop}%
\bibitem [{\citenamefont {Mehtar-Tani}\ \emph {et~al.}(2013)\citenamefont
  {Mehtar-Tani}, \citenamefont {Milhano},\ and\ \citenamefont
  {Tywoniuk}}]{Mehtar-Tani:2013pia}%
  \BibitemOpen
  \bibfield  {author} {\bibinfo {author} {\bibfnamefont {Y.}~\bibnamefont
  {Mehtar-Tani}}, \bibinfo {author} {\bibfnamefont {J.~G.}\ \bibnamefont
  {Milhano}}, \ and\ \bibinfo {author} {\bibfnamefont {K.}~\bibnamefont
  {Tywoniuk}},\ }\href {\doibase 10.1142/S0217751X13400137} {\bibfield
  {journal} {\bibinfo  {journal} {Int. J. Mod. Phys. A}\ }\textbf {\bibinfo
  {volume} {28}},\ \bibinfo {pages} {1340013} (\bibinfo {year} {2013})},\
  \Eprint {http://arxiv.org/abs/1302.2579} {arXiv:1302.2579 [hep-ph]}
  \BibitemShut {NoStop}%
\bibitem [{\citenamefont {Sirunyan}\ \emph
  {et~al.}(2018{\natexlab{a}})\citenamefont {Sirunyan} \emph
  {et~al.}}]{CMS:2017qlm}%
  \BibitemOpen
  \bibfield  {author} {\bibinfo {author} {\bibfnamefont {A.~M.}\ \bibnamefont
  {Sirunyan}} \emph {et~al.} (\bibinfo {collaboration} {CMS}),\ }\href
  {\doibase 10.1103/PhysRevLett.120.142302} {\bibfield  {journal} {\bibinfo
  {journal} {Phys. Rev. Lett.}\ }\textbf {\bibinfo {volume} {120}},\ \bibinfo
  {pages} {142302} (\bibinfo {year} {2018}{\natexlab{a}})},\ \Eprint
  {http://arxiv.org/abs/1708.09429} {arXiv:1708.09429 [nucl-ex]} \BibitemShut
  {NoStop}%
\bibitem [{\citenamefont {Acharya}\ \emph {et~al.}(2020)\citenamefont {Acharya}
  \emph {et~al.}}]{ALICE:2019ykw}%
  \BibitemOpen
  \bibfield  {author} {\bibinfo {author} {\bibfnamefont {S.}~\bibnamefont
  {Acharya}} \emph {et~al.} (\bibinfo {collaboration} {ALICE}),\ }\href
  {\doibase 10.1016/j.physletb.2020.135227} {\bibfield  {journal} {\bibinfo
  {journal} {Phys. Lett. B}\ }\textbf {\bibinfo {volume} {802}},\ \bibinfo
  {pages} {135227} (\bibinfo {year} {2020})},\ \Eprint
  {http://arxiv.org/abs/1905.02512} {arXiv:1905.02512 [nucl-ex]} \BibitemShut
  {NoStop}%
\bibitem [{\citenamefont {Acharya}\ \emph {et~al.}(2022)\citenamefont {Acharya}
  \emph {et~al.}}]{ALargeIonColliderExperiment:2021mqf}%
  \BibitemOpen
  \bibfield  {author} {\bibinfo {author} {\bibfnamefont {S.}~\bibnamefont
  {Acharya}} \emph {et~al.} (\bibinfo {collaboration} {A Large Ion Collider
  Experiment, ALICE}),\ }\href {\doibase 10.1103/PhysRevLett.128.102001}
  {\bibfield  {journal} {\bibinfo  {journal} {Phys. Rev. Lett.}\ }\textbf
  {\bibinfo {volume} {128}},\ \bibinfo {pages} {102001} (\bibinfo {year}
  {2022})},\ \Eprint {http://arxiv.org/abs/2107.12984} {arXiv:2107.12984
  [nucl-ex]} \BibitemShut {NoStop}%
\bibitem [{\citenamefont {Cunqueiro}\ and\ \citenamefont
  {Sickles}(2022)}]{Cunqueiro:2021wls}%
  \BibitemOpen
  \bibfield  {author} {\bibinfo {author} {\bibfnamefont {L.}~\bibnamefont
  {Cunqueiro}}\ and\ \bibinfo {author} {\bibfnamefont {A.~M.}\ \bibnamefont
  {Sickles}},\ }\href {\doibase 10.1016/j.ppnp.2022.103940} {\bibfield
  {journal} {\bibinfo  {journal} {Prog. Part. Nucl. Phys.}\ }\textbf {\bibinfo
  {volume} {124}},\ \bibinfo {pages} {103940} (\bibinfo {year} {2022})},\
  \Eprint {http://arxiv.org/abs/2110.14490} {arXiv:2110.14490 [nucl-ex]}
  \BibitemShut {NoStop}%
\bibitem [{\citenamefont {Spousta}\ and\ \citenamefont
  {Cole}(2016)}]{Spousta:2015fca}%
  \BibitemOpen
  \bibfield  {author} {\bibinfo {author} {\bibfnamefont {M.}~\bibnamefont
  {Spousta}}\ and\ \bibinfo {author} {\bibfnamefont {B.}~\bibnamefont {Cole}},\
  }\href {\doibase 10.1140/epjc/s10052-016-3896-0} {\bibfield  {journal}
  {\bibinfo  {journal} {Eur. Phys. J. C}\ }\textbf {\bibinfo {volume} {76}},\
  \bibinfo {pages} {50} (\bibinfo {year} {2016})},\ \Eprint
  {http://arxiv.org/abs/1504.05169} {arXiv:1504.05169 [hep-ph]} \BibitemShut
  {NoStop}%
\bibitem [{\citenamefont {Ringer}\ \emph {et~al.}(2020)\citenamefont {Ringer},
  \citenamefont {Xiao},\ and\ \citenamefont {Yuan}}]{Ringer:2019rfk}%
  \BibitemOpen
  \bibfield  {author} {\bibinfo {author} {\bibfnamefont {F.}~\bibnamefont
  {Ringer}}, \bibinfo {author} {\bibfnamefont {B.-W.}\ \bibnamefont {Xiao}}, \
  and\ \bibinfo {author} {\bibfnamefont {F.}~\bibnamefont {Yuan}},\ }\href
  {\doibase 10.1016/j.physletb.2020.135634} {\bibfield  {journal} {\bibinfo
  {journal} {Phys. Lett. B}\ }\textbf {\bibinfo {volume} {808}},\ \bibinfo
  {pages} {135634} (\bibinfo {year} {2020})},\ \Eprint
  {http://arxiv.org/abs/1907.12541} {arXiv:1907.12541 [hep-ph]} \BibitemShut
  {NoStop}%
\bibitem [{\citenamefont {Qiu}\ \emph {et~al.}(2019)\citenamefont {Qiu},
  \citenamefont {Ringer}, \citenamefont {Sato},\ and\ \citenamefont
  {Zurita}}]{Qiu:2019sfj}%
  \BibitemOpen
  \bibfield  {author} {\bibinfo {author} {\bibfnamefont {J.-W.}\ \bibnamefont
  {Qiu}}, \bibinfo {author} {\bibfnamefont {F.}~\bibnamefont {Ringer}},
  \bibinfo {author} {\bibfnamefont {N.}~\bibnamefont {Sato}}, \ and\ \bibinfo
  {author} {\bibfnamefont {P.}~\bibnamefont {Zurita}},\ }\href {\doibase
  10.1103/PhysRevLett.122.252301} {\bibfield  {journal} {\bibinfo  {journal}
  {Phys. Rev. Lett.}\ }\textbf {\bibinfo {volume} {122}},\ \bibinfo {pages}
  {252301} (\bibinfo {year} {2019})},\ \Eprint
  {http://arxiv.org/abs/1903.01993} {arXiv:1903.01993 [hep-ph]} \BibitemShut
  {NoStop}%
\bibitem [{\citenamefont {Brewer}\ \emph {et~al.}(2021)\citenamefont {Brewer},
  \citenamefont {Thaler},\ and\ \citenamefont {Turner}}]{Brewer:2020och}%
  \BibitemOpen
  \bibfield  {author} {\bibinfo {author} {\bibfnamefont {J.}~\bibnamefont
  {Brewer}}, \bibinfo {author} {\bibfnamefont {J.}~\bibnamefont {Thaler}}, \
  and\ \bibinfo {author} {\bibfnamefont {A.~P.}\ \bibnamefont {Turner}},\
  }\href {\doibase 10.1103/PhysRevC.103.L021901} {\bibfield  {journal}
  {\bibinfo  {journal} {Phys. Rev. C}\ }\textbf {\bibinfo {volume} {103}},\
  \bibinfo {pages} {L021901} (\bibinfo {year} {2021})},\ \Eprint
  {http://arxiv.org/abs/2008.08596} {arXiv:2008.08596 [hep-ph]} \BibitemShut
  {NoStop}%
\bibitem [{\citenamefont {Sirunyan}\ \emph {et~al.}(2020)\citenamefont
  {Sirunyan} \emph {et~al.}}]{CMS:2020plq}%
  \BibitemOpen
  \bibfield  {author} {\bibinfo {author} {\bibfnamefont {A.~M.}\ \bibnamefont
  {Sirunyan}} \emph {et~al.} (\bibinfo {collaboration} {CMS}),\ }\href
  {\doibase 10.1007/JHEP07(2020)115} {\bibfield  {journal} {\bibinfo  {journal}
  {JHEP}\ }\textbf {\bibinfo {volume} {07}},\ \bibinfo {pages} {115} (\bibinfo
  {year} {2020})},\ \Eprint {http://arxiv.org/abs/2004.00602} {arXiv:2004.00602
  [hep-ex]} \BibitemShut {NoStop}%
\bibitem [{\citenamefont {Li}\ and\ \citenamefont {Vitev}(2020)}]{Li:2019dre}%
  \BibitemOpen
  \bibfield  {author} {\bibinfo {author} {\bibfnamefont {H.~T.}\ \bibnamefont
  {Li}}\ and\ \bibinfo {author} {\bibfnamefont {I.}~\bibnamefont {Vitev}},\
  }\href {\doibase 10.1103/PhysRevD.101.076020} {\bibfield  {journal} {\bibinfo
   {journal} {Phys. Rev. D}\ }\textbf {\bibinfo {volume} {101}},\ \bibinfo
  {pages} {076020} (\bibinfo {year} {2020})},\ \Eprint
  {http://arxiv.org/abs/1908.06979} {arXiv:1908.06979 [hep-ph]} \BibitemShut
  {NoStop}%
\bibitem [{\citenamefont {Apolin\'ario}\ \emph {et~al.}(2020)\citenamefont
  {Apolin\'ario}, \citenamefont {Barata},\ and\ \citenamefont
  {Milhano}}]{Apolinario:2020nyw}%
  \BibitemOpen
  \bibfield  {author} {\bibinfo {author} {\bibfnamefont {L.}~\bibnamefont
  {Apolin\'ario}}, \bibinfo {author} {\bibfnamefont {J.~a.}\ \bibnamefont
  {Barata}}, \ and\ \bibinfo {author} {\bibfnamefont {G.}~\bibnamefont
  {Milhano}},\ }\href {\doibase 10.1140/epjc/s10052-020-8133-1} {\bibfield
  {journal} {\bibinfo  {journal} {Eur. Phys. J. C}\ }\textbf {\bibinfo {volume}
  {80}},\ \bibinfo {pages} {586} (\bibinfo {year} {2020})},\ \Eprint
  {http://arxiv.org/abs/2003.02893} {arXiv:2003.02893 [hep-ph]} \BibitemShut
  {NoStop}%
\bibitem [{\citenamefont {Mehtar-Tani}\ \emph {et~al.}(2011)\citenamefont
  {Mehtar-Tani}, \citenamefont {Salgado},\ and\ \citenamefont
  {Tywoniuk}}]{Mehtar-Tani:2010ebp}%
  \BibitemOpen
  \bibfield  {author} {\bibinfo {author} {\bibfnamefont {Y.}~\bibnamefont
  {Mehtar-Tani}}, \bibinfo {author} {\bibfnamefont {C.~A.}\ \bibnamefont
  {Salgado}}, \ and\ \bibinfo {author} {\bibfnamefont {K.}~\bibnamefont
  {Tywoniuk}},\ }\href {\doibase 10.1103/PhysRevLett.106.122002} {\bibfield
  {journal} {\bibinfo  {journal} {Phys. Rev. Lett.}\ }\textbf {\bibinfo
  {volume} {106}},\ \bibinfo {pages} {122002} (\bibinfo {year} {2011})},\
  \Eprint {http://arxiv.org/abs/1009.2965} {arXiv:1009.2965 [hep-ph]}
  \BibitemShut {NoStop}%
\bibitem [{\citenamefont {Mehtar-Tani}\ \emph
  {et~al.}(2012{\natexlab{a}})\citenamefont {Mehtar-Tani}, \citenamefont
  {Salgado},\ and\ \citenamefont {Tywoniuk}}]{Mehtar-Tani:2011hma}%
  \BibitemOpen
  \bibfield  {author} {\bibinfo {author} {\bibfnamefont {Y.}~\bibnamefont
  {Mehtar-Tani}}, \bibinfo {author} {\bibfnamefont {C.~A.}\ \bibnamefont
  {Salgado}}, \ and\ \bibinfo {author} {\bibfnamefont {K.}~\bibnamefont
  {Tywoniuk}},\ }\href {\doibase 10.1016/j.physletb.2011.12.042} {\bibfield
  {journal} {\bibinfo  {journal} {Phys. Lett. B}\ }\textbf {\bibinfo {volume}
  {707}},\ \bibinfo {pages} {156} (\bibinfo {year} {2012}{\natexlab{a}})},\
  \Eprint {http://arxiv.org/abs/1102.4317} {arXiv:1102.4317 [hep-ph]}
  \BibitemShut {NoStop}%
\bibitem [{\citenamefont {Casalderrey-Solana}\ and\ \citenamefont
  {Iancu}(2011)}]{Casalderrey-Solana:2011ule}%
  \BibitemOpen
  \bibfield  {author} {\bibinfo {author} {\bibfnamefont {J.}~\bibnamefont
  {Casalderrey-Solana}}\ and\ \bibinfo {author} {\bibfnamefont
  {E.}~\bibnamefont {Iancu}},\ }\href {\doibase 10.1007/JHEP08(2011)015}
  {\bibfield  {journal} {\bibinfo  {journal} {JHEP}\ }\textbf {\bibinfo
  {volume} {08}},\ \bibinfo {pages} {015} (\bibinfo {year} {2011})},\ \Eprint
  {http://arxiv.org/abs/1105.1760} {arXiv:1105.1760 [hep-ph]} \BibitemShut
  {NoStop}%
\bibitem [{\citenamefont {Mehtar-Tani}\ and\ \citenamefont
  {Tywoniuk}(2013)}]{Mehtar-Tani:2011vlz}%
  \BibitemOpen
  \bibfield  {author} {\bibinfo {author} {\bibfnamefont {Y.}~\bibnamefont
  {Mehtar-Tani}}\ and\ \bibinfo {author} {\bibfnamefont {K.}~\bibnamefont
  {Tywoniuk}},\ }\href {\doibase 10.1007/JHEP01(2013)031} {\bibfield  {journal}
  {\bibinfo  {journal} {JHEP}\ }\textbf {\bibinfo {volume} {01}},\ \bibinfo
  {pages} {031} (\bibinfo {year} {2013})},\ \Eprint
  {http://arxiv.org/abs/1105.1346} {arXiv:1105.1346 [hep-ph]} \BibitemShut
  {NoStop}%
\bibitem [{\citenamefont {Casalderrey-Solana}\ \emph
  {et~al.}(2013)\citenamefont {Casalderrey-Solana}, \citenamefont
  {Mehtar-Tani}, \citenamefont {Salgado},\ and\ \citenamefont
  {Tywoniuk}}]{Casalderrey-Solana:2012evi}%
  \BibitemOpen
  \bibfield  {author} {\bibinfo {author} {\bibfnamefont {J.}~\bibnamefont
  {Casalderrey-Solana}}, \bibinfo {author} {\bibfnamefont {Y.}~\bibnamefont
  {Mehtar-Tani}}, \bibinfo {author} {\bibfnamefont {C.~A.}\ \bibnamefont
  {Salgado}}, \ and\ \bibinfo {author} {\bibfnamefont {K.}~\bibnamefont
  {Tywoniuk}},\ }\href {\doibase 10.1016/j.physletb.2013.07.046} {\bibfield
  {journal} {\bibinfo  {journal} {Phys. Lett. B}\ }\textbf {\bibinfo {volume}
  {725}},\ \bibinfo {pages} {357} (\bibinfo {year} {2013})},\ \Eprint
  {http://arxiv.org/abs/1210.7765} {arXiv:1210.7765 [hep-ph]} \BibitemShut
  {NoStop}%
\bibitem [{\citenamefont {Apolin\'ario}\ \emph {et~al.}(2015)\citenamefont
  {Apolin\'ario}, \citenamefont {Armesto}, \citenamefont {Milhano},\ and\
  \citenamefont {Salgado}}]{Apolinario:2014csa}%
  \BibitemOpen
  \bibfield  {author} {\bibinfo {author} {\bibfnamefont {L.}~\bibnamefont
  {Apolin\'ario}}, \bibinfo {author} {\bibfnamefont {N.}~\bibnamefont
  {Armesto}}, \bibinfo {author} {\bibfnamefont {J.~G.}\ \bibnamefont
  {Milhano}}, \ and\ \bibinfo {author} {\bibfnamefont {C.~A.}\ \bibnamefont
  {Salgado}},\ }\href {\doibase 10.1007/JHEP02(2015)119} {\bibfield  {journal}
  {\bibinfo  {journal} {JHEP}\ }\textbf {\bibinfo {volume} {02}},\ \bibinfo
  {pages} {119} (\bibinfo {year} {2015})},\ \Eprint
  {http://arxiv.org/abs/1407.0599} {arXiv:1407.0599 [hep-ph]} \BibitemShut
  {NoStop}%
\bibitem [{\citenamefont {Wiedemann}(2000)}]{Wiedemann:2000za}%
  \BibitemOpen
  \bibfield  {author} {\bibinfo {author} {\bibfnamefont {U.~A.}\ \bibnamefont
  {Wiedemann}},\ }\href {\doibase 10.1016/S0550-3213(00)00457-0} {\bibfield
  {journal} {\bibinfo  {journal} {Nucl. Phys. B}\ }\textbf {\bibinfo {volume}
  {588}},\ \bibinfo {pages} {303} (\bibinfo {year} {2000})},\ \Eprint
  {http://arxiv.org/abs/hep-ph/0005129} {arXiv:hep-ph/0005129} \BibitemShut
  {NoStop}%
\bibitem [{\citenamefont {Gyulassy}\ \emph {et~al.}(2000)\citenamefont
  {Gyulassy}, \citenamefont {Levai},\ and\ \citenamefont
  {Vitev}}]{Gyulassy:2000fs}%
  \BibitemOpen
  \bibfield  {author} {\bibinfo {author} {\bibfnamefont {M.}~\bibnamefont
  {Gyulassy}}, \bibinfo {author} {\bibfnamefont {P.}~\bibnamefont {Levai}}, \
  and\ \bibinfo {author} {\bibfnamefont {I.}~\bibnamefont {Vitev}},\ }\href
  {\doibase 10.1103/PhysRevLett.85.5535} {\bibfield  {journal} {\bibinfo
  {journal} {Phys. Rev. Lett.}\ }\textbf {\bibinfo {volume} {85}},\ \bibinfo
  {pages} {5535} (\bibinfo {year} {2000})},\ \Eprint
  {http://arxiv.org/abs/nucl-th/0005032} {arXiv:nucl-th/0005032} \BibitemShut
  {NoStop}%
\bibitem [{\citenamefont {Sievert}\ and\ \citenamefont
  {Vitev}(2018)}]{Sievert:2018imd}%
  \BibitemOpen
  \bibfield  {author} {\bibinfo {author} {\bibfnamefont {M.~D.}\ \bibnamefont
  {Sievert}}\ and\ \bibinfo {author} {\bibfnamefont {I.}~\bibnamefont
  {Vitev}},\ }\href {\doibase 10.1103/PhysRevD.98.094010} {\bibfield  {journal}
  {\bibinfo  {journal} {Phys. Rev. D}\ }\textbf {\bibinfo {volume} {98}},\
  \bibinfo {pages} {094010} (\bibinfo {year} {2018})},\ \Eprint
  {http://arxiv.org/abs/1807.03799} {arXiv:1807.03799 [hep-ph]} \BibitemShut
  {NoStop}%
\bibitem [{\citenamefont {Mehtar-Tani}\ \emph
  {et~al.}(2012{\natexlab{b}})\citenamefont {Mehtar-Tani}, \citenamefont
  {Salgado},\ and\ \citenamefont {Tywoniuk}}]{Mehtar-Tani:2011lic}%
  \BibitemOpen
  \bibfield  {author} {\bibinfo {author} {\bibfnamefont {Y.}~\bibnamefont
  {Mehtar-Tani}}, \bibinfo {author} {\bibfnamefont {C.~A.}\ \bibnamefont
  {Salgado}}, \ and\ \bibinfo {author} {\bibfnamefont {K.}~\bibnamefont
  {Tywoniuk}},\ }\href {\doibase 10.1007/JHEP04(2012)064} {\bibfield  {journal}
  {\bibinfo  {journal} {JHEP}\ }\textbf {\bibinfo {volume} {04}},\ \bibinfo
  {pages} {064} (\bibinfo {year} {2012}{\natexlab{b}})},\ \Eprint
  {http://arxiv.org/abs/1112.5031} {arXiv:1112.5031 [hep-ph]} \BibitemShut
  {NoStop}%
\bibitem [{\citenamefont {Casalderrey-Solana}\ \emph
  {et~al.}(2016{\natexlab{a}})\citenamefont {Casalderrey-Solana}, \citenamefont
  {Pablos},\ and\ \citenamefont {Tywoniuk}}]{Casalderrey-Solana:2015bww}%
  \BibitemOpen
  \bibfield  {author} {\bibinfo {author} {\bibfnamefont {J.}~\bibnamefont
  {Casalderrey-Solana}}, \bibinfo {author} {\bibfnamefont {D.}~\bibnamefont
  {Pablos}}, \ and\ \bibinfo {author} {\bibfnamefont {K.}~\bibnamefont
  {Tywoniuk}},\ }\href {\doibase 10.1007/JHEP11(2016)174} {\bibfield  {journal}
  {\bibinfo  {journal} {JHEP}\ }\textbf {\bibinfo {volume} {11}},\ \bibinfo
  {pages} {174} (\bibinfo {year} {2016}{\natexlab{a}})},\ \Eprint
  {http://arxiv.org/abs/1512.07561} {arXiv:1512.07561 [hep-ph]} \BibitemShut
  {NoStop}%
\bibitem [{\citenamefont {Caucal}\ \emph {et~al.}(2018)\citenamefont {Caucal},
  \citenamefont {Iancu}, \citenamefont {Mueller},\ and\ \citenamefont
  {Soyez}}]{Caucal:2018dla}%
  \BibitemOpen
  \bibfield  {author} {\bibinfo {author} {\bibfnamefont {P.}~\bibnamefont
  {Caucal}}, \bibinfo {author} {\bibfnamefont {E.}~\bibnamefont {Iancu}},
  \bibinfo {author} {\bibfnamefont {A.~H.}\ \bibnamefont {Mueller}}, \ and\
  \bibinfo {author} {\bibfnamefont {G.}~\bibnamefont {Soyez}},\ }\href
  {\doibase 10.1103/PhysRevLett.120.232001} {\bibfield  {journal} {\bibinfo
  {journal} {Phys. Rev. Lett.}\ }\textbf {\bibinfo {volume} {120}},\ \bibinfo
  {pages} {232001} (\bibinfo {year} {2018})},\ \Eprint
  {http://arxiv.org/abs/1801.09703} {arXiv:1801.09703 [hep-ph]} \BibitemShut
  {NoStop}%
\bibitem [{\citenamefont {Caucal}\ \emph {et~al.}(2019)\citenamefont {Caucal},
  \citenamefont {Iancu},\ and\ \citenamefont {Soyez}}]{Caucal:2019uvr}%
  \BibitemOpen
  \bibfield  {author} {\bibinfo {author} {\bibfnamefont {P.}~\bibnamefont
  {Caucal}}, \bibinfo {author} {\bibfnamefont {E.}~\bibnamefont {Iancu}}, \
  and\ \bibinfo {author} {\bibfnamefont {G.}~\bibnamefont {Soyez}},\ }\href
  {\doibase 10.1007/JHEP10(2019)273} {\bibfield  {journal} {\bibinfo  {journal}
  {JHEP}\ }\textbf {\bibinfo {volume} {10}},\ \bibinfo {pages} {273} (\bibinfo
  {year} {2019})},\ \Eprint {http://arxiv.org/abs/1907.04866} {arXiv:1907.04866
  [hep-ph]} \BibitemShut {NoStop}%
\bibitem [{\citenamefont {Casalderrey-Solana}\ \emph
  {et~al.}(2014)\citenamefont {Casalderrey-Solana}, \citenamefont {Gulhan},
  \citenamefont {Milhano}, \citenamefont {Pablos},\ and\ \citenamefont
  {Rajagopal}}]{Casalderrey-Solana:2014bpa}%
  \BibitemOpen
  \bibfield  {author} {\bibinfo {author} {\bibfnamefont {J.}~\bibnamefont
  {Casalderrey-Solana}}, \bibinfo {author} {\bibfnamefont {D.~C.}\ \bibnamefont
  {Gulhan}}, \bibinfo {author} {\bibfnamefont {J.~G.}\ \bibnamefont {Milhano}},
  \bibinfo {author} {\bibfnamefont {D.}~\bibnamefont {Pablos}}, \ and\ \bibinfo
  {author} {\bibfnamefont {K.}~\bibnamefont {Rajagopal}},\ }\href {\doibase
  10.1007/JHEP09(2015)175} {\bibfield  {journal} {\bibinfo  {journal} {JHEP}\
  }\textbf {\bibinfo {volume} {10}},\ \bibinfo {pages} {019} (\bibinfo {year}
  {2014})},\ \bibinfo {note} {[Erratum: JHEP 09, 175 (2015)]},\ \Eprint
  {http://arxiv.org/abs/1405.3864} {arXiv:1405.3864 [hep-ph]} \BibitemShut
  {NoStop}%
\bibitem [{\citenamefont {Hulcher}\ \emph {et~al.}(2018)\citenamefont
  {Hulcher}, \citenamefont {Pablos},\ and\ \citenamefont
  {Rajagopal}}]{Hulcher:2017cpt}%
  \BibitemOpen
  \bibfield  {author} {\bibinfo {author} {\bibfnamefont {Z.}~\bibnamefont
  {Hulcher}}, \bibinfo {author} {\bibfnamefont {D.}~\bibnamefont {Pablos}}, \
  and\ \bibinfo {author} {\bibfnamefont {K.}~\bibnamefont {Rajagopal}},\ }\href
  {\doibase 10.1007/JHEP03(2018)010} {\bibfield  {journal} {\bibinfo  {journal}
  {JHEP}\ }\textbf {\bibinfo {volume} {03}},\ \bibinfo {pages} {010} (\bibinfo
  {year} {2018})},\ \Eprint {http://arxiv.org/abs/1707.05245} {arXiv:1707.05245
  [hep-ph]} \BibitemShut {NoStop}%
\bibitem [{\citenamefont {Casalderrey-Solana}\ \emph
  {et~al.}(2020)\citenamefont {Casalderrey-Solana}, \citenamefont {Milhano},
  \citenamefont {Pablos},\ and\ \citenamefont
  {Rajagopal}}]{Casalderrey-Solana:2019ubu}%
  \BibitemOpen
  \bibfield  {author} {\bibinfo {author} {\bibfnamefont {J.}~\bibnamefont
  {Casalderrey-Solana}}, \bibinfo {author} {\bibfnamefont {G.}~\bibnamefont
  {Milhano}}, \bibinfo {author} {\bibfnamefont {D.}~\bibnamefont {Pablos}}, \
  and\ \bibinfo {author} {\bibfnamefont {K.}~\bibnamefont {Rajagopal}},\ }\href
  {\doibase 10.1007/JHEP01(2020)044} {\bibfield  {journal} {\bibinfo  {journal}
  {JHEP}\ }\textbf {\bibinfo {volume} {01}},\ \bibinfo {pages} {044} (\bibinfo
  {year} {2020})},\ \Eprint {http://arxiv.org/abs/1907.11248} {arXiv:1907.11248
  [hep-ph]} \BibitemShut {NoStop}%
\bibitem [{\citenamefont {Bossi}(2022)}]{Bossi:2022fpc}%
  \BibitemOpen
  \bibfield  {author} {\bibinfo {author} {\bibfnamefont {H.}~\bibnamefont
  {Bossi}} (\bibinfo {collaboration} {ALICE}),\ }in\ \href@noop {} {\emph
  {\bibinfo {booktitle} {{29th International Conference on Ultra-relativistic
  Nucleus-Nucleus Collisions}}}}\ (\bibinfo {year} {2022})\ \Eprint
  {http://arxiv.org/abs/2208.14492} {arXiv:2208.14492 [nucl-ex]} \BibitemShut
  {NoStop}%
\bibitem [{\citenamefont {Aaboud}\ \emph {et~al.}(2018)\citenamefont {Aaboud}
  \emph {et~al.}}]{ATLAS:2018bvp}%
  \BibitemOpen
  \bibfield  {author} {\bibinfo {author} {\bibfnamefont {M.}~\bibnamefont
  {Aaboud}} \emph {et~al.} (\bibinfo {collaboration} {ATLAS}),\ }\href
  {\doibase 10.1103/PhysRevC.98.024908} {\bibfield  {journal} {\bibinfo
  {journal} {Phys. Rev. C}\ }\textbf {\bibinfo {volume} {98}},\ \bibinfo
  {pages} {024908} (\bibinfo {year} {2018})},\ \Eprint
  {http://arxiv.org/abs/1805.05424} {arXiv:1805.05424 [nucl-ex]} \BibitemShut
  {NoStop}%
\bibitem [{\citenamefont {Sirunyan}\ \emph {et~al.}(2019)\citenamefont
  {Sirunyan} \emph {et~al.}}]{CMS:2018yhi}%
  \BibitemOpen
  \bibfield  {author} {\bibinfo {author} {\bibfnamefont {A.~M.}\ \bibnamefont
  {Sirunyan}} \emph {et~al.} (\bibinfo {collaboration} {CMS}),\ }\href
  {\doibase 10.1007/JHEP05(2019)043} {\bibfield  {journal} {\bibinfo  {journal}
  {JHEP}\ }\textbf {\bibinfo {volume} {05}},\ \bibinfo {pages} {043} (\bibinfo
  {year} {2019})},\ \Eprint {http://arxiv.org/abs/1812.01691} {arXiv:1812.01691
  [hep-ex]} \BibitemShut {NoStop}%
\bibitem [{\citenamefont {Khachatryan}\ \emph {et~al.}(2021)\citenamefont
  {Khachatryan} \emph {et~al.}}]{CMS:2020ldm}%
  \BibitemOpen
  \bibfield  {author} {\bibinfo {author} {\bibfnamefont {V.}~\bibnamefont
  {Khachatryan}} \emph {et~al.} (\bibinfo {collaboration} {CMS}),\ }\href
  {\doibase 10.1088/1748-0221/16/02/P02010} {\bibfield  {journal} {\bibinfo
  {journal} {JINST}\ }\textbf {\bibinfo {volume} {16}},\ \bibinfo {pages}
  {P02010} (\bibinfo {year} {2021})},\ \Eprint
  {http://arxiv.org/abs/2011.01185} {arXiv:2011.01185 [physics.ins-det]}
  \BibitemShut {NoStop}%
\bibitem [{\citenamefont {Citron}\ \emph {et~al.}(2019)\citenamefont {Citron}
  \emph {et~al.}}]{Citron:2018lsq}%
  \BibitemOpen
  \bibfield  {author} {\bibinfo {author} {\bibfnamefont {Z.}~\bibnamefont
  {Citron}} \emph {et~al.},\ }\href {\doibase 10.23731/CYRM-2019-007.1159}
  {\bibfield  {journal} {\bibinfo  {journal} {CERN Yellow Rep. Monogr.}\
  }\textbf {\bibinfo {volume} {7}},\ \bibinfo {pages} {1159} (\bibinfo {year}
  {2019})},\ \Eprint {http://arxiv.org/abs/1812.06772} {arXiv:1812.06772
  [hep-ph]} \BibitemShut {NoStop}%
\bibitem [{\citenamefont {ALICE}(2022{\natexlab{a}})}]{ALICE:2803563}%
  \BibitemOpen
  \bibfield  {author} {\bibinfo {author} {\bibnamefont {ALICE}},\ }\href
  {https://cds.cern.ch/record/2803563} {\emph {\bibinfo {title} {{Letter of
  intent for ALICE 3: A next generation heavy-ion experiment at the LHC}}}},\
  \bibinfo {type} {Tech. Rep.}\ (\bibinfo  {institution} {CERN},\ \bibinfo
  {address} {Geneva},\ \bibinfo {year} {2022})\BibitemShut {NoStop}%
\bibitem [{\citenamefont {CMS}(2017)}]{CMS:2017jpq}%
  \BibitemOpen
  \bibfield  {author} {\bibinfo {author} {\bibnamefont {CMS}},\ }\href
  {https://cds.cern.ch/record/2293646} {\emph {\bibinfo {title} {{The Phase-2
  Upgrade of the CMS Endcap Calorimeter}}}},\ \bibinfo {type} {Tech. Rep.}\
  (\bibinfo {year} {2017})\BibitemShut {NoStop}%
\bibitem [{\citenamefont {ATLAS}(2021)}]{ATLAS:2021yvc}%
  \BibitemOpen
  \bibfield  {author} {\bibinfo {author} {\bibnamefont {ATLAS}},\ }\href
  {http://cds.cern.ch/record/2776651} {\emph {\bibinfo {title} {{Expected
  tracking and related performance with the updated ATLAS Inner Tracker layout
  at the High-Luminosity LHC}}}},\ \bibinfo {type} {Tech. Rep.}\ (\bibinfo
  {year} {2021})\BibitemShut {NoStop}%
\bibitem [{\citenamefont {Bierlich}\ \emph {et~al.}(2022)\citenamefont
  {Bierlich} \emph {et~al.}}]{Bierlich:2022pfr}%
  \BibitemOpen
  \bibfield  {author} {\bibinfo {author} {\bibfnamefont {C.}~\bibnamefont
  {Bierlich}} \emph {et~al.},\ }\href@noop {} {\  (\bibinfo {year} {2022})},\
  \Eprint {http://arxiv.org/abs/2203.11601} {arXiv:2203.11601 [hep-ph]}
  \BibitemShut {NoStop}%
\bibitem [{\citenamefont {Caucal}\ \emph {et~al.}(2022)\citenamefont {Caucal},
  \citenamefont {Soto-Ontoso},\ and\ \citenamefont {Takacs}}]{Caucal:2021cfb}%
  \BibitemOpen
  \bibfield  {author} {\bibinfo {author} {\bibfnamefont {P.}~\bibnamefont
  {Caucal}}, \bibinfo {author} {\bibfnamefont {A.}~\bibnamefont {Soto-Ontoso}},
  \ and\ \bibinfo {author} {\bibfnamefont {A.}~\bibnamefont {Takacs}},\ }\href
  {\doibase 10.1103/PhysRevD.105.114046} {\bibfield  {journal} {\bibinfo
  {journal} {Phys. Rev. D}\ }\textbf {\bibinfo {volume} {105}},\ \bibinfo
  {pages} {114046} (\bibinfo {year} {2022})},\ \Eprint
  {http://arxiv.org/abs/2111.14768} {arXiv:2111.14768 [hep-ph]} \BibitemShut
  {NoStop}%
\bibitem [{\citenamefont {Cacciari}\ \emph {et~al.}(2008)\citenamefont
  {Cacciari}, \citenamefont {Salam},\ and\ \citenamefont
  {Soyez}}]{Cacciari:2008gp}%
  \BibitemOpen
  \bibfield  {author} {\bibinfo {author} {\bibfnamefont {M.}~\bibnamefont
  {Cacciari}}, \bibinfo {author} {\bibfnamefont {G.~P.}\ \bibnamefont {Salam}},
  \ and\ \bibinfo {author} {\bibfnamefont {G.}~\bibnamefont {Soyez}},\ }\href
  {\doibase 10.1088/1126-6708/2008/04/063} {\bibfield  {journal} {\bibinfo
  {journal} {JHEP}\ }\textbf {\bibinfo {volume} {04}},\ \bibinfo {pages} {063}
  (\bibinfo {year} {2008})},\ \Eprint {http://arxiv.org/abs/0802.1189}
  {arXiv:0802.1189 [hep-ph]} \BibitemShut {NoStop}%
\bibitem [{\citenamefont {Cacciari}\ \emph {et~al.}(2012)\citenamefont
  {Cacciari}, \citenamefont {Salam},\ and\ \citenamefont
  {Soyez}}]{Cacciari:2011ma}%
  \BibitemOpen
  \bibfield  {author} {\bibinfo {author} {\bibfnamefont {M.}~\bibnamefont
  {Cacciari}}, \bibinfo {author} {\bibfnamefont {G.~P.}\ \bibnamefont {Salam}},
  \ and\ \bibinfo {author} {\bibfnamefont {G.}~\bibnamefont {Soyez}},\ }\href
  {\doibase 10.1140/epjc/s10052-012-1896-2} {\bibfield  {journal} {\bibinfo
  {journal} {Eur. Phys. J. C}\ }\textbf {\bibinfo {volume} {72}},\ \bibinfo
  {pages} {1896} (\bibinfo {year} {2012})},\ \Eprint
  {http://arxiv.org/abs/1111.6097} {arXiv:1111.6097 [hep-ph]} \BibitemShut
  {NoStop}%
\bibitem [{\citenamefont {Mehtar-Tani}\ \emph {et~al.}(2020)\citenamefont
  {Mehtar-Tani}, \citenamefont {Soto-Ontoso},\ and\ \citenamefont
  {Tywoniuk}}]{Mehtar-Tani:2019rrk}%
  \BibitemOpen
  \bibfield  {author} {\bibinfo {author} {\bibfnamefont {Y.}~\bibnamefont
  {Mehtar-Tani}}, \bibinfo {author} {\bibfnamefont {A.}~\bibnamefont
  {Soto-Ontoso}}, \ and\ \bibinfo {author} {\bibfnamefont {K.}~\bibnamefont
  {Tywoniuk}},\ }\href {\doibase 10.1103/PhysRevD.101.034004} {\bibfield
  {journal} {\bibinfo  {journal} {Phys. Rev. D}\ }\textbf {\bibinfo {volume}
  {101}},\ \bibinfo {pages} {034004} (\bibinfo {year} {2020})},\ \Eprint
  {http://arxiv.org/abs/1911.00375} {arXiv:1911.00375 [hep-ph]} \BibitemShut
  {NoStop}%
\bibitem [{\citenamefont {Caucal}\ \emph
  {et~al.}(2021{\natexlab{a}})\citenamefont {Caucal}, \citenamefont
  {Soto-Ontoso},\ and\ \citenamefont {Takacs}}]{Caucal:2021bae}%
  \BibitemOpen
  \bibfield  {author} {\bibinfo {author} {\bibfnamefont {P.}~\bibnamefont
  {Caucal}}, \bibinfo {author} {\bibfnamefont {A.}~\bibnamefont {Soto-Ontoso}},
  \ and\ \bibinfo {author} {\bibfnamefont {A.}~\bibnamefont {Takacs}},\ }\href
  {\doibase 10.1007/JHEP07(2021)020} {\bibfield  {journal} {\bibinfo  {journal}
  {JHEP}\ }\textbf {\bibinfo {volume} {07}},\ \bibinfo {pages} {020} (\bibinfo
  {year} {2021}{\natexlab{a}})},\ \Eprint {http://arxiv.org/abs/2103.06566}
  {arXiv:2103.06566 [hep-ph]} \BibitemShut {NoStop}%
\bibitem [{\citenamefont {ALICE}(2022{\natexlab{b}})}]{ALICE:2022hyz}%
  \BibitemOpen
  \bibfield  {author} {\bibinfo {author} {\bibnamefont {ALICE}},\ }\href@noop
  {} {\  (\bibinfo {year} {2022}{\natexlab{b}})},\ \Eprint
  {http://arxiv.org/abs/2204.10246} {arXiv:2204.10246 [nucl-ex]} \BibitemShut
  {NoStop}%
\bibitem [{\citenamefont {Dokshitzer}\ \emph {et~al.}(1997)\citenamefont
  {Dokshitzer}, \citenamefont {Leder}, \citenamefont {Moretti},\ and\
  \citenamefont {Webber}}]{Dokshitzer:1997in}%
  \BibitemOpen
  \bibfield  {author} {\bibinfo {author} {\bibfnamefont {Y.~L.}\ \bibnamefont
  {Dokshitzer}}, \bibinfo {author} {\bibfnamefont {G.~D.}\ \bibnamefont
  {Leder}}, \bibinfo {author} {\bibfnamefont {S.}~\bibnamefont {Moretti}}, \
  and\ \bibinfo {author} {\bibfnamefont {B.~R.}\ \bibnamefont {Webber}},\
  }\href {\doibase 10.1088/1126-6708/1997/08/001} {\bibfield  {journal}
  {\bibinfo  {journal} {JHEP}\ }\textbf {\bibinfo {volume} {08}},\ \bibinfo
  {pages} {001} (\bibinfo {year} {1997})},\ \Eprint
  {http://arxiv.org/abs/hep-ph/9707323} {arXiv:hep-ph/9707323} \BibitemShut
  {NoStop}%
\bibitem [{\citenamefont {Eskola}\ \emph {et~al.}(2017)\citenamefont {Eskola},
  \citenamefont {Paakkinen}, \citenamefont {Paukkunen},\ and\ \citenamefont
  {Salgado}}]{Eskola:2016oht}%
  \BibitemOpen
  \bibfield  {author} {\bibinfo {author} {\bibfnamefont {K.~J.}\ \bibnamefont
  {Eskola}}, \bibinfo {author} {\bibfnamefont {P.}~\bibnamefont {Paakkinen}},
  \bibinfo {author} {\bibfnamefont {H.}~\bibnamefont {Paukkunen}}, \ and\
  \bibinfo {author} {\bibfnamefont {C.~A.}\ \bibnamefont {Salgado}},\ }\href
  {\doibase 10.1140/epjc/s10052-017-4725-9} {\bibfield  {journal} {\bibinfo
  {journal} {Eur. Phys. J. C}\ }\textbf {\bibinfo {volume} {77}},\ \bibinfo
  {pages} {163} (\bibinfo {year} {2017})},\ \Eprint
  {http://arxiv.org/abs/1612.05741} {arXiv:1612.05741 [hep-ph]} \BibitemShut
  {NoStop}%
\bibitem [{\citenamefont {Casalderrey-Solana}\ \emph
  {et~al.}(2016{\natexlab{b}})\citenamefont {Casalderrey-Solana}, \citenamefont
  {Gulhan}, \citenamefont {Milhano}, \citenamefont {Pablos},\ and\
  \citenamefont {Rajagopal}}]{Casalderrey-Solana:2015vaa}%
  \BibitemOpen
  \bibfield  {author} {\bibinfo {author} {\bibfnamefont {J.}~\bibnamefont
  {Casalderrey-Solana}}, \bibinfo {author} {\bibfnamefont {D.~C.}\ \bibnamefont
  {Gulhan}}, \bibinfo {author} {\bibfnamefont {J.~G.}\ \bibnamefont {Milhano}},
  \bibinfo {author} {\bibfnamefont {D.}~\bibnamefont {Pablos}}, \ and\ \bibinfo
  {author} {\bibfnamefont {K.}~\bibnamefont {Rajagopal}},\ }\href {\doibase
  10.1007/JHEP03(2016)053} {\bibfield  {journal} {\bibinfo  {journal} {JHEP}\
  }\textbf {\bibinfo {volume} {03}},\ \bibinfo {pages} {053} (\bibinfo {year}
  {2016}{\natexlab{b}})},\ \Eprint {http://arxiv.org/abs/1508.00815}
  {arXiv:1508.00815 [hep-ph]} \BibitemShut {NoStop}%
\bibitem [{\citenamefont {Baier}\ \emph {et~al.}(2001)\citenamefont {Baier},
  \citenamefont {Dokshitzer}, \citenamefont {Mueller},\ and\ \citenamefont
  {Schiff}}]{Baier:2001yt}%
  \BibitemOpen
  \bibfield  {author} {\bibinfo {author} {\bibfnamefont {R.}~\bibnamefont
  {Baier}}, \bibinfo {author} {\bibfnamefont {Y.~L.}\ \bibnamefont
  {Dokshitzer}}, \bibinfo {author} {\bibfnamefont {A.~H.}\ \bibnamefont
  {Mueller}}, \ and\ \bibinfo {author} {\bibfnamefont {D.}~\bibnamefont
  {Schiff}},\ }\href {\doibase 10.1088/1126-6708/2001/09/033} {\bibfield
  {journal} {\bibinfo  {journal} {JHEP}\ }\textbf {\bibinfo {volume} {09}},\
  \bibinfo {pages} {033} (\bibinfo {year} {2001})},\ \Eprint
  {http://arxiv.org/abs/hep-ph/0106347} {arXiv:hep-ph/0106347} \BibitemShut
  {NoStop}%
\bibitem [{\citenamefont {D'Eramo}\ \emph {et~al.}(2019)\citenamefont
  {D'Eramo}, \citenamefont {Rajagopal},\ and\ \citenamefont
  {Yin}}]{DEramo:2018eoy}%
  \BibitemOpen
  \bibfield  {author} {\bibinfo {author} {\bibfnamefont {F.}~\bibnamefont
  {D'Eramo}}, \bibinfo {author} {\bibfnamefont {K.}~\bibnamefont {Rajagopal}},
  \ and\ \bibinfo {author} {\bibfnamefont {Y.}~\bibnamefont {Yin}},\ }\href
  {\doibase 10.1007/JHEP01(2019)172} {\bibfield  {journal} {\bibinfo  {journal}
  {JHEP}\ }\textbf {\bibinfo {volume} {01}},\ \bibinfo {pages} {172} (\bibinfo
  {year} {2019})},\ \Eprint {http://arxiv.org/abs/1808.03250} {arXiv:1808.03250
  [hep-ph]} \BibitemShut {NoStop}%
\bibitem [{\citenamefont {Ehlers}(2021)}]{Ehlers:2020piz}%
  \BibitemOpen
  \bibfield  {author} {\bibinfo {author} {\bibfnamefont {R.}~\bibnamefont
  {Ehlers}} (\bibinfo {collaboration} {ALICE}),\ }\href {\doibase
  10.22323/1.387.0146} {\bibfield  {journal} {\bibinfo  {journal} {PoS}\
  }\textbf {\bibinfo {volume} {HardProbes2020}},\ \bibinfo {pages} {146}
  (\bibinfo {year} {2021})},\ \Eprint {http://arxiv.org/abs/2009.12247}
  {arXiv:2009.12247 [nucl-ex]} \BibitemShut {NoStop}%
\bibitem [{\citenamefont {Hulcher}\ \emph {et~al.}(2022)\citenamefont
  {Hulcher}, \citenamefont {Pablos},\ and\ \citenamefont
  {Rajagopal}}]{Hulcher:2022kmn}%
  \BibitemOpen
  \bibfield  {author} {\bibinfo {author} {\bibfnamefont {Z.}~\bibnamefont
  {Hulcher}}, \bibinfo {author} {\bibfnamefont {D.}~\bibnamefont {Pablos}}, \
  and\ \bibinfo {author} {\bibfnamefont {K.}~\bibnamefont {Rajagopal}},\ }in\
  \href@noop {} {\emph {\bibinfo {booktitle} {{29th International Conference on
  Ultra-relativistic Nucleus-Nucleus Collisions}}}}\ (\bibinfo {year} {2022})\
  \Eprint {http://arxiv.org/abs/2208.13593} {arXiv:2208.13593 [hep-ph]}
  \BibitemShut {NoStop}%
\bibitem [{\citenamefont {Mehtar-Tani}\ and\ \citenamefont
  {Tywoniuk}(2018)}]{Mehtar-Tani:2017web}%
  \BibitemOpen
  \bibfield  {author} {\bibinfo {author} {\bibfnamefont {Y.}~\bibnamefont
  {Mehtar-Tani}}\ and\ \bibinfo {author} {\bibfnamefont {K.}~\bibnamefont
  {Tywoniuk}},\ }\href {\doibase 10.1103/PhysRevD.98.051501} {\bibfield
  {journal} {\bibinfo  {journal} {Phys. Rev. D}\ }\textbf {\bibinfo {volume}
  {98}},\ \bibinfo {pages} {051501} (\bibinfo {year} {2018})},\ \Eprint
  {http://arxiv.org/abs/1707.07361} {arXiv:1707.07361 [hep-ph]} \BibitemShut
  {NoStop}%
\bibitem [{\citenamefont {Barata}\ \emph {et~al.}(2022)\citenamefont {Barata},
  \citenamefont {Caucal}, \citenamefont {Isaaksen}, \citenamefont
  {Soto-Ontoso}, \citenamefont {Takacs},\ and\ \citenamefont
  {Tywoniuk}}]{dyg-ioe:paper}%
  \BibitemOpen
  \bibfield  {author} {\bibinfo {author} {\bibfnamefont {J.}~\bibnamefont
  {Barata}}, \bibinfo {author} {\bibfnamefont {P.}~\bibnamefont {Caucal}},
  \bibinfo {author} {\bibfnamefont {J.}~\bibnamefont {Isaaksen}}, \bibinfo
  {author} {\bibfnamefont {A.}~\bibnamefont {Soto-Ontoso}}, \bibinfo {author}
  {\bibfnamefont {A.}~\bibnamefont {Takacs}}, \ and\ \bibinfo {author}
  {\bibfnamefont {K.}~\bibnamefont {Tywoniuk}},\ }\href@noop {} {\bibfield
  {journal} {\bibinfo  {journal} {in preparation}\ } (\bibinfo {year}
  {2022})}\BibitemShut {NoStop}%
\bibitem [{\citenamefont {Chesler}\ and\ \citenamefont
  {Rajagopal}(2014)}]{Chesler:2014jva}%
  \BibitemOpen
  \bibfield  {author} {\bibinfo {author} {\bibfnamefont {P.~M.}\ \bibnamefont
  {Chesler}}\ and\ \bibinfo {author} {\bibfnamefont {K.}~\bibnamefont
  {Rajagopal}},\ }\href {\doibase 10.1103/PhysRevD.90.025033} {\bibfield
  {journal} {\bibinfo  {journal} {Phys. Rev. D}\ }\textbf {\bibinfo {volume}
  {90}},\ \bibinfo {pages} {025033} (\bibinfo {year} {2014})},\ \Eprint
  {http://arxiv.org/abs/1402.6756} {arXiv:1402.6756 [hep-th]} \BibitemShut
  {NoStop}%
\bibitem [{\citenamefont {Chesler}\ and\ \citenamefont
  {Rajagopal}(2016)}]{Chesler:2015nqz}%
  \BibitemOpen
  \bibfield  {author} {\bibinfo {author} {\bibfnamefont {P.~M.}\ \bibnamefont
  {Chesler}}\ and\ \bibinfo {author} {\bibfnamefont {K.}~\bibnamefont
  {Rajagopal}},\ }\href {\doibase 10.1007/JHEP05(2016)098} {\bibfield
  {journal} {\bibinfo  {journal} {JHEP}\ }\textbf {\bibinfo {volume} {05}},\
  \bibinfo {pages} {098} (\bibinfo {year} {2016})},\ \Eprint
  {http://arxiv.org/abs/1511.07567} {arXiv:1511.07567 [hep-th]} \BibitemShut
  {NoStop}%
\bibitem [{\citenamefont {Casalderrey-Solana}\ \emph
  {et~al.}(2019)\citenamefont {Casalderrey-Solana}, \citenamefont {Hulcher},
  \citenamefont {Milhano}, \citenamefont {Pablos},\ and\ \citenamefont
  {Rajagopal}}]{Casalderrey-Solana:2018wrw}%
  \BibitemOpen
  \bibfield  {author} {\bibinfo {author} {\bibfnamefont {J.}~\bibnamefont
  {Casalderrey-Solana}}, \bibinfo {author} {\bibfnamefont {Z.}~\bibnamefont
  {Hulcher}}, \bibinfo {author} {\bibfnamefont {G.}~\bibnamefont {Milhano}},
  \bibinfo {author} {\bibfnamefont {D.}~\bibnamefont {Pablos}}, \ and\ \bibinfo
  {author} {\bibfnamefont {K.}~\bibnamefont {Rajagopal}},\ }\href {\doibase
  10.1103/PhysRevC.99.051901} {\bibfield  {journal} {\bibinfo  {journal} {Phys.
  Rev. C}\ }\textbf {\bibinfo {volume} {99}},\ \bibinfo {pages} {051901}
  (\bibinfo {year} {2019})},\ \Eprint {http://arxiv.org/abs/1808.07386}
  {arXiv:1808.07386 [hep-ph]} \BibitemShut {NoStop}%
\bibitem [{\citenamefont {Casalderrey-Solana}\ \emph
  {et~al.}(2017)\citenamefont {Casalderrey-Solana}, \citenamefont {Gulhan},
  \citenamefont {Milhano}, \citenamefont {Pablos},\ and\ \citenamefont
  {Rajagopal}}]{Casalderrey-Solana:2016jvj}%
  \BibitemOpen
  \bibfield  {author} {\bibinfo {author} {\bibfnamefont {J.}~\bibnamefont
  {Casalderrey-Solana}}, \bibinfo {author} {\bibfnamefont {D.}~\bibnamefont
  {Gulhan}}, \bibinfo {author} {\bibfnamefont {G.}~\bibnamefont {Milhano}},
  \bibinfo {author} {\bibfnamefont {D.}~\bibnamefont {Pablos}}, \ and\ \bibinfo
  {author} {\bibfnamefont {K.}~\bibnamefont {Rajagopal}},\ }\href {\doibase
  10.1007/JHEP03(2017)135} {\bibfield  {journal} {\bibinfo  {journal} {JHEP}\
  }\textbf {\bibinfo {volume} {03}},\ \bibinfo {pages} {135} (\bibinfo {year}
  {2017})},\ \Eprint {http://arxiv.org/abs/1609.05842} {arXiv:1609.05842
  [hep-ph]} \BibitemShut {NoStop}%
\bibitem [{\citenamefont {Casalderrey-Solana}\ \emph
  {et~al.}(2021{\natexlab{a}})\citenamefont {Casalderrey-Solana}, \citenamefont
  {Milhano}, \citenamefont {Pablos}, \citenamefont {Rajagopal},\ and\
  \citenamefont {Yao}}]{Casalderrey-Solana:2020rsj}%
  \BibitemOpen
  \bibfield  {author} {\bibinfo {author} {\bibfnamefont {J.}~\bibnamefont
  {Casalderrey-Solana}}, \bibinfo {author} {\bibfnamefont {J.~G.}\ \bibnamefont
  {Milhano}}, \bibinfo {author} {\bibfnamefont {D.}~\bibnamefont {Pablos}},
  \bibinfo {author} {\bibfnamefont {K.}~\bibnamefont {Rajagopal}}, \ and\
  \bibinfo {author} {\bibfnamefont {X.}~\bibnamefont {Yao}},\ }\href {\doibase
  10.1007/JHEP05(2021)230} {\bibfield  {journal} {\bibinfo  {journal} {JHEP}\
  }\textbf {\bibinfo {volume} {05}},\ \bibinfo {pages} {230} (\bibinfo {year}
  {2021}{\natexlab{a}})},\ \Eprint {http://arxiv.org/abs/2010.01140}
  {arXiv:2010.01140 [hep-ph]} \BibitemShut {NoStop}%
\bibitem [{\citenamefont {Tachibana}\ \emph {et~al.}(2017)\citenamefont
  {Tachibana}, \citenamefont {Chang},\ and\ \citenamefont
  {Qin}}]{Tachibana:2017syd}%
  \BibitemOpen
  \bibfield  {author} {\bibinfo {author} {\bibfnamefont {Y.}~\bibnamefont
  {Tachibana}}, \bibinfo {author} {\bibfnamefont {N.-B.}\ \bibnamefont
  {Chang}}, \ and\ \bibinfo {author} {\bibfnamefont {G.-Y.}\ \bibnamefont
  {Qin}},\ }\href {\doibase 10.1103/PhysRevC.95.044909} {\bibfield  {journal}
  {\bibinfo  {journal} {Phys. Rev. C}\ }\textbf {\bibinfo {volume} {95}},\
  \bibinfo {pages} {044909} (\bibinfo {year} {2017})},\ \Eprint
  {http://arxiv.org/abs/1701.07951} {arXiv:1701.07951 [nucl-th]} \BibitemShut
  {NoStop}%
\bibitem [{\citenamefont {Pablos}(2020)}]{Pablos:2019ngg}%
  \BibitemOpen
  \bibfield  {author} {\bibinfo {author} {\bibfnamefont {D.}~\bibnamefont
  {Pablos}},\ }\href {\doibase 10.1103/PhysRevLett.124.052301} {\bibfield
  {journal} {\bibinfo  {journal} {Phys. Rev. Lett.}\ }\textbf {\bibinfo
  {volume} {124}},\ \bibinfo {pages} {052301} (\bibinfo {year} {2020})},\
  \Eprint {http://arxiv.org/abs/1907.12301} {arXiv:1907.12301 [hep-ph]}
  \BibitemShut {NoStop}%
\bibitem [{\citenamefont {Pablos}(2021)}]{Pablos:2020wnp}%
  \BibitemOpen
  \bibfield  {author} {\bibinfo {author} {\bibfnamefont {D.}~\bibnamefont
  {Pablos}},\ }\href {\doibase 10.22323/1.387.0147} {\bibfield  {journal}
  {\bibinfo  {journal} {PoS}\ }\textbf {\bibinfo {volume} {HardProbes2020}},\
  \bibinfo {pages} {147} (\bibinfo {year} {2021})},\ \Eprint
  {http://arxiv.org/abs/2009.02202} {arXiv:2009.02202 [hep-ph]} \BibitemShut
  {NoStop}%
\bibitem [{\citenamefont {Mehtar-Tani}\ \emph {et~al.}(2021)\citenamefont
  {Mehtar-Tani}, \citenamefont {Pablos},\ and\ \citenamefont
  {Tywoniuk}}]{Mehtar-Tani:2021fud}%
  \BibitemOpen
  \bibfield  {author} {\bibinfo {author} {\bibfnamefont {Y.}~\bibnamefont
  {Mehtar-Tani}}, \bibinfo {author} {\bibfnamefont {D.}~\bibnamefont {Pablos}},
  \ and\ \bibinfo {author} {\bibfnamefont {K.}~\bibnamefont {Tywoniuk}},\
  }\href {\doibase 10.1103/PhysRevLett.127.252301} {\bibfield  {journal}
  {\bibinfo  {journal} {Phys. Rev. Lett.}\ }\textbf {\bibinfo {volume} {127}},\
  \bibinfo {pages} {252301} (\bibinfo {year} {2021})},\ \Eprint
  {http://arxiv.org/abs/2101.01742} {arXiv:2101.01742 [hep-ph]} \BibitemShut
  {NoStop}%
\bibitem [{\citenamefont {Casalderrey-Solana}\ \emph
  {et~al.}(2021{\natexlab{b}})\citenamefont {Casalderrey-Solana}, \citenamefont
  {Milhano}, \citenamefont {Pablos},\ and\ \citenamefont
  {Rajagopal}}]{Casalderrey-Solana:2020jbx}%
  \BibitemOpen
  \bibfield  {author} {\bibinfo {author} {\bibfnamefont {J.}~\bibnamefont
  {Casalderrey-Solana}}, \bibinfo {author} {\bibfnamefont {G.}~\bibnamefont
  {Milhano}}, \bibinfo {author} {\bibfnamefont {D.}~\bibnamefont {Pablos}}, \
  and\ \bibinfo {author} {\bibfnamefont {K.}~\bibnamefont {Rajagopal}},\ }\href
  {\doibase 10.1016/j.nuclphysa.2020.121904} {\bibfield  {journal} {\bibinfo
  {journal} {Nucl. Phys. A}\ }\textbf {\bibinfo {volume} {1005}},\ \bibinfo
  {pages} {121904} (\bibinfo {year} {2021}{\natexlab{b}})},\ \Eprint
  {http://arxiv.org/abs/2002.09193} {arXiv:2002.09193 [hep-ph]} \BibitemShut
  {NoStop}%
\bibitem [{\citenamefont {Rajagopal}\ \emph {et~al.}(2016)\citenamefont
  {Rajagopal}, \citenamefont {Sadofyev},\ and\ \citenamefont {van~der
  Schee}}]{Rajagopal:2016uip}%
  \BibitemOpen
  \bibfield  {author} {\bibinfo {author} {\bibfnamefont {K.}~\bibnamefont
  {Rajagopal}}, \bibinfo {author} {\bibfnamefont {A.~V.}\ \bibnamefont
  {Sadofyev}}, \ and\ \bibinfo {author} {\bibfnamefont {W.}~\bibnamefont
  {van~der Schee}},\ }\href {\doibase 10.1103/PhysRevLett.116.211603}
  {\bibfield  {journal} {\bibinfo  {journal} {Phys. Rev. Lett.}\ }\textbf
  {\bibinfo {volume} {116}},\ \bibinfo {pages} {211603} (\bibinfo {year}
  {2016})},\ \Eprint {http://arxiv.org/abs/1602.04187} {arXiv:1602.04187
  [nucl-th]} \BibitemShut {NoStop}%
\bibitem [{\citenamefont {Brewer}\ \emph {et~al.}(2022)\citenamefont {Brewer},
  \citenamefont {Brodsky},\ and\ \citenamefont {Rajagopal}}]{Brewer:2021hmh}%
  \BibitemOpen
  \bibfield  {author} {\bibinfo {author} {\bibfnamefont {J.}~\bibnamefont
  {Brewer}}, \bibinfo {author} {\bibfnamefont {Q.}~\bibnamefont {Brodsky}}, \
  and\ \bibinfo {author} {\bibfnamefont {K.}~\bibnamefont {Rajagopal}},\ }\href
  {\doibase 10.1007/JHEP02(2022)175} {\bibfield  {journal} {\bibinfo  {journal}
  {JHEP}\ }\textbf {\bibinfo {volume} {02}},\ \bibinfo {pages} {175} (\bibinfo
  {year} {2022})},\ \Eprint {http://arxiv.org/abs/2110.13159} {arXiv:2110.13159
  [hep-ph]} \BibitemShut {NoStop}%
\bibitem [{\citenamefont {Du}\ \emph {et~al.}(2020)\citenamefont {Du},
  \citenamefont {Pablos},\ and\ \citenamefont {Tywoniuk}}]{Du:2020pmp}%
  \BibitemOpen
  \bibfield  {author} {\bibinfo {author} {\bibfnamefont {Y.-L.}\ \bibnamefont
  {Du}}, \bibinfo {author} {\bibfnamefont {D.}~\bibnamefont {Pablos}}, \ and\
  \bibinfo {author} {\bibfnamefont {K.}~\bibnamefont {Tywoniuk}},\ }\href
  {\doibase 10.1007/JHEP03(2021)206} {\bibfield  {journal} {\bibinfo  {journal}
  {JHEP}\ }\textbf {\bibinfo {volume} {21}},\ \bibinfo {pages} {206} (\bibinfo
  {year} {2020})},\ \Eprint {http://arxiv.org/abs/2012.07797} {arXiv:2012.07797
  [hep-ph]} \BibitemShut {NoStop}%
\bibitem [{\citenamefont {Milhano}\ \emph {et~al.}(2018)\citenamefont
  {Milhano}, \citenamefont {Wiedemann},\ and\ \citenamefont
  {Zapp}}]{Milhano:2017nzm}%
  \BibitemOpen
  \bibfield  {author} {\bibinfo {author} {\bibfnamefont {G.}~\bibnamefont
  {Milhano}}, \bibinfo {author} {\bibfnamefont {U.~A.}\ \bibnamefont
  {Wiedemann}}, \ and\ \bibinfo {author} {\bibfnamefont {K.~C.}\ \bibnamefont
  {Zapp}},\ }\href {\doibase 10.1016/j.physletb.2018.01.029} {\bibfield
  {journal} {\bibinfo  {journal} {Phys. Lett. B}\ }\textbf {\bibinfo {volume}
  {779}},\ \bibinfo {pages} {409} (\bibinfo {year} {2018})},\ \Eprint
  {http://arxiv.org/abs/1707.04142} {arXiv:1707.04142 [hep-ph]} \BibitemShut
  {NoStop}%
\bibitem [{\citenamefont {Gubser}\ \emph {et~al.}(2008)\citenamefont {Gubser},
  \citenamefont {Gulotta}, \citenamefont {Pufu},\ and\ \citenamefont
  {Rocha}}]{Gubser:2008as}%
  \BibitemOpen
  \bibfield  {author} {\bibinfo {author} {\bibfnamefont {S.~S.}\ \bibnamefont
  {Gubser}}, \bibinfo {author} {\bibfnamefont {D.~R.}\ \bibnamefont {Gulotta}},
  \bibinfo {author} {\bibfnamefont {S.~S.}\ \bibnamefont {Pufu}}, \ and\
  \bibinfo {author} {\bibfnamefont {F.~D.}\ \bibnamefont {Rocha}},\ }\href
  {\doibase 10.1088/1126-6708/2008/10/052} {\bibfield  {journal} {\bibinfo
  {journal} {JHEP}\ }\textbf {\bibinfo {volume} {10}},\ \bibinfo {pages} {052}
  (\bibinfo {year} {2008})},\ \Eprint {http://arxiv.org/abs/0803.1470}
  {arXiv:0803.1470 [hep-th]} \BibitemShut {NoStop}%
\bibitem [{\citenamefont {He}\ \emph {et~al.}(2019)\citenamefont {He},
  \citenamefont {Cao}, \citenamefont {Chen}, \citenamefont {Luo}, \citenamefont
  {Pang},\ and\ \citenamefont {Wang}}]{He:2018xjv}%
  \BibitemOpen
  \bibfield  {author} {\bibinfo {author} {\bibfnamefont {Y.}~\bibnamefont
  {He}}, \bibinfo {author} {\bibfnamefont {S.}~\bibnamefont {Cao}}, \bibinfo
  {author} {\bibfnamefont {W.}~\bibnamefont {Chen}}, \bibinfo {author}
  {\bibfnamefont {T.}~\bibnamefont {Luo}}, \bibinfo {author} {\bibfnamefont
  {L.-G.}\ \bibnamefont {Pang}}, \ and\ \bibinfo {author} {\bibfnamefont
  {X.-N.}\ \bibnamefont {Wang}},\ }\href {\doibase 10.1103/PhysRevC.99.054911}
  {\bibfield  {journal} {\bibinfo  {journal} {Phys. Rev. C}\ }\textbf {\bibinfo
  {volume} {99}},\ \bibinfo {pages} {054911} (\bibinfo {year} {2019})},\
  \Eprint {http://arxiv.org/abs/1809.02525} {arXiv:1809.02525 [nucl-th]}
  \BibitemShut {NoStop}%
\bibitem [{\citenamefont {Caucal}\ \emph
  {et~al.}(2021{\natexlab{b}})\citenamefont {Caucal}, \citenamefont {Iancu},\
  and\ \citenamefont {Soyez}}]{Caucal:2020uic}%
  \BibitemOpen
  \bibfield  {author} {\bibinfo {author} {\bibfnamefont {P.}~\bibnamefont
  {Caucal}}, \bibinfo {author} {\bibfnamefont {E.}~\bibnamefont {Iancu}}, \
  and\ \bibinfo {author} {\bibfnamefont {G.}~\bibnamefont {Soyez}},\ }\href
  {\doibase 10.1007/JHEP04(2021)209} {\bibfield  {journal} {\bibinfo  {journal}
  {JHEP}\ }\textbf {\bibinfo {volume} {04}},\ \bibinfo {pages} {209} (\bibinfo
  {year} {2021}{\natexlab{b}})},\ \Eprint {http://arxiv.org/abs/2012.01457}
  {arXiv:2012.01457 [hep-ph]} \BibitemShut {NoStop}%
\bibitem [{\citenamefont {Takacs}\ and\ \citenamefont
  {Tywoniuk}(2021)}]{Takacs:2021bpv}%
  \BibitemOpen
  \bibfield  {author} {\bibinfo {author} {\bibfnamefont {A.}~\bibnamefont
  {Takacs}}\ and\ \bibinfo {author} {\bibfnamefont {K.}~\bibnamefont
  {Tywoniuk}},\ }\href {\doibase 10.1007/JHEP10(2021)038} {\bibfield  {journal}
  {\bibinfo  {journal} {JHEP}\ }\textbf {\bibinfo {volume} {10}},\ \bibinfo
  {pages} {038} (\bibinfo {year} {2021})},\ \Eprint
  {http://arxiv.org/abs/2103.14676} {arXiv:2103.14676 [hep-ph]} \BibitemShut
  {NoStop}%
\bibitem [{\citenamefont {Chatrchyan}\ \emph {et~al.}(2014)\citenamefont
  {Chatrchyan} \emph {et~al.}}]{CMS:2013qak}%
  \BibitemOpen
  \bibfield  {author} {\bibinfo {author} {\bibfnamefont {S.}~\bibnamefont
  {Chatrchyan}} \emph {et~al.} (\bibinfo {collaboration} {CMS}),\ }\href
  {\doibase 10.1103/PhysRevLett.113.132301} {\bibfield  {journal} {\bibinfo
  {journal} {Phys. Rev. Lett.}\ }\textbf {\bibinfo {volume} {113}},\ \bibinfo
  {pages} {132301} (\bibinfo {year} {2014})},\ \bibinfo {note} {[Erratum:
  Phys.Rev.Lett. 115, 029903 (2015)]},\ \Eprint
  {http://arxiv.org/abs/1312.4198} {arXiv:1312.4198 [nucl-ex]} \BibitemShut
  {NoStop}%
\bibitem [{\citenamefont {ALICE}(2022{\natexlab{c}})}]{ALICE:2022phr}%
  \BibitemOpen
  \bibfield  {author} {\bibinfo {author} {\bibnamefont {ALICE}},\ }\href@noop
  {} {\  (\bibinfo {year} {2022}{\natexlab{c}})},\ \Eprint
  {http://arxiv.org/abs/2208.04857} {arXiv:2208.04857 [nucl-ex]} \BibitemShut
  {NoStop}%
\bibitem [{\citenamefont {Li}\ and\ \citenamefont {Vitev}(2019)}]{Li:2017wwc}%
  \BibitemOpen
  \bibfield  {author} {\bibinfo {author} {\bibfnamefont {H.~T.}\ \bibnamefont
  {Li}}\ and\ \bibinfo {author} {\bibfnamefont {I.}~\bibnamefont {Vitev}},\
  }\href {\doibase 10.1016/j.physletb.2019.04.052} {\bibfield  {journal}
  {\bibinfo  {journal} {Phys. Lett. B}\ }\textbf {\bibinfo {volume} {793}},\
  \bibinfo {pages} {259} (\bibinfo {year} {2019})},\ \Eprint
  {http://arxiv.org/abs/1801.00008} {arXiv:1801.00008 [hep-ph]} \BibitemShut
  {NoStop}%
\bibitem [{\citenamefont {Gallicchio}\ and\ \citenamefont
  {Schwartz}(2011)}]{Gallicchio:2011xc}%
  \BibitemOpen
  \bibfield  {author} {\bibinfo {author} {\bibfnamefont {J.}~\bibnamefont
  {Gallicchio}}\ and\ \bibinfo {author} {\bibfnamefont {M.~D.}\ \bibnamefont
  {Schwartz}},\ }\href {\doibase 10.1007/JHEP10(2011)103} {\bibfield  {journal}
  {\bibinfo  {journal} {JHEP}\ }\textbf {\bibinfo {volume} {10}},\ \bibinfo
  {pages} {103} (\bibinfo {year} {2011})},\ \Eprint
  {http://arxiv.org/abs/1104.1175} {arXiv:1104.1175 [hep-ph]} \BibitemShut
  {NoStop}%
\bibitem [{\citenamefont {Wang}\ and\ \citenamefont
  {Zhu}(2013)}]{Wang:2013cia}%
  \BibitemOpen
  \bibfield  {author} {\bibinfo {author} {\bibfnamefont {X.-N.}\ \bibnamefont
  {Wang}}\ and\ \bibinfo {author} {\bibfnamefont {Y.}~\bibnamefont {Zhu}},\
  }\href {\doibase 10.1103/PhysRevLett.111.062301} {\bibfield  {journal}
  {\bibinfo  {journal} {Phys. Rev. Lett.}\ }\textbf {\bibinfo {volume} {111}},\
  \bibinfo {pages} {062301} (\bibinfo {year} {2013})},\ \Eprint
  {http://arxiv.org/abs/1302.5874} {arXiv:1302.5874 [hep-ph]} \BibitemShut
  {NoStop}%
\bibitem [{\citenamefont {Aaboud}\ \emph
  {et~al.}(2019{\natexlab{b}})\citenamefont {Aaboud} \emph
  {et~al.}}]{ATLAS:2019dsv}%
  \BibitemOpen
  \bibfield  {author} {\bibinfo {author} {\bibfnamefont {M.}~\bibnamefont
  {Aaboud}} \emph {et~al.} (\bibinfo {collaboration} {ATLAS}),\ }\href
  {\doibase 10.1103/PhysRevLett.123.042001} {\bibfield  {journal} {\bibinfo
  {journal} {Phys. Rev. Lett.}\ }\textbf {\bibinfo {volume} {123}},\ \bibinfo
  {pages} {042001} (\bibinfo {year} {2019}{\natexlab{b}})},\ \Eprint
  {http://arxiv.org/abs/1902.10007} {arXiv:1902.10007 [nucl-ex]} \BibitemShut
  {NoStop}%
\bibitem [{\citenamefont {Sirunyan}\ \emph
  {et~al.}(2018{\natexlab{b}})\citenamefont {Sirunyan} \emph
  {et~al.}}]{CMS:2018mqn}%
  \BibitemOpen
  \bibfield  {author} {\bibinfo {author} {\bibfnamefont {A.~M.}\ \bibnamefont
  {Sirunyan}} \emph {et~al.} (\bibinfo {collaboration} {CMS}),\ }\href
  {\doibase 10.1103/PhysRevLett.121.242301} {\bibfield  {journal} {\bibinfo
  {journal} {Phys. Rev. Lett.}\ }\textbf {\bibinfo {volume} {121}},\ \bibinfo
  {pages} {242301} (\bibinfo {year} {2018}{\natexlab{b}})},\ \Eprint
  {http://arxiv.org/abs/1801.04895} {arXiv:1801.04895 [hep-ex]} \BibitemShut
  {NoStop}%
\bibitem [{\citenamefont {Kang}\ \emph {et~al.}(2017)\citenamefont {Kang},
  \citenamefont {Vitev},\ and\ \citenamefont {Xing}}]{Kang:2017xnc}%
  \BibitemOpen
  \bibfield  {author} {\bibinfo {author} {\bibfnamefont {Z.-B.}\ \bibnamefont
  {Kang}}, \bibinfo {author} {\bibfnamefont {I.}~\bibnamefont {Vitev}}, \ and\
  \bibinfo {author} {\bibfnamefont {H.}~\bibnamefont {Xing}},\ }\href {\doibase
  10.1103/PhysRevC.96.014912} {\bibfield  {journal} {\bibinfo  {journal} {Phys.
  Rev. C}\ }\textbf {\bibinfo {volume} {96}},\ \bibinfo {pages} {014912}
  (\bibinfo {year} {2017})},\ \Eprint {http://arxiv.org/abs/1702.07276}
  {arXiv:1702.07276 [hep-ph]} \BibitemShut {NoStop}%
\bibitem [{\citenamefont {Yang}\ \emph {et~al.}(2021)\citenamefont {Yang},
  \citenamefont {Chen}, \citenamefont {He}, \citenamefont {Ke}, \citenamefont
  {Pang},\ and\ \citenamefont {Wang}}]{Yang:2021qtl}%
  \BibitemOpen
  \bibfield  {author} {\bibinfo {author} {\bibfnamefont {Z.}~\bibnamefont
  {Yang}}, \bibinfo {author} {\bibfnamefont {W.}~\bibnamefont {Chen}}, \bibinfo
  {author} {\bibfnamefont {Y.}~\bibnamefont {He}}, \bibinfo {author}
  {\bibfnamefont {W.}~\bibnamefont {Ke}}, \bibinfo {author} {\bibfnamefont
  {L.}~\bibnamefont {Pang}}, \ and\ \bibinfo {author} {\bibfnamefont {X.-N.}\
  \bibnamefont {Wang}},\ }\href {\doibase 10.1103/PhysRevLett.127.082301}
  {\bibfield  {journal} {\bibinfo  {journal} {Phys. Rev. Lett.}\ }\textbf
  {\bibinfo {volume} {127}},\ \bibinfo {pages} {082301} (\bibinfo {year}
  {2021})},\ \Eprint {http://arxiv.org/abs/2101.05422} {arXiv:2101.05422
  [hep-ph]} \BibitemShut {NoStop}%
\bibitem [{\citenamefont {Attems}\ \emph {et~al.}(2022)\citenamefont {Attems},
  \citenamefont {Brewer}, \citenamefont {Innocenti}, \citenamefont
  {Mazeliauskas}, \citenamefont {Park}, \citenamefont {van~der Schee},\ and\
  \citenamefont {Wiedemann}}]{Attems:2022otp}%
  \BibitemOpen
  \bibfield  {author} {\bibinfo {author} {\bibfnamefont {M.}~\bibnamefont
  {Attems}}, \bibinfo {author} {\bibfnamefont {J.}~\bibnamefont {Brewer}},
  \bibinfo {author} {\bibfnamefont {G.~M.}\ \bibnamefont {Innocenti}}, \bibinfo
  {author} {\bibfnamefont {A.}~\bibnamefont {Mazeliauskas}}, \bibinfo {author}
  {\bibfnamefont {S.}~\bibnamefont {Park}}, \bibinfo {author} {\bibfnamefont
  {W.}~\bibnamefont {van~der Schee}}, \ and\ \bibinfo {author} {\bibfnamefont
  {U.}~\bibnamefont {Wiedemann}},\ }\href@noop {} {\  (\bibinfo {year}
  {2022})},\ \Eprint {http://arxiv.org/abs/2209.13600} {arXiv:2209.13600
  [hep-ph]} \BibitemShut {NoStop}%
\bibitem [{\citenamefont {Sirunyan}\ \emph {et~al.}(2022)\citenamefont
  {Sirunyan} \emph {et~al.}}]{CMS:2021otx}%
  \BibitemOpen
  \bibfield  {author} {\bibinfo {author} {\bibfnamefont {A.~M.}\ \bibnamefont
  {Sirunyan}} \emph {et~al.} (\bibinfo {collaboration} {CMS}),\ }\href
  {\doibase 10.1103/PhysRevLett.128.122301} {\bibfield  {journal} {\bibinfo
  {journal} {Phys. Rev. Lett.}\ }\textbf {\bibinfo {volume} {128}},\ \bibinfo
  {pages} {122301} (\bibinfo {year} {2022})},\ \Eprint
  {http://arxiv.org/abs/2103.04377} {arXiv:2103.04377 [hep-ex]} \BibitemShut
  {NoStop}%
\end{thebibliography}%

\end{document}